\documentstyle[12pt,bezier]{article}
\def\FKS{F_{\rm KS}}

\def\setfonts{%
  \font\frbig=eufm10 scaled\magstep1
  \font\frscr=eufm10
  \font\frscrscr=eufm8
  \newfam\frfam
  \textfont\frfam=\frbig
  \scriptfont\frfam=\frscr
  \scriptscriptfont\frfam=\frscrscr
  \def\fr{\fam\frfam}

  \font\openbig=msbm10 scaled\magstep1
  \font\openscr=msbm10
  \font\openscrscr=msbm8
  \newfam\openfam
  \textfont\openfam=\openbig
  \scriptfont\openfam=\openscr
  \scriptscriptfont\openfam=\openscrscr
  \def\open{\fam\openfam}

  \font\ssfbig=cmss10 scaled\magstep1
  \font\ssfscr=cmss10 
  \font\ssfscrscr=cmss8
  \newfam\ssffam
  \textfont\ssffam=\ssfbig
  \scriptfont\ssffam=\ssfscr
  \scriptscriptfont\ssffam=\ssfscrscr
  \def\ssf{\fam\ssffam}
  }

\makeatletter

\def\tableofcontents{
  \@starttoc{toc}}
\@addtoreset{equation}{section}
\newdimen\normalarrayskip
\newdimen\minarrayskip
\normalarrayskip\baselineskip
\minarrayskip\jot
\newif\ifold \oldtrue \def\new{\oldfalse}
\def\arraymode{\ifold\relax\else\displaystyle\fi}

\def\@arrayskip{\ifold\baselineskip\z@\lineskip\z@
  \else
  \baselineskip\minarrayskip\lineskip2\minarrayskip\fi}
\def\@arrayclassz{\ifcase \@lastchclass \@acolampacol \or
  \@ampacol \or \or \or \@addamp \or
  \@acolampacol \or \@firstampfalse \@acol \fi
  \edef\@preamble{\@preamble
    \ifcase \@chnum
    \hfil$\relax\arraymode\@sharp$\hfil
    \or $\relax\arraymode\@sharp$\hfil
    \or \hfil$\relax\arraymode\@sharp$\fi}}
\def\@array[#1]#2{\setbox\@arstrutbox=\hbox{\vrule
    height\arraystretch \ht\strutbox
    depth\arraystretch \dp\strutbox
    width\z@}\@mkpream{#2}\edef\@preamble{\halign \noexpand\@halignto
    \bgroup \tabskip\z@ \@arstrut \@preamble \tabskip\z@ \cr}%
  \let\@startpbox\@@startpbox \let\@endpbox\@@endpbox
  \if #1t\vtop \else \if#1b\vbox \else \vcenter \fi\fi
  \bgroup \let\par\relax
  \let\@sharp##\let\protect\relax
  \@arrayskip\@preamble}

\makeatother

\setfonts

\def\lvm{\leavevmode\hbox to\parindent{\hfill}}
\def\req#1{\/{\rm(\ref{#1})}}

\def\commut#1#2{\left[{#1},\,{#2}\right]}
\def\jrc{j_{\rm rc}}
\def\Lambdarc{\Lambda_{\rm rc}}
\def\jcc{j_{\rm cc}}
\def\Lambdacc{\Lambda_{\rm cc}}

\def\rrangle{\rangle\kern-2pt\rangle} \def\mmid{{\mid}\kern-1pt{\mid}}
\def\kket#1{\mmid #1\rrangle}

\def\BE{\begin{equation}}
  \def\EE{\end{equation} }
\def\BA{\begin{array}} 
  \def\EA{\end{array}}
\def\L{\left}
  \def\R{\right}

\def\bar{\overline}
\def\frac#1#2{\mathchoice{%
    {\textstyle{{#1}\over{#2}}}}{{#1\over#2}}{{#1\over#2}}{{#1\over#2}}}
\def\ket#1{\mathchoice{%
    {\left|{#1}\right\rangle}}{|{#1}\rangle}{|{#1}\rangle}{|{#1}\rangle}}
\def\kettop#1{\left|{#1}\right\rangle_{\rm top}}
\def\ketSL#1{\left|{#1}\right\rangle_{\SL2}}
\def\ketGH#1{\left|{#1}\right\rangle_{\rm GH}}

\def\semi{\mathop{\rlap{\raisebox{1.5pt}{\tiny
        ${\mid}$}}\kern-.5pt\mbox{\large$\times$}}}
\def\Tr{\mathop{\rm Tr}\nolimits}

\def\tmu{\widetilde{\mu}}

\def\d{\partial}
\def\spsi{\psi^*}

\def\N#1{N\!=\!#1}
\def\ssr#1{{\ssf r}_{\alpha_{#1}}}
\def\tSl#1{{\widehat{s\ell}}(#1)}
\def\tSL#1{{\widehat{s\ell}}(#1)}
\def\SL#1{s\ell(#1)}

\def\SSL#1#2{s\ell(#1|#2)}
\def\Jplus{J^+}
\def\Jminus{J^-}
\def\Jnaught{J^0}
\def\Lambdach{\Lambda_{\rm ch}}

\def\upleadsto{%
  \begin{picture}(2,8)
    \unitlength=1pt
    \bezier{20}(0,-5)(-3,-2)(0,-1)
    \bezier{20}(0,-1)(3,1)(0,3)
    \bezier{20}(0,3)(-3,5)(0,7)
    \bezier{10}(0,7)(1,8)(.5,9.5)
    \put(1,11.3){\vector(0,1){2}}
  \end{picture}%
  }

\def\half{\frac{1}{2}}
\def\fourth{\frac{1}{4}}

\def\cA{{\cal A}}

\def\cE{{\cal E}}

\def\cN{{\cal N}}
\def\cO{{\cal O}}
\def\cP{{\cal P}}

\def\cU{{\cal U}}

\def\oA{{\open A}}
\def\oB{{\open B}}
\def\oD{{\open D}}
\def\oN{{\open N}}
\def\oC{{\open C}}
\def\oQ{{\open Q}}

\def\oX{{\open X}}
\def\oY{{\open Y}}
\def\oZ{{\open Z}}

\def\ctop{{\ssf c}}
\def\Ctop{{\ssf C}}
\def\htop{{\ssf h}}

\def\hplus{{\ssf h}^+}
\def\hminus{{\ssf h}^-}
\def\jplus{{\ssf j}^+}
\def\jpm{{\ssf j}^\pm}
\def\jtop{{\ssf j}}
\def\jminus{{\ssf j}^-}
\def\theel{{\ssf l}}

\def\tensor{\otimes}

\def\const{{\rm const}}

\def\frh{{\fr h}}

\def\mm{\cal}
\def\smm{\fr}

\def\mC{{\mm C}}
\def\mM{{\mm M}}
\def\mN{{\mm N}}
\def\mH{{\mm H}}
\def\mR{{\mm R}}

\def\mU{{\mm U}}
\def\mV{{\mm V}}

\def\smM{{\smm M}}

\def\smU{{\smm U}}
\def\smV{{\smm V}}
\def\smR{{\smm R}}
\def\smW{{\smm W}}

\def\CVER{{\cal CVER}}
\def\VER{{\cal VER}}
\def\CTVER{{\cal CTVER}}
\def\TVER{{\cal TVER}}
\def\CHW{{\cal CHW}}
\def\HW{{\cal HW}}
\def\CRHW{{\cal CRHW}}
\def\RHW{{\cal RHW}}
\def\CTOP{{\cal CTOP}}
\def\TOP{{\cal TOP}}
\def\CMHW{{\cal CMHW}}
\def\MHW{{\cal MHW}}
\def\CRVER{{\cal CRVER}}
\def\RVER{{\cal RVER}}
\def\CMVER{{\cal CMVER}}
\def\MVER{{\cal MVER}}

\newtheorem{lemma}{Lemma}[section]

\newtheorem{thm}[lemma]{Theorem}

\newtheorem{dfn}[lemma]{Definition}

\newenvironment{rem}{%
  \smallskip\stepcounter{lemma}\noindent{\bf Remark~\thelemma} \ }%
{\par\medskip}

\newenvironment{prf}{%
  \noindent{\sc Proof} \ }%
{\noindent$\Box$\par\medskip} 

\def\emt{energy-momentum tensor}
\def\hw{highest-weight}

\def\NPB{Nucl.\ Phys.\ B}
\def\PLB{Phys.\ Lett.\ B}
\def\MPLA{Mod.\ Phys.\ Lett.\ A}

\def\IJMPA{Int.\ J.\ Mod.\ Phys.\ A}

\begin{document}
\hfuzz=1.5pt
\addtolength{\baselineskip}{6pt}

\setcounter{page}{1}

\begin{flushright}
  {\tt hep-th/9701043}
\end{flushright}

\pagestyle{myheadings} \markboth{B Feigin, A Semikhatov, I
  Tipunin}{Equivalence between $\tSL2$ and $N=2$ representations}
\thispagestyle{empty}

\begin{center}
  {\large{\sc Equivalence between Chain Categories of Representations
      of Affine $\SL2$ and $N=2$ Superconformal Algebras}}\\[16pt]
  {B.~L.~Feigin${}^{\rm a}\!\!$, \ A.~M.~Semikhatov${}^{\rm b}\!\!$, \
    and \ I.~Yu.~Tipunin${}^{\rm b}$}\\[16pt]
  \parbox{.9\textwidth}{ \footnotesize \sl ${}^{\rm a}$\,Landau
    Institute for Theoretical Physics, Russian Academy of
    Sciences\\[4pt]
    ${}^{\rm b}$\,Tamm Theory Division, Lebedev Physics Institute,
    Russian Academy of Sciences }\\[12pt]
  \parbox{.95\textwidth}{
    {\footnotesize Highest-weight type representation theories of the
      affine $\SL2$ and $\N2$ superconformal algebras are shown to be
      equivalent modulo the respective spectral flows.}}

\medskip

\end{center}
{\addtolength{\parskip}{-1pt}\footnotesize
  \tableofcontents
  }

\newpage

\section{Introduction}\lvm
In this paper, we study \hw{} type modules over the $\N2$
superconformal algebra and the affine $\SL2$ algebra and demonstrate
the equivalence of the two representation theories modulo the
respective spectral flows.  The motivation comes partly from the
previous analysis of $\tSL2$ fusion rules~\cite{[AY]}, which suggests
the occurrence of more general modules than the usual Verma
modules~\cite{[FM-fusion]}.  The $\N2$ superconformal
algebra~\cite{[Ade]}, which is essential in a number of conformal
field-theoretic/string-theoretic constructions is known to have
various relations to~$\tSL2$. A prerequisite for constructing complete
$\N2$ fusion rules is a better understanding of modules over the $\N2$
algebra, and in particular, in view of a recent progress in
constructing $\tSL2$ conformal blocks~\cite{[Andreev],[PRY],[FGP]}, of
their relation to $\tSL2$ modules.

In what follows, we `compare' $\tSL2$- and $\N2$ representation
theories and then find certain categories built out of representations
of these algebras that are equivalent. In particular, we give a
complete proof of the claim of~\cite{[S-sing]} that singular vectors
in the `topological' (${}\equiv{}$`chiral') Verma modules over the
$\N2$ algebra are isomorphic to singular vectors in $\tSL2$ Verma
modules.  We extend this result along two directions.  First, the
relation between topological $\N2$ Verma modules and the standard
$\tSL2$ Verma modules is extended to the equivalence of certain
categories of representations of the two algebras, which are
essentially the respective {\it twisted\/} $\cO$-categories
(see~\cite{[TheBook]}), in fact the categories of chains of such
modules related by the respective spectral flows.  Hence the
importance of twisted ($\equiv$~spectral-flow-transformed) modules; on
the $\N2$ side, such modules have been around for some
time~\cite{[ST2],[ST3]}, while for the $\tSL2$ algebra they do not
seem to have previously been studied in any detail (see,
however,~\cite{[BH]}). In general, the spectral flow transformations
and the canonical involution constitute the automorphism group of
$\tSL2$ as well as of the $\N2$ algebra; the action of {\it
  automorphisms\/} on representations can be defined in a natural way,
hence the appearance of twisted modules.

Second, we extend the correspondence between $\tSL2$ and $\N2$ Verma
modules to representations of a different \hw{} type.  For the $\N2$
algebra, one should distinguish between two types of Verma-like
objects: the topological Verma modules mentioned above, and the
so-called {\it massive\/} $\N2$ Verma modules. The latter are often
considered~\cite{[BFK]} as the `standard $\N2$ Verma modules'$\!$.
Their $\tSL2$ counterpart, however, is not so standard: these are the
{\it relaxed Verma modules\/}, which differ from the usual $\tSL2$
Verma modules by somewhat `relaxed' \hw{} conditions, and as a result
possess infinitely many highest-weight vectors. On the $\tSL2$ side,
the standard $\tSl2$ Verma modules may appear as submodules of the
relaxed Verma modules. On the $\N2$ side, the situation is similar: a
massive Verma module may contain submodules that are topological Verma
modules (in fact, {\it twisted\/} topological Verma modules; such are,
in particular, the submodules generated from the {\it charged\/} $\N2$
singular vectors).

It thus turns out that the modules looking `standard' on the $\N2$
side correspond to less standard $\tSL2$ modules, and vice versa,
which is one of the reasons for the appearance of many types of
modules we consider in what follows.

A proper identification of relaxed Verma modules may be important for
constructing the complete, $\SSL21_q$-symmetric, fusion rules.  The
$\tSL2$ modules appearing in the fusion constructions and in the
constructions of conformal
blocks~\cite{[AY],[FM-fusion],[FGP],[Andreev]} are related to the
relaxed Verma modules introduced in this paper.  The relaxed Verma
modules can be defined as the induced modules from the representations
of the {\it Lie\/} algebra $\SL2$ of neither the highest-weight nor
the lowest-weight type.  In the modules used previously, however,
there occur several `Wakimoto-type' effects when some singular vectors
vanish and cosingular vectors appear instead (this applies to the
charged singular vectors).

The main results of this paper are \ (i)~the structural theory of the
relaxed $\tSL2$ Verma modules, and \ (ii)~the equivalence of
categories.  As regards the relaxed Verma modules, we classify
possible degenerations, explicitly construct singular vectors in the
framework of the `complex-power' approach~\cite{[MFF]}, and describe
the nature of the corresponding submodules. The structure of relaxed
Verma modules, as regards the appearance of submodules, is identical
to that of $\N2$ massive Verma modules; see however~\cite{[ST4]}
for the analysis in intrinsic $\N2$ terms.

As to the equivalence of categories, we establish the pairwise
equivalence of the following categories of $\tSL2$ modules on the one
hand and $\N2$ modules on the other hand (the respective corners of
the squares):
\begin{equation}\new
  \begin{array}{ccc}
    {\CHW}&\leadsto&{\CRHW}\\
    \upleadsto&&\upleadsto\\
    {\CVER}&\leadsto&{\CRVER}\\
    {}&{\tSL2}&
  \end{array}\quad{\rm and}\quad\new\begin{array}{ccc}
    {\cal CTOP}&\leadsto&{\CMHW}\\
    \upleadsto&{}&\upleadsto\\
    {\CTVER}&\leadsto&{\CMVER}\\
    {}&{\N2}&
  \end{array}\label{square}
\end{equation}
In this diagram, `${\cal C}$' always stands for {\it chains\/}, which
we define below, of modules from the corresponding categories, while
the categories themselves are as follows.

In the $\tSL2$ square, $\VER$ consists of the usual $\tSL2$ Verma
modules together with all their images under the spectral flow (i.e.,
all the twisted Verma modules, which we also define below).
Introducing the spectral-flow-transformed $\tSL2$ modules is essential
for establishing the relation with $\N2$ modules, where the role of
the spectral flow has been appreciated for some
time~\cite{[SS],[LVW]}. \ $\RVER$~consists of (twisted) relaxed Verma
modules, which differ from the modules from $\VER$ by one missing
annihilation condition satisfied by the highest-weight vector: as a
result, relaxed Verma modules acquire infinitely many
`almost-highest-weight' vectors. Further, in the top line of the
$\tSL2$ square we have the respective \hw-\/{\it type\/} modules,
obtained from the bottom line by taking all possible quotient modules
and gluing different modules to each other. These `$\cO$'-categories
include twisted modules as well; in what follows, we give an intrinsic
definition of these categories using a criterion which is invariant
under twisting.\footnote{The standard definition of the $\cO$ category
  is less convenient in the case involving twisted modules, because
  the separation of operators from the algebra into creation and
  annihilation ones depends on the twist.} \ We call $\HW$ and $\RHW$
the categories of (twisted)~\hw- and relaxed-\hw-\/{\it type\/}
modules.  For the category $\HW$, therefore, the ``universal'' role of
Verma modules in producing irreducible representations (by taking
quotients) is played by the standard $\tSL2$ Verma modules~$\VER$. As
regards~$\RHW$, on the other hand, distinguished in this way are the
relaxed Verma modules~$\RVER$.  As a terminological remark, however,
let us note that speaking of {\it Verma\/} modules in the $\tSL2$
context, we will always mean only the standard Verma modules~$\VER$,
as opposed to the {\it relaxed Verma\/} modules~$\RVER$.

On the $\N2$ side, $\TVER$ are the topological Verma modules and all
their spectral flow transforms, while $\TOP$ are the corresponding
topological-highest-weight type modules, constructed from modules from
$\TVER$ by taking quotients and gluing. We will also give a
spectral-flow invariant criterion that characterizes modules from
$\TOP$, irrespective of their twists.  Further, $\MVER$ are the twisted
`massive' Verma modules; those with a zero twist are commonly viewed
as the `standard' $\N2$ Verma modules.  Finally, $\MHW$ is made up of
modules of the same highest-weight type as the twisted massive Verma
modules, but contains many more modules, all of which can be
constructed out of modules from $\MVER$ by taking quotients and gluing.
These, too, will be defined in `intrinsic' terms (using the condition
of terminating fermionic chains).

The {\it chains\/} of the corresponding modules have to be introduced
since the mapping that we use to establish equivalence of categories
maps a given $\tSL2$ module into a sum involving $\N2$ modules twisted
by all integers.  Taking chains of modules makes the respective
corners of the two squares in~\req{square} equivalent.  In application
to Verma-like modules, a part of this statement amounts to the
isomorphism between singular vectors in the respective modules.  As
another consequence, one can further derive identities between
characters of certain $\tSL2$ and $\N2$ modules, as we will do in this
paper for the `Verma' case, and carry over the fusion rules from
$\tSL2$ to $\N2$ theories (which is left for the future).

\medskip

The tools essential for the analysis of $\tSL2$ and $\N2$ modules
include diagrams of {\it extremal vectors\/}.  The idea to consider
extremal vectors was put forward in~\cite{[FS]}, where it was observed
that a number of representation-theoretic problems can naturally be
reformulated, and quite often, solved, in terms of extremal vectors.
An independent construction of~\cite{[ST3]} (and a similar
one,~\cite{[S-sl21sing]}) can be considered as a manifestation of this
general observation.

As regards singular vectors, we borrow from~\cite{[ST3]} the
construction of $\N2$ singular vectors in terms of continued fermionic
operators. We also develop a parallel construction of singular vectors
in the relaxed Verma modules over the affine $\SL2$.  The latter
construction is formulated in quite different terms than the former,
however the two constructions turn out to be equivalent and, moreover,
allow direct mappings between each other.  An important point is that
the isomorphism between singular vectors requires that one consider
the $\N2$ singular vectors that satisfy {\it twisted\/} \hw{}
conditions.  Allowing the singular vectors to satisfy twisted \hw{}
conditions makes it possible to choose those representatives among
singular vectors with different twists that generate maximal
submodules (see~\cite{[ST4]} as to how this `eliminates' the
superfluous $\N2$ subsingular vectors).  An essential point that we
emphasize is the necessity to consider the entire extremal diagrams,
rather than their fixed representatives, in order to correctly
describe the structure of submodules.

\medskip

In Section~\ref{sec:2}, we introduce the $\tSL2$ and $\N2$
superconformal algebras, review some of their properties (in
particular, the corresponding spectral flows), investigate the
structure of the corresponding Verma modules, and define different
categories of such modules.  We also introduce auxiliary free-field
systems.  In Section~\ref{sec:3}, we recall the simplest
Kazama--Suzuki (KS) mapping~\cite{[DvPYZ],[KS]} and construct another
mapping, which is essentially inverse to KS even though it acts
between somewhat different spaces; it will be called the anti-KS
mapping.  These are used in Section~\ref{sec:4} to prove equivalence
between the respective categories of $\tSL2$ and $\N2$
representations.

\medskip

The paper is rather long, which may be partly excused by a large
number of modules that we have to introduce in order to keep balance
between the~$\tSL2$ and~$\N2$ sides. We omit some of the proofs.

\section{The $\tSL2$ and $\N2$ algebras, their automorphisms and
  Verma modules\label{sec:2}}\lvm
In this section, we introduce the $\tSL2$ and the $\N2$ superconformal
algebras.  We recall their properties, define several types of modules
and give theorems describing the structure of singular vectors in
these modules. We also introduce free-field theories that will be
needed for a constructive realization of the mapping between
representations of these two algebras.

\subsection{$\tSL2$}\lvm
Some of the facts given below about the affine $\SL2$ algebra are well
known, yet we hope to avoid being too boring. However, we try to give
more details about twisted and `relaxed' modules. In
section~\ref{subsub:relaxed}, we describe singular vectors in relaxed
Verma modules, where the situation is more involved than with the
standard Verma modules, and give structural theorems about this class
of $\tSl2$ modules.

\subsubsection{$\tSL2$ Verma modules, spectral flow transform,
  and extremal diagrams}\lvm The level-$k$ algebra is defined as
\begin{equation}\new
  \begin{array}{rcl}
    {[}J^0_m,\,J^\pm_n]&=&{}\pm J^\pm_{m+n}\,,\qquad
    [J^0_m,\,J^0_n]~=~{}\frac{K}{2}\,m\,\delta_{m+n,0}\,,\\
    {[}J^+_m,\,J^-_n]&=&{}K\,m\,\delta_{m+n,0} + 2J^0_{m+n}
  \end{array}
  \label{sl2modes}
\end{equation}
(where $K$ is the central element, which we will not distinguish from
its value $k$ in representations). Commutation relations \req{sl2modes}
correspond to the following operator product expansions of currents:
\begin{equation}\new
  \begin{array}{rcl}
    J^0(z)J^\pm(w)&=&{}\pm{J^\pm\over
      z-w}\,,\qquad J^0(z)J^0(w)~=~{}{k/2\over(z-w)^2}\,,\\
    J^+(z)J^-(w)&=&{}{k\over(z-w)^2}+{2J^0\over z-w}\,.
  \end{array}\label{sl2algebra}
\end{equation}
In terms of the currents, the Sugawara \emt\ has the
form
\begin{equation}
  T^{\rm Sug}=\frac{1}{k+2}
  \L(J^0J^0+\half(J^+J^-+J^-J^+)\R)\,.
  \label{twistedS}
\end{equation}
Here and in what follows, we assume $k\neq-2$.  The usual mode
expansion of the \emt s is $T^{\rm Sug}(z)=\sum_{n\in\oZ}L^{\rm
  Sug}_n\,z^{-n-2}$, and similarly for the other \emt s that we work
with below.

\medskip

We now define the Verma module $\mM_{j,k}$ with the highest-weight
vector $\ketSL{j,k}$ in the standard manner, by imposing the following
annihilation conditions on the \hw{} vector:
\begin{equation}\label{sl2hig}\new
  \begin{array}{rcl}
    J^+_{\geq0}\,\ketSL{j,k}&=&J^0_{\geq1}\,\ketSL{j,k}=
    J^-_{\geq1}\,\ketSL{j,k}=0\,,\\
    J^0_{0}\,\ketSL{j,k}&=&j\,\ketSL{j,k}\,,\qquad K\,\ketSL{j,k}=
    k\,\ketSL{j,k},
  \end{array}~
  j,\,k\in\oC\,.
\end{equation}
The module is freely generated from $\ketSL{j,k}$ by the remaining
modes of $J^+$, $J^-$, and $J^0$; thus the character of $\mM_{j,k}$,
is given by
\begin{equation}
  \chi_{j,k}^{\SL2}(z,q)=
  \Tr_{\mM_{j,k}}^{\phantom{y}}(q^{L^{\rm Sug}_0}\,z^{J^0_0})=
  {z^j\,q^{\Delta_j}\over
    \prod\limits_{i=1}^\infty(1-zq^i)\,
    \prod\limits_{i=0}^\infty(1-z^{-1}q^i)\,
    \prod\limits_{i=1}^\infty(1-q^i)}\,,
\end{equation}
where $\Delta_j=j(j+1)/(k+2)$ is the Sugawara dimension of the \hw\
state.

The $\tSL2$ Verma modules can conveniently be described with the help
of the {\it extremal diagram\/}
\begin{equation}
  \unitlength=1pt
  \begin{picture}(250,80)
    \put(-35,62){\Huge $\ldots$}
    \put(0,60){$\bullet$}
    \put(15,65){${}^{J^-_0}$}
    \put(28,63){\vector(-1,0){22}}
    \put(30,60){$\bullet$}
    \put(45,65){${}^{J^-_0}$}
    \put(58,63){\vector(-1,0){22}}
    \put(60,60){$\bullet$}
    \put(75,65){${}^{J^-_0}$}
    \put(88,63){\vector(-1,0){22}}
    \put(90,60){$\circ$}
    \put(103,64){${}_{J^+_{-1}}$}
    \put(97,60){\vector(2,-1){17}}
    \put(115,47){$\bullet$}
    \put(24,-13){%
      \put(103,64){${}_{J^+_{-1}}$}
      \put(97,60){\vector(2,-1){17}}
      \put(115,47){$\bullet$}
      }
    \put(48,-26){%
      \put(103,64){${}_{J^+_{-1}}$}
      \put(97,60){\vector(2,-1){17}}
      \put(115,47){$\bullet$}
      }
    \put(72,-39){%
      \put(103,50){\Huge $\cdot$}
      \put(109,47){\Huge $\cdot$}
      \put(115,44){\Huge $\cdot$}
      }
  \end{picture}
  \label{Vermaextr}
\end{equation}
which expresses the fact that $\Jplus_{-1}$ and $\Jminus_0$ are the
highest-level operators that do not yet annihilate the \hw{} state
$\circ$.  All the other states of the module, according to their
(charge, level) bigrading, lie in the interior of the angle in the
diagram. In these conventions, e.g., $\Jnaught_{-1}$ is represented by
a downward vertical arrow.

\medskip

We will need automorphisms of the affine $\SL2$ algebra. These are the
canonical involution and the spectral flow transform~\cite{[BH]}
\begin{equation}
  \cU_\theta:\quad
  J^+_n\mapsto J^+_{n+\theta}\,,\qquad
  J^-_n\mapsto J^-_{n-\theta}\,,\qquad
  J^0_n\mapsto J^0_n+\frac{k}{2}\theta\delta_{n,0}\,,\qquad\theta\in\oZ\,.
  \label{spectralsl2}
\end{equation}
Spectral flow transformations with $\theta\in2\oZ$ are the
continuation to the entire algebra of the affine Weyl group action.
In general, $\tSL2$-modules are not invariant under the spectral
flow\,\footnote{A notable exception is provided, for $k\in\oN$, by the
  integrable representations. These are invariant under the action of
  spectral flow transformations with $\theta\in2\oZ$, however the
  spectral flow transformations with $\theta\in2\oZ+1$ map one
  integrable representation into another.} and are mapped into
`twisted' modules.

\begin{dfn}
  A twisted Verma module $\smM_{j,k,\theta}$ is freely generated by
  $J^+_{\leq\theta-1}$, $J^-_{\leq-\theta}$, and $J^0_{\leq-1}$ from a
  twisted highest-weight vector $\ketSL{j,k;\theta}$ defined by the
  conditions
  \begin{equation}\new
    \begin{array}{l}
      J^+_{\geq\theta}\,\ketSL{j,k;\theta}=J^0_{\geq1}\,\ketSL{j,k;\theta}=
      J^-_{\geq-\theta+1}\,\ketSL{j,k;\theta}=0\,,\\
      \left(J^0_{0}+\frac{k}{2}\theta\right)\,\ketSL{j,k;\theta}=
      j\,\ketSL{j,k;\theta}\,.
    \end{array}
    \label{sl2higgeneral}
  \end{equation}
\end{dfn}
The respective extremal diagrams are obvious `rotations'
of~\req{Vermaextr}. We identify $\ketSL{j,k}=\ketSL{j,k;0}$.

All possible (integral) twists of Verma modules constitute the
category~$\VER$. This category is very `small': already in the
untwisted case, many (if not all) interesting representations are {\it
  not} Verma modules, but rather can be obtained from Verma modules by
taking quotients and `gluing'.  This gives the category $\cO$
(see~\cite{[TheBook]}), in which every irreducible representation is a
quotient of a Verma module.  An important feature, which in fact {\it
  characterizes\/} the modules from category $\cO$ is that the action
of all the annihilation operators on any vector from a given module
spans out a finite-dimensional space (and the modules must be
finite-generated). In terms of extremal diagrams, this means simply
that the annihilators act into the interior of the angle
\begin{equation}
  \unitlength=1pt
  \begin{picture}(250,20)
    \put(100,-2.5){$\cdot$}
    \put(105,0){\vector(1,0){22}}
    \put(109,-13){${}^{\Jplus_0}$}
    \put(98,2){\vector(-2,1){17}}
    \put(82,-8){${}^{\Jminus_1}$}
    \thicklines
    \bezier{7}(85,15)(117,22)(135,5)
  \end{picture}
  \label{annihil}
\end{equation}

\medskip

\noindent
and therefore continuing to act with the annihilators on a state from
a module from $\cO$, one inevitably reaches the `edge', at which the
annihilators act by zero.

We need an $\cO$-type category that would include the twisted modules.
A remarkable fact is that it does not have to be defined by
`enumerating' different twistings of $\cO$, as
$$
\bigcup_{\theta\in\oZ}\,\cO_\theta=
\bigcup_{\theta\in\oZ}\,\cU_\theta\cO\,;
$$
instead, there exists an intrinsic definition. Consider first

\begin{dfn}
  Let $\ket{X}$ be an element of a module over the affine $\SL2$
  algebra and let us fix an integer~$\theta$. For $J$ being either
  $\Jplus$ or $\Jminus$, we say that the $J_\theta$-chain terminates
  on $\ket{X}$, and write $(J_\theta)^{+\infty}\,\ket{X}=0$ if
  $$
  \exists N\in\oZ,\quad n\geq N~:~ (J_\theta)^n\,\ket{X}=0\,.
  $$
\end{dfn}

Now, the property discussed before Eq.~\req{annihil} is invariant
under the twists. This underlies the next definition.  Before giving
it, let us assume once and for all that all the modules under
consideration are finite-generated (this requirement is often imposed
in the form of the condition that the modules be graded with respect
to an appropriate Virasoro generator $L_0$). Then,

\begin{dfn}
  Objects of the category $\HW$ of $\tSL2$ twisted \hw-type
  representations are modules $\smU$ over the affine $\SL2$ algebra
  such that the following conditions are satisfied:\nopagebreak
  \begin{enumerate}
    \addtolength{\parskip}{-6pt}
  \item $J^0_0$ is a diagonalizable operator, i.e. $\smU$ can be
    decomposed into a direct sum of $J^0_0$-eigenspaces;
  \item The action with all the operators $J^0_p$, $p\in\oN$, on any
    element $\ket{X}\in\smU$ spans a finite-dimensional space;
  \item for any element $\ket{X}$ of $\smU$,
    \begin{equation}
      \forall n\in\oZ\qquad
      \new\begin{array}{ll}
        {\it either}&(J^+_{n})^{+\infty}\,\ket X=0\\
        {\it or}    &(J^-_{-n})^{+\infty}\,\ket X=0
      \end{array}\label{Jterminate}
    \end{equation}
  \end{enumerate}
  Morphisms are standard homomorphisms between $\tSL2$-modules.
\end{dfn}

Thus conditions \req{Jterminate} do not distinguish between different
$\theta$-twists of category $\cO$.  Heuristically, these conditions
say that each straight line (in the conventions of \req{Vermaextr}) of
$\tSL2$ operators acting in a given module from the category
intersects the boundary of the module, or simply that the extremal
diagram, even though not of the simple form \req{Vermaextr}, is
nevertheless angle-shaped, the angle being {\it strictly less\/} than
180 degrees.

\subsubsection{Singular vectors in $\tSL2$ Verma modules}\lvm
A singular vector in a Verma module $\mM_{j,k}$ is a vector that is
not proportional to the \hw\ vector and nevertheless satisfies the
same annihilation conditions~\req{sl2hig} as the highest-weight vector
does.

For $\alpha\in\oC$, we introduce the objects $(J^+_{-1})^\alpha$ and
$(J^-_0)^\alpha$ that implement the standard action of generators of
the affine Weyl group on the space of highest-weights, see
\cite{[MFF]} for the details.  These objects correspond to reflections
with respect to two positive simple roots of $\tSL2$.  In fact
$(J^+_{-1})^\alpha$ and $(J^-_0)^\alpha$ define the following Weyl
group action on the line $k=\const$ in the $kj$ plane of highest
weights:
\begin{equation}\new\begin{array}{l}
    (J^-_0)^{2j+1}\,:\,\ketSL{j,k}\rightarrow\ketSL{-1-j,k}\,,\\
    (J^+_{-1})^{k+1-2j}\,:\,\ketSL{j,k}\rightarrow\ketSL{k+1-j,k}\,.
  \end{array}\label{sl2weylaction}
\end{equation}

The action of $(J^+_{-1})^\alpha$ and $(J^-_0)^\alpha$ can be extended
from the highest-weight vectors to the Verma modules over these
vectors.  The extension is given by the following formulas:
\begin{equation}\!\!\!\!\new
  \begin{array}{rcl}
    (\Jminus_0)^\alpha\,\Jplus_m &=&
    -\alpha (\alpha - 1) \Jminus_m(\Jminus_0)^{\alpha-2} -
    2 \alpha \Jnaught_m\,(\Jminus_0)^{\alpha-1} +
    \Jplus_m\,(\Jminus_0)^{\alpha}
    \,,\\
    (\Jminus_0)^{\alpha}\,\Jnaught_m &=& \alpha \Jminus_m(\Jminus_0)^{\alpha-1} +
    \Jnaught_m\,(\Jminus_0)^{\alpha}
    \,,\\
    (\Jplus_{-1})^\alpha\,\Jminus_m &=&
    -\alpha (\alpha - 1) \Jplus_{m-2}(\Jplus_{-1})^{\alpha-2} -
    k\,\alpha\,\delta_{m - 1, 0} (\Jplus_{-1})^{\alpha-1} +
    2 \alpha \Jnaught_{m-1}\,(\Jplus_{-1})^{\alpha-1} +
    \Jminus_m\, (\Jplus_{-1})^{\alpha}\,,\\
    (\Jplus_{-1})^{\alpha}\,\Jnaught_m &=&
    -\alpha \Jplus_{m-1}(\Jplus_{-1})^{\alpha-1} +
    \Jnaught_m\,(\Jplus_{-1})^{\alpha}\,,
  \end{array}
  \label{properties}
\end{equation}
which can be derived for $\alpha $ being a positive integer and then
postulated for an arbitrary complex~$\alpha $. Then,
\begin{thm}\mbox{}\nopagebreak

  {\rm I. (\cite{[KK]})} A singular vector exists in the Verma module
  $\mM_{j,k}$ over the affine $\SL2$ algebra if and only if
  $j=\jplus(r,s,k)$ or $j=\jminus(r,s,k)$, where
  \begin{equation}\left.\new
      \begin{array}{l}
        \jplus(r,s,k)=\frac{r-1}{2}-(k+2)\frac{s-1}{2}\\
        \jminus(r,s,k)=-\frac{r+1}{2}+(k+2)\frac{s}{2}
      \end{array}\right\}
    \quad
    \begin{array}{l}
      r,\,s\in\oN\,,\\
      k\in\oC\,.
    \end{array}
    \label{sl2singcond}
  \end{equation}

  {\rm II. (\cite{[MFF]})} All singular vectors in the module
  $\mM_{\jtop^\pm(r,s,k),k}$ are given by the explicit constructions:
  \begin{equation}
    \new\begin{array}{rcl}
      \ket{{\rm MFF}^+(r,s,k)}\kern-6pt&=&\kern-6pt
      (J^-_0)^{r+(s-1)(k+2)}(J^+_{-1})^{r+(s-2)(k+2)}(J^-_0)^{r+(s-3)(k+2)}
      \ldots\\
      {}&{}&\qquad{}\cdot(J^+_{-1})^{r-(s-2)(k+2)}
      (J^-_0)^{r-(s-1)(k+2)}\ket{\jplus(r,s,k),k}_{\SL2}\,,\\
      \ket{{\rm MFF}^-(r,s,k)}\kern-6pt&=&\kern-6pt
      (J^+_{-1})^{r+(s-1)(k+2)}(J^-_0)^{r+(s-2)(k+2)}
      (J^+_{-1})^{r+(s-3)(k+2)}\ldots\\
      {}&{}&\qquad{}\cdot(J^-_0)^{r-(s-2)(k+2)}
      (J^+_{-1})^{r-(s-1)(k+2)}\ket{\jminus(r,s,k),k}_{\SL2}\,.
    \end{array}
    \label{mffminus}
    \label{mffplus}
  \end{equation}
\end{thm}

\begin{rem}
  As is well known \cite{[MFF]}, the factors in~\req{mffminus}
  correspond to the affine Weyl group reflections with respect to the
  simple roots. When there are several singular vectors in a given
  Verma module, one can have different solutions to
  equations~\req{sl2singcond} for $(r,s)$, each such pair yielding a
  singular vector as given by Eqs.~\req{mffminus}.  Whenever we say
  ``all singular vectors'', we mean, in accordance with~\cite{[Mal]},
  that the singular vectors are constructed either directly as
  in~\req{mffminus} or---which may be the case for rational $k$---by
  composing the factors from two such formulas (the latter being
  equivalent to truncating a formula from~\req{mffminus} as soon as
  the corresponding Weyl reflections produce a highest weight that
  differs from the original one by a negative integral multiple of a
  positive root).
\end{rem}

Clearly, singular vectors in the twisted modules $\smM_{j,k;\theta}$
follow from the standard singular vectors \req{mffplus} by applying
the spectral flow transform.

\subsubsection{Relaxed Verma modules}\lvm
A different class of $\tSL2$ modules can be introduced by relaxing
annihilation conditions~\req{sl2higgeneral}:
\begin{dfn}
  For $\theta\in\oZ$, the relaxed twisted Verma module
  $\smR_{j,\Lambda,k;\theta}$ is generated from the state
  $\ketSL{j,\Lambda,k;\theta}$ that satisfies the annihilation
  conditions
  \begin{equation}
    J^+_{\geq\theta+1}\,\ketSL{j,\Lambda,k;\theta}=J^0_{\geq1}\,
    \ketSL{j,\Lambda,k;\theta}=
    J^-_{\geq-\theta+1}\,\ketSL{j,\Lambda,k;\theta}=0
    \label{floorhw}
  \end{equation}
  by a free action of the operators $J^+_{\leq\theta-1}$,
  $J^-_{\leq-\theta-1}$, and $J^0_{\leq-1}$ \ and by the action of
  operators $J^+_{\theta}$ and $J^-_{-\theta}$ subject to the
  constraint
  \begin{equation}
    J^-_{-\theta}J^+_\theta \,
    \ketSL{j,\Lambda,k;\theta}=\Lambda\,\ketSL{j,\Lambda,k;\theta}\,.
  \end{equation}
  In addition, the \hw{} state $\ketSL{j,\Lambda,k;\theta}$ satisfies
  \begin{equation}
    \left(J^0_0+\frac{k}{2}\theta\right)\,
    \ketSL{j,\Lambda,k;\theta}=j\,\ketSL{j,\Lambda,k;\theta}\,.
  \end{equation}
\end{dfn}
The corresponding extremal diagram opens up to the straight angle; in
the case of $\theta=0$ it thus becomes
\begin{equation}
  \unitlength=1pt
  \begin{picture}(250,20)
    \put(-35,2){\Huge $\ldots$}
    \put(0,0){$\bullet$}
    \put(15,5){${}^{J^-_0}$}
    \put(28,3){\vector(-1,0){22}}
    \put(30,0){$\bullet$}
    \put(45,5){${}^{J^-_0}$}
    \put(58,3){\vector(-1,0){22}}
    \put(60,0){$\bullet$}
    \put(75,5){${}^{J^-_0}$}
    \put(88,3){\vector(-1,0){22}}
    \put(90,0){$\star$}
    \put(100,5){${}^{J^+_0}$}
    \put(97,3){\vector(1,0){22}}
    \put(120,0){$\bullet$}
    \put(130,5){${}^{J^+_0}$}
    \put(127,3){\vector(1,0){22}}
    \put(150,0){$\bullet$}
    \put(160,5){${}^{J^+_0}$}
    \put(157,3){\vector(1,0){22}}
    \put(180,0){$\bullet$}
    \put(193,2){\Huge $\ldots$}
  \end{picture}
  \label{floor}
\end{equation}
The state marked with $\star$ is the above
$\ketSL{j,\Lambda,k;\theta}$, characterized, besides the level, by two
quantum numbers $j$ and~$\Lambda$.  The other states
$\ketSL{j,\Lambda,k;\theta|n}$, $n\in\oZ$, in the top floor are
\begin{equation}
  \ketSL{j,\Lambda,k;\theta|n}=\left\{\kern-4pt\new\begin{array}{ll}
      (J^-_{-\theta})^{-n}\,\ketSL{j,\Lambda,k;\theta}\,,&n<0\,,\\
      (J^+_\theta)^{n}\,\ketSL{j,\Lambda,k;\theta}\,,&n>0\,,
    \end{array}\right.
  \label{theother}
\end{equation}
with $\ketSL{j,\Lambda,k;\theta|0}=\ketSL{j,\Lambda,k;\theta}$.  We
also set $\ketSL{j,\Lambda,k|n}=\ketSL{j,\Lambda,k;0|n}$.  The
Sugawara dimension of $\ket{j,\Lambda,k;\theta|0}_{\SL2}$ is
\begin{equation}
  \Delta={j^2+j+\Lambda\over k+2}-\theta j + \frac{k}{4}\theta^2\,.
\end{equation}

\begin{rem}As we have already remarked, speaking of {\it Verma\/}
  modules in the $\tSL2$ context, we will always mean only the
  standard Verma modules, as opposed to the relaxed Verma modules; the
  same will apply to {\it Verma\/} \hw{} conditions, by which we will
  mean (possibly spectral-flow-transformed) \hw{} conditions in
  $\smM_{j,k;\theta}$, rather than~Eqs.~\req{floorhw}.
\end{rem}

By the {\it relative charge\/} of a state
$\ket{v}\in\mR_{j,\Lambda,k}$ we understand $x=y-j$, where
$J^0_0\,\ket{v}=y\,\ket{v}$ (and
$J^0_0\,\ket{j,\Lambda,k}=j\,\ket{j,\Lambda,k}$). Thus, states with a
negative (positive) relative charge are on the left (resp., on the
right) of~$\star=\ket{j,\Lambda,k}$.

In the generic case, one can travel both ways along the extremal
diagram: when $\theta=0$, for example, `untwisted' diagram \req{floor}
acquires a `fat' form
\begin{equation}
  \unitlength=1pt
  \begin{picture}(250,30)
    \put(-35,12){\Huge $\ldots$}
    \put(0,10){$\bullet$}
    \put(15,17){${}^{J^-_0}$}
    \put(28,15){\vector(-1,0){22}}
    \put(7,11){\vector(1,0){22}}
    \put(15,3){${}_{J^+_0}$}
    \put(30,10){$\bullet$}
    \put(45,17){${}^{J^-_0}$}
    \put(58,15){\vector(-1,0){22}}
    \put(37,11){\vector(1,0){22}}
    \put(45,3){${}_{J^+_0}$}
    \put(60,10){$\bullet$}
    \put(75,17){${}^{J^-_0}$}
    \put(88,15){\vector(-1,0){22}}
    \put(67,11){\vector(1,0){22}}
    \put(75,3){${}_{J^+_0}$}
    \put(90,10){$\star$}
    \put(105,17){${}^{J^-_0}$}
    \put(118,15){\vector(-1,0){22}}
    \put(97,11){\vector(1,0){22}}
    \put(105,3){${}_{J^+_0}$}
    \put(120,10){$\bullet$}
    \put(135,17){${}^{J^-_0}$}
    \put(148,15){\vector(-1,0){22}}
    \put(127,11){\vector(1,0){22}}
    \put(135,3){${}_{J^+_0}$}
    \put(150,10){$\bullet$}
    \put(165,17){${}^{J^-_0}$}
    \put(178,15){\vector(-1,0){22}}
    \put(157,11){\vector(1,0){22}}
    \put(165,3){${}_{J^+_0}$}
    \put(180,10){$\bullet$}
    \put(193,12){\Huge $\ldots$}
  \end{picture}
  \label{bothways}
\end{equation}
while the one with $\theta=-1$,
\begin{equation}
  \unitlength=1pt
  \begin{picture}(250,110)
    \put(-107,55){%
      \put(103,50){\Huge $\cdot$}
      \put(109,47){\Huge $\cdot$}
      \put(115,44){\Huge $\cdot$}
      }
    \put(-72,39){%
      \put(92,40){${}^{J^+_{-1}}$}
      \put(95,58){\vector(2,-1){17}}
      \put(101,57.5){${}^{J^-_{1}}$}
      \put(114,53){\vector(-2,1){17}}
      \put(89,59){$\bullet$}
      }
    \put(-48,26){%
      \put(92,40){${}^{J^+_{-1}}$}
      \put(95,58){\vector(2,-1){17}}
      \put(101,57.5){${}^{J^-_{1}}$}
      \put(114,53){\vector(-2,1){17}}
      \put(89,59){$\bullet$}
      }
    \put(-24,13){%
      \put(92,40){${}^{J^+_{-1}}$}
      \put(95,58){\vector(2,-1){17}}
      \put(101,57.5){${}^{J^-_{1}}$}
      \put(114,53){\vector(-2,1){17}}
      \put(89,59){$\bullet$}
      }
    \put(90,60){$\star$}
    \put(92,40){${}^{J^+_{-1}}$}
    \put(95,58){\vector(2,-1){17}}
    \put(101,57.5){${}^{J^-_{1}}$}
    \put(114,53){\vector(-2,1){17}}
    \put(115,47){$\bullet$}
    \put(24,-13){%
      \put(92,40){${}^{J^+_{-1}}$}
      \put(95,58){\vector(2,-1){17}}
      \put(101,57.5){${}^{J^-_{1}}$}
      \put(114,53){\vector(-2,1){17}}
      \put(115,47){$\bullet$}
      }
    \put(48,-26){%
      \put(92,40){${}^{J^+_{-1}}$}
      \put(95,58){\vector(2,-1){17}}
      \put(106,55){${}^{J^-_{1}}$}
      \put(114,53){\vector(-2,1){17}}
      \put(115,47){$\bullet$}
      }
    \put(72,-39){%
      \put(103,50){\Huge $\cdot$}
      \put(109,47){\Huge $\cdot$}
      \put(115,44){\Huge $\cdot$}
      }
  \end{picture}
  \label{bothwaystw}
\end{equation}
etc.  Thus all the points are equivalent ($\star\leadsto\bullet$),
since the composition of the direct and the inverse arrows does in
each case result only in a factor. However, this factor may vanish for
some values of the parameters, which gives rise to the ordinary {\it
  Verma\/} modules.  Namely, in the untwisted case~\req{floor}, we
have
\begin{equation}\new
  \begin{array}{lrcl}
    n\leq0\,{:}\qquad&
    J^-_0\,\ketSL{j,\Lambda,k|n}&=&\ketSL{j,\Lambda,k|n-1}\,,\\
    {}&  J^+_0\,\ketSL{j,\Lambda,k|n-1}&=&(\Lambda-n(n-1)-2(n-1)j)\,
    \ketSL{j,\Lambda,k|n}\,,\\
    n\geq0\,{:}\qquad&
    J^+_0\,\ketSL{j,\Lambda,k|n}&=&\ketSL{j,\Lambda,k|n+1}\,,\\
    {}&  J^-_0\,\ketSL{j,\Lambda,k|n+1}&=&(\Lambda-n(n+1)-2nj)\,
    \ketSL{j,\Lambda,k|n}
  \end{array}
  \label{somevanish}
\end{equation}
and similarly for non-zero twists $\theta\neq0$.  Whenever the
parameters are such that, e.\,g., $J^+_0\approx0$ at a certain step,
one cannot come back to the~$\star$ state by acting with~$J^+_0$:
\begin{equation}
  {n\leq-1\,{:}\mathstrut}\qquad
  \Lambda=n(n+1)+2nj\quad\Longrightarrow\quad J^+_0\ketSL{j,\Lambda,k|n}=0\,.
  \label{Vermaneg}
\end{equation}
One keeps on acting with $J^+_{-1}$ instead (one mode down), and thus
the extremal diagram branches as
\begin{equation}
  \unitlength=1pt
  \begin{picture}(250,80)
    \put(-35,62){\Huge $\ldots$}
    \put(0,60){$\bullet$}
    \put(15,67){${}^{J^-_0}$}
    \put(28,65){\vector(-1,0){22}}
    \put(7,61){\vector(1,0){22}}
    \put(15,55){${}_{J^+_0}$}
    \put(30,60){$\bullet$}
    \put(45,67){${}^{J^-_0}$}
    \put(58,65){\vector(-1,0){22}}
    \put(37,61){\vector(1,0){22}}
    \put(45,55){${}_{J^+_0}$}
    \put(60,60){$\bullet$}
    \put(75,67){${}^{J^-_0}$}
    \put(88,65){\vector(-1,0){22}}
    \put(67,61){\vector(1,0){22}}
    \put(75,55){${}_{J^+_0}$}
    \put(90,60){$\circ$}
    \put(105,67){${}^{J^-_0}$}
    \put(118,65){\vector(-1,0){22}}
    \put(120,60){\raisebox{1.5pt}{${\scriptscriptstyle\odot}$}}
    \put(135,67){${}^{J^-_0}$}
    \put(148,65){\vector(-1,0){22}}
    \put(127,61){\vector(1,0){22}}
    \put(137,55){${}_{J^+_0}$}
    \put(150,60){$\bullet$}
    \put(40,0){
      \put(150,60){$\bullet$}
      \put(165,67){${}^{J^-_0}$}
      \put(178,65){\vector(-1,0){22}}
      \put(157,61){\vector(1,0){22}}
      \put(165,55){${}_{J^+_0}$}
      \put(180,60){$\star$}
      }
    \put(163,62){\Large$\ldots$}
    \put(233,62){\Large$\ldots$}
    \put(92,40){${}^{J^+_{-1}}$}
    \put(95,58){\vector(2,-1){17}}
    \put(114,53){\vector(-2,1){17}}
    \put(115,47){$\bullet$}
    \put(24,-13){%
      \put(92,40){${}^{J^+_{-1}}$}
      \put(95,58){\vector(2,-1){17}}
      \put(101,57.5){${}^{J^-_{1}}$}
      \put(114,53){\vector(-2,1){17}}
      \put(115,47){$\bullet$}
      }
    \put(48,-26){%
      \put(92,40){${}^{J^+_{-1}}$}
      \put(95,58){\vector(2,-1){17}}
      \put(106,55){${}^{J^-_{1}}$}
      \put(114,53){\vector(-2,1){17}}
      \put(115,47){$\bullet$}
      }
    \put(72,-39){%
      \put(103,50){\Huge $\cdot$}
      \put(109,47){\Huge $\cdot$}
      \put(115,44){\Huge $\cdot$}
      }
  \end{picture}
  \label{withVerma}
\end{equation}
Therefore, any Verma module can be thought of as a {\it submodule\/}
of a relaxed Verma module.  Taking the quotient with respect to this
Verma submodule (or, in the notation of~\req{withVerma}, with respect
to $\circ$), we see that the state~$\odot$ would satisfy Verma \hw{}
conditions~\req{sl2higgeneral} with~$\theta=1$.  Hence, each Verma
module is at the same time a {\it quotient module\/} of the
appropriately twisted relaxed Verma module.

The situation is similar when $J^-_0\approx0$ at a certain stage in
diagram~\req{bothways}:
\begin{equation}
  {n\geq1\,{:}\mathstrut}\qquad
  \Lambda=n(n-1)+2(n-1)j\quad\Longrightarrow\quad
  J^-_0\ketSL{j,\Lambda,k|n}=0\,.
  \label{Vermapos}
\end{equation}
Then the branching of the extremal diagram is a mirror image
of~\req{withVerma}, and the submodule is given by the spectral flow
transform with $\theta=1$ of a standard Verma module.  In both
cases~\req{Vermaneg} and \req{Vermapos}, we have
\begin{equation}
  \Jnaught_0\,\ketSL{j,\Lambda,k|n}=(j+n)\,\ketSL{j,\Lambda,k|n}\,,
  \quad n\in\oZ\,.
\end{equation}
Hence the Verma-module spin is $j_{\rm Verma}=j+n$ for $n<0$ (where
the Verma submodule in the relaxed module is untwisted), and $j_{\rm
  Verma}=j+n+\frac{k}{2}$ for $n>0$ (where the Verma module is twisted
by~$\theta=1$).

\medskip

By a straightforward application of the spectral flow, the above
considerations translate into those for the {\it twisted\/} relaxed
Verma modules; diagram \req{withVerma} changes accordingly.

\medskip

As in the Verma case, we now extend the category $\RVER$ of all
twisted relaxed Verma modules to a larger category $\RHW$ of arbitrary
(twisted) relaxed-highest-weight-{\it type\/} modules.  We use the
same method as in the case of the $\HW$ category.  Now that the angle
in the extremal diagrams has opened up to the straight angle, the
extremal diagrams certainly contains straight lines that are infinite
on both sides.  However, a condition on the class of modules can still
be given in the form of the requirement that, starting with any vector
from a given module, the action with any of the bent $J^\pm$-chains
$$
\unitlength=1pt
\begin{picture}(200,18)
  \put(100,-2.5){$\cdot$}
  \put(105,1.5){\vector(4,1){20}}
  \put(98,2){\vector(-2,1){17}}
  \thicklines
\end{picture}
$$
terminates:
\begin{dfn}
  The objects of category $\RHW$ are $\tSL2$ modules $\smU$ such that
  the following conditions are satisfied:
  \begin{enumerate}
    \addtolength{\parskip}{-6pt}
  \item $J^0_0$ is a diagonalizable operator;\pagebreak[3]
  \item The action with all $J^0_p$, $p\in\oN$, on any element
    $\ket{X}\in\smU$ spans a finite-dimensional space;
  \item for any element $\ket{X}$ of $\smU$,
    \begin{equation}
      \forall \theta\in\oZ\qquad
      \new\begin{array}{ll}
        {\it either}&(J^+_{\theta})^{+\infty}\,\ket X=0\\
        {\it or}    &(J^-_{-\theta+1})^{+\infty}\,\ket X=0
      \end{array}
    \end{equation}
  \end{enumerate}
  Morphisms are standard homomorphisms between $\tSL2$-modules.
\end{dfn}
Again, this condition is invariant under the spectral flow, and thus
selects all the twisted modules, none of which can be `overrelaxed' in
the sense of its extremal diagram occupying {\it more\/} than half a
plane.

\subsubsection{`Charged' singular vectors}\lvm
The `Verma points' encountered in the top floor of an extremal diagram
can of course be viewed as singular vectors in the corresponding
relaxed Verma module. The construction of these singular vectors is
now obvious: we define
\begin{equation}
  \Lambdach(p,j)=p(p+1)+2pj\,,\qquad p\in\oZ\,,
  \label{Lambdach}
\end{equation}
then the state
\begin{equation}
  \ketSL{C(p,j,k)}=\left\{\kern-4pt\new\begin{array}{ll}
      (J^-_0)^{-p}\,\ketSL{j,\Lambdach(p,j),k}\,,&p\leq-1\,,\\
      (J^+_0)^{p+1}\,\ketSL{j,\Lambdach(p,j),k}\,,&p\geq0
    \end{array}\right.
  \label{chargedsl2}
\end{equation}
satisfies the Verma \hw{} conditions for $p\leq-1$ and the twisted
Verma \hw{} conditions with the twist parameter $\theta=1$ for
$p\geq1$.  We would like to stress that \req{chargedsl2} follows
immediately from the analysis of extremal diagrams. In fact, these
singular vectors may not even deserve a special name, and we mention
them specifically in order to make the presentation parallel to the
one for the $\N2$ algebra, where `charged' singular vectors are
traditionally considered as a separate notion~\cite{[BFK]} (although
they are given by an equivalently simple construction~\cite{[ST3]},
see~\req{thirdE}).

\subsubsection{Relation with the previous constructions of some $\tSL2$
  modules}\lvm It is of some interest to see how the $\tSL2$ modules
constructed using the `functional realization' of $\SL2$, which have
appeared in the study of fusion, are related to relaxed Verma modules
introduced in this paper.  In the untwisted case, consider top-level
states~\req{floor} and assume that $\Lambda$ is expressed through a
superficial variable $\mu$ as
$$
\Lambda=\mu(\mu-1-2j)
$$
(given $\Lambda$ and $j$, we would in general have two different
values of $\mu$, but we do not discuss now how they can be
distinguished). Then, {\it assuming all the relevant factors
  non-vanishing\/}, renormalize states~\req{theother} into
\begin{equation}
  \ketSL{j,\Lambda(\mu,j),k|n}=\left\{\kern-4pt\new\begin{array}{ll}
      f_-(n,\mu)\,\ket{j,\Lambda(\mu,j),k|n}^*\,,&n\leq-1\\
      \ket{j,\Lambda(\mu,j),k|0}^*\,,&n=0\\
      f_+(n,j,\mu)\,\ket{j,\Lambda(\mu,j),k|n}^*\,,&n\geq1
    \end{array}\right.
  \label{relations}
\end{equation}
where
\begin{equation}\new\begin{array}{rcll}
    f_-(n,\mu)&=&(-1)^n\,{\Gamma(1 - \mu - n)\over\Gamma(1 - \mu)}\,,&n\leq-1\\
    f_+(n,j,\mu)&=&(-1)^n\,{\Gamma(2j + 1 - \mu + n)\over\Gamma(2j + 1 - \mu)}\,,
    &n\geq1
  \end{array}
\end{equation}
Now, we parametrize $\mu$ and $j$ as
\begin{equation}\begin{array}{rcl}
    j&=&\tmu\,,\\
    \mu&=&\tmu-s\,,
  \end{array}
\end{equation}
and introduce a simplified notation
\begin{equation}
  \ket{s,\tmu,k,n}^\#\equiv
  \ket{\tmu,(s-\tmu)(s+\tmu+1),k|n}^*.
\end{equation}
Then Eqs.~\req{somevanish} rewrite as
\begin{equation}\new\begin{array}{rcl}
    J^+_0\,\ket{s,\tmu,k,n}^\#&=&
    -(\tmu+n+s+1)\,\ket{s,\tmu,k,n+1}^\#\,,\\
    J^0_0\,\ket{s,\tmu,k,n}^\#&=&
    (\tmu+n)\,\ket{s,\tmu,k,n}^\#\,,\\
    J^-_0\,\ket{s,\tmu,k,n}^\#&=&
    (\tmu+n-s-1)\,\ket{s,\tmu,k,n-1}^\#\,.
  \end{array}\quad n\in\oZ
  \label{threefactors}
\end{equation}

We can consider the set of states $\ket{s,\tmu,k,\,\cdot\,}^\#$ as
representing a state
$$
\kket{s,\tmu,k}^\# = \left\{\ket{s,\tmu,k,\,\cdot\,}^\#\right\}\,,
$$
which can be written in the `$x$-representation' as
\begin{equation}
  \langle x\kket{s,\tmu,k}^\#=
  \sum_{n\in\oZ}(-1)^n x^{\tmu+s+n}\,\ket{s,\tmu,k,n}^\#,
  \label{x}
\end{equation}
so that the action of $J^{0,\pm}_0$ is represented by the right action
of the generators
\begin{equation}\new\begin{array}{rcl}
    S^+&=&\frac{\d}{\d x}\,,\\
    S^0&=&x\frac{\d}{\d x}-s\,,\\
    S^-&=&-x^2\frac{\d}{\d x} + 2 s x\,.
  \end{array}
\end{equation}
Denote by $P_{s,\tmu,k}$ the module freely generated from the states
$\ket{s,\tmu,k,n}^\#$, $n\in\oZ$, by $J^{\pm,0}_{\leq-1}$.

We see from \req{Vermaneg} and \req{Vermapos} that a Verma \hw{} state
(a charged singular vector) occurs in the extremal diagram whenever
\begin{equation}
  \mu=2 j + p\,,\quad p\in\oZ\,,\quad{\rm or}\quad \mu=p\,,\quad p\in\oZ\,.
\end{equation}
Precisely at these values of $\mu$, one of the functions $f_+$ or
$f_-$ either has a pole or is identically zero, which invalidates
relations~\req{relations}.  Then the charged singular vectors
disappear and cosingular vectors appear instead.

\subsubsection{Singular vectors in relaxed Verma
  modules\label{subsub:relaxed}}\lvm We are going to describe singular
vectors in relaxed Verma modules only in the `untwisted' case
$\theta=0$; similar results for modules with $\theta\neq0$ can be
obtained immediately by applying the spectral flow transform.

Our strategy is to reduce the relaxed case to the case of ordinary
Verma modules. Given a relaxed Verma module $\mR_{j,\Lambda, k}$, we
can arrive at an ordinary Verma module by the following trick.  The
states $(\Jminus_0)^{-\mu}\,\ket{j,\Lambda, k}$ and
$(\Jplus_0)^{\mu+1}\,\ket{j,\Lambda, k}$ satisfy Verma \hw{}
conditions~\req{sl2higgeneral} with $\theta=0$ or~$1$ respectively
whenever $\mu$ satisfies
\begin{equation}
  \mu^2+(2j+1)\mu-\Lambda=0\,.
  \label{mucondition}
\end{equation}
Therefore, the states
\begin{equation}\new
  \begin{array}{llc}
    \ketSL{j_-,k;0}=(\Jminus_0)^{-\mu}\,\ket{j,\Lambda, k}\,,&
    \ketSL{j'_-,k;0}=(\Jminus_0)^{-\mu'}\,\ket{j,\Lambda, k}\,,&
    \mu'=-\mu-2j-1\,,\\
    \ketSL{j_+,k;1}=(\Jplus_0)^{\mu+1}\,\ket{j,\Lambda, k}\,,&
    \ketSL{j'_+,k;1}=(\Jplus_0)^{\mu'+1}\,\ket{j,\Lambda, k}\,,&
    \mu'=-\mu-2j-1\,,
    \label{toverma}
  \end{array}
\end{equation}
which, in general, {\it do not belong to $\mR_{j,\Lambda,k}$\/},
represent Verma \hw\ vectors with the respective parameters
\begin{equation}\new
  \begin{array}{ll}
    j_-=j+\mu\,,&j'_-=-j-\mu-1\,,\\
    j_+=j+\mu+1+\frac{k}{2}\,,&j'_+=-j-\mu+\frac{k}{2}\,.
    \label{hwparameters}
  \end{array}
\end{equation}
We now require that a singular vector exist in one of the `continued'
modules with the \hw{} states from~\req{toverma}.
Formulas~\req{hwparameters} immediately give the following Lemma:
\begin{lemma}\label{equivalences}
  The following conditions are equivalent:
  \begin{equation}\new
    \begin{array}{rclcrclrrcl}
      j_-&=&\jtop^\pm(r,s,k)&\Longleftrightarrow&j'_+&=&\jtop^\mp(r,s,k)&
      \Longleftrightarrow&j'_-&=&\jtop^\mp(r,s \mp 1,k)\,,\\
      j'_-&=&\jtop^\mp(r,s,k)&\Longleftrightarrow&j_+&=&\jtop^\pm(r,s,k)&
      \Longleftrightarrow&j'_+&=&\jtop^\mp(r,s \pm 1,k)\,.
    \end{array}
  \end{equation}
\end{lemma}

The existence of singular vectors in $\mR_{j,\Lambda,k}$ is determined
by whether and how many of states~\req{toverma} admit a singular
vector and/or belong to $\mR_{j,\Lambda,k}$. Whenever one of these
states belongs to the relaxed Verma module $\mR_{j,\Lambda,k}$ --- or,
equivalently, $\mu$ or $\mu'$ is an integer of the appropriate sign
--- we arrive at precisely the charged singular vector considered
above (observe that \req{Lambdach} is nothing but~\req{mucondition}
with~$\mu=p$). As regards singular vectors that might exist on
states~\req{toverma}, these must be the MFF vectors, which reduces the
problem of explicitly constructing singular vectors in
$\mR_{j,\Lambda,k}$ to the MFF singular vectors.

We see from~\req{hwparameters} and~\req{mucondition} that the
condition for $\mR_{j,\Lambda,k}$ to contain a relaxed Verma submodule
reads as $\Lambda=\Lambda(r, s, j, k)$, where
\begin{equation}
  \Lambda(r, s, j, k)=-(j-\jminus(r,s,k))(j-\jplus(r,s+1,k))\,.
  \label{masslambda}
\end{equation}
Then, the state $(\Jminus_0)^{-\mu}\,\ketSL{j, \Lambda(r, s, j, k),
  k}$, which satisfies the Verma \hw{} conditions, has the
spin~$\jminus(r,s,k)$.  Therefore the singular vector constructed on
this state reads
\begin{equation}
  {\cal MFF}^-(r, s, k)\,(\Jminus_0)^{-\jminus(r, s, k) + j}\,
  \ketSL{j, \Lambda(r, s, j, k),k}\,,
  \label{work10}
\end{equation}
where ${\cal MFF}^-$ is the singular vector {\it operator\/} (read off
by dropping the \hw{} state in~\req{mffminus}). This Verma-module
singular vector has to be mapped back to the original relaxed Verma
module~$\mR_{j,\Lambda(r, s, j, k), k}$, where it will become a
relaxed singular vector. In particular, no non-integral powers of
$\Jminus_0$ should remain, which is achieved by acting on \req{work10}
with $(\Jminus_0)^{\jminus(r, s, k) - j + N}$, where $N$ is an
integer, and making use of~\req{properties}.  However, to be left with
only {\it positive\/} integral powers after the rearrangements, the
integer $N$ has to be $\geq r+rs$. We thus define
\begin{equation}
  \ket{\Sigma^-(r,s,j,k)}=
  (\Jminus_0)^{\jminus(r, s, k) - j + r + r s}\,
  {\cal MFF}^-(r, s, k)\,(\Jminus_0)^{-\jminus(r, s, k) + j}\,
  \ketSL{j, \Lambda(r, s, j, k), k}
  \label{sigmaminus}
\end{equation}
as a `canonical' representative of the relaxed singular vector
in~$\mR_{j,\Lambda(r, s, j, k), k}$.

Similarly, the state $(\Jplus_0)^{\mu+1}\,\ket{j,\Lambda(r,s,j,k), k}$
with $\mu$ as in~\req{hwparameters}, i.e., $(\Jplus_0)^{1 + \jplus(r,
  s + 1, k) - j}\cdot \ketSL{j, \Lambda(r, s, j, k), k}$, is a Verma
\hw{} state twisted by the spectral flow transform with $\theta=1$. We
then construct the corresponding MFF singular vector and finally map
it back to the relaxed Verma module. In this way, the singular vector
in the relaxed Verma module becomes
\begin{eqnarray}
    &&\ket{\Sigma^+(r,s,j,k)}={}\label{sigmaplus}\\
    &&\qquad{}=
    (\Jplus_0)^{-1 - \jplus(r, s + 1, k) + j + r + r s}\,
    {\cal MFF}^{+, 1}(r, s, k)\,
    (\Jplus_0)^{1 + \jplus(r, s + 1, k) - j}\,
    \ketSL{j, \Lambda(r, s, j, k), k}\,,\nonumber
\end{eqnarray}
where ${\cal MFF}^{+, \theta}$ is the spectral flow transform
\req{spectralsl2} of the singular vector operator~${\cal MFF}^+$.

The construction can be illustrated in the following extremal diagram:
\begin{equation}
  \unitlength=1pt
  \begin{picture}(250,60)
    \put(30,0){
      {\linethickness{.7pt}
        \bezier{70}(87.5,45)(22.5,12)(-60,41)
        \bezier{30}(-78,26)(-22.5,21.5)(0,5)
        \bezier{70}(96.5,45)(163.5,12)(244,40)
        \bezier{30}(262,26)(206.5,19.5)(184,5)
        }
      \put(-60,41){\vector(-3,1){1}}
      \put(0,5){\vector(3,-2){1}}
      \put(244,40){\vector(3,1){1}}
      \put(186,6){\vector(-3,-2){1}}
      \put(-66.5,38.5){$\bullet$}
      \put(246.5,38.5){$\bullet$}
      {\linethickness{0.5pt}
        \put(251.5,41.5){\line(1,0){20}}
        \put(247.7,40.3){\line(-1,-1){15}}
        }
      \put(20,-14){\put(246.5,38.5){$\bullet$}
        {\linethickness{0.5pt}
          \put(251.5,41.5){\line(1,0){20}}
          \put(247.7,40.3){\line(-1,-1){10}}
          }}
      {\linethickness{0.5pt}
        \put(-86.5,41.5){\line(1,0){20}}
        \put(-61.7,40.3){\line(1,-1){15}}
        }
      \put(-20,-15){\put(-66.5,38.5){$\bullet$}
        {\linethickness{0.5pt}
          \put(-86.5,41.5){\line(1,0){20}}
          \put(-61.7,40.3){\line(1,-1){15}}
          } }
      \put(0,36){${}_{{}{(\Jminus_0)^{-\mu}}}$}
      \put(164,36){${}_{{}{(\Jplus_0)^{\mu+1}}}$}
      \put(-60,13){${}_{{}^{(\Jminus_0)^{\mu+r+rs}}}$}
      \put(208,13){${}_{{}^{(\Jplus_0)^{-\mu-1+r+rs}}}$}
      \put(0,45){
        \put(-35,2){\Large $\ldots$}
        \put(0,0){$\bullet$}
        \put(15,5){${}^{J^-_0}$}
        \put(28,3){\vector(-1,0){22}}
        \put(30,0){$\bullet$}
        \put(45,5){${}^{J^-_0}$}
        \put(58,3){\vector(-1,0){22}}
        \put(60,0){$\bullet$}
        \put(75,5){${}^{J^-_0}$}
        \put(88,3){\vector(-1,0){22}}
        \put(90,0){$\star$}
        \put(100,5){${}^{J^+_0}$}
        \put(97,3){\vector(1,0){22}}
        \put(120,0){$\bullet$}
        \put(130,5){${}^{J^+_0}$}
        \put(127,3){\vector(1,0){22}}
        \put(150,0){$\bullet$}
        \put(160,5){${}^{J^+_0}$}
        \put(157,3){\vector(1,0){22}}
        \put(180,0){$\bullet$}
        \put(193,2){\Large $\ldots$}
        }
      \put(-35,2){\Large $\ldots$}
      \put(0,0){$\bullet$}
      \put(15,5){${}^{J^-_0}$}
      \put(28,3){\vector(-1,0){22}}
      \put(30,0){$\bullet$}
      \put(45,5){${}^{J^-_0}$}
      \put(58,3){\vector(-1,0){22}}
      \put(60,0){$\bullet$}
      \put(75,5){${}^{J^-_0}$}
      \put(88,3){\vector(-1,0){22}}
      \put(90,0){$\star$}
      \put(100,5){${}^{J^+_0}$}
      \put(97,3){\vector(1,0){22}}
      \put(120,0){$\bullet$}
      \put(130,5){${}^{J^+_0}$}
      \put(127,3){\vector(1,0){22}}
      \put(150,0){$\bullet$}
      \put(160,5){${}^{J^+_0}$}
      \put(157,3){\vector(1,0){22}}
      \put(180,0){$\bullet$}
      \put(193,2){\Large $\ldots$}
      }
  \end{picture}
  \label{twofloors}
\end{equation}
Every point in the lower floor is annihilated by the operators
$J^{\pm,0}_{\geq1}$ and, thus, satisfies the relaxed \hw{} conditions.
The Verma module shown on the left contains a submodule.  The Verma
module and its submodule on the right are `rotated' by the spectral
flow transform with~$\theta=1$. The Verma modules are to be thought of
as disconnected from the extremal diagram of the relaxed Verma
module~$\mR_{j,\Lambda(r,s,j,k),k}$, since they do not, in general,
belong to~$\mR_{j,\Lambda(r,s,j,k),k}$ (the mappings shown in dotted
lines are given by complex powers of the generators).

Expressions~\req{sigmaplus}, \req{sigmaminus}, and~\req{chargedsl2}
give explicit constructions for singular vectors that can appear
in relaxed Verma modules.  In order to describe all degenerate cases
with several singular vectors in $\mR_{j,\Lambda,k}$ appearing
simultaneously, we first classify the different degeneration
patterns.  Recall that, while the module $\mR_{j,\Lambda,k}$ is
irreducible for general $j$, $\Lambda$, and $k$, singular vectors
appear in codimension~1, when these parameters are related by 1
equation. In the general case in codimension~1, $j$ and $k$ are still
arbitrary. To describe degeneration patterns in higher codimensions,
we divide the space of the \hw s $O\equiv\{(j,\Lambda,k)\}$ into a
union $\cup_iO_i=O$ of such $O_i$ that $O_i\cap O_j=\emptyset$ and all
modules $\mR_{j,\Lambda,k}$ with $(j,\Lambda,k)\in O_i$ have the same
type of degeneration for every $i$.

For brevity, we will say that a \hw{} state admits a singular vector
whenever the Verma module generated from that state contains this
singular vector and that a \hw{} state admits no singular
vectors if no singular vectors exist in the Verma module.  The sets
$O_1$, $O_2$, \ldots, $O_n$ with the above properties can be singled
out by requiring that a certain number of the states
from~\req{toverma} belong to the relaxed Verma module or/and admit
singular vectors.  The list of all possible cases is as follows:
\begin{itemize}
\item codimension~0:

\item[$O_\emptyset$] None of states~\req{toverma} belong to the
  module~$\mR_{j,\Lambda,k}$ and at least one of states~\req{toverma}
  admits no singular vectors.

\item codimension~1:

\item[$O_c$] Exactly one of states~\req{toverma} belongs to the
  module~$\mR_{j,\Lambda,k}$ and none of states~\req{toverma} admit a
  singular vector.

\item[$O_r$] One of states~\req{toverma} admits precisely one singular
  vector, each of the other states~\req{toverma} admits at least one
  singular vector and none of states~\req{toverma} belong to the
  module~$\mR_{j,\Lambda,k}$.

\item codimension~2:

\item[$O_{rr}$] Each of states~\req{toverma} admits at least two
  singular vectors and none of the states from~\req{toverma} belong
  to~$\mR_{j,\Lambda,k}$.

\item[$O_{cc}$] Precisely one of the states from each
  line in~\req{toverma} belongs to~$\mR_{j,\Lambda,k}$ and
  none of these two states admit a singular vector;

\item[$O_{rc}$] One of the states from~\req{toverma} belongs
  to~$\mR_{j,\Lambda,k}$ and admits precisely one singular vector;
  none of states~\req{toverma} admit two different singular vectors;
  no two states from different lines in~\req{toverma} belong
  to~$\mR_{j,\Lambda,k}$.

\item codimension~3:

\item[$O_{rrc}$] One of the states from~\req{toverma}
  belongs to~$\mR_{j,\Lambda,k}$ and admits at least two singular
  vectors; no two states from different lines
  in~\req{toverma} belong to~$\mR_{j,\Lambda,k}$.

\item[$O_{rcc}$] Precisely one of the states from each line
  in~\req{toverma} belongs to~$\mR_{j,\Lambda,k}$; each of these two
  states admits at least one singular vector.

\end{itemize}
The sets $O_i$ from this list can be singled out by systems of
equations.  Thus, the $O_i$ of codimension~1 can be thought of as
hyperplanes with certain lines eliminated, while the $O_i$ of
codimension~2 are lines with some (rational) points eliminated, and
the $O_i$ sets of codimension~3 are discrete collections of points.

We will describe degenerate cases in the order given in the above
list.  To begin with, observe that the completeness of the above list
implies that {\it the relaxed Verma module $\mR_{j,\Lambda,k}$ is
  irreducible if and only if $(j,\Lambda,k)\in O_\emptyset$}.
Further, the cases where the relaxed Verma module $\mR_{j,\Lambda,k}$
contains precisely one submodule are described in the following
theorem:
\begin{thm}\label{thm:codim1}\mbox{}\nopagebreak

  {\rm I.}~The \hw{} $(j,\Lambda,k)$ of the relaxed Verma
  module~$\mR_{j, \Lambda, k}$ belongs to the set $O_c$ if and only if
  $\Lambda=\Lambdach(n,j)$ and
  \begin{equation}
    (n,j,k)\in(\oZ\times\oC\times\oC)\setminus
    \Bigl\{(n,\half(r-2n-1)-\half(k+2)s,k)\bigm|r,s\in\oZ,~r\neq0,~
    r\cdot s\geq0,~n\in\oZ,~k\in\oC\Bigr\}\,.
  \end{equation}
  The \hw{} of the relaxed Verma module~$\mR_{j, \Lambda, k}$ belongs
  to the set~$O_r$ if and only if $\Lambda=\Lambda(r,s,j,k)$ with
  $$\new
    (r,s,j,k)\in
    \Bigl(\oN\times\oN\times\oC\times(\oC\setminus\oQ)
    \bigcup\,\oY\Bigr) \setminus\,\oX(\oC)\,,
  $$
  where\footnote{Whenever we write $k+2=\frac{p}{q}$ with $p\in\oZ$,
    $q\in\oN$, we always assume $p$ and $q$ to be coprime.}
  \begin{equation}
    \label{oY}
    \new\begin{array}{rcl}
      \oY&=&\Bigl\{(r,s,j,-\frac{p}{q}-2)\!\Bigm|\!1\leq r\leq p,~
      1\leq s\leq q,~p,q\in\oN,~j\in\oC\Bigr\}\bigcup{}\\
      {}&{}&
      \qquad\Bigl\{(r,1,j,-\frac{p}{q}-2)\!\Bigm|\!p+1\leq r\leq 2p,~
      p,q\in\oN,~j\in\oC\Bigr\}\,,
    \end{array}
  \end{equation}
  and, for any set $\cP\subset\oC$, we have defined
  \begin{equation}
    \oX(\cP)=\Bigl\{(r,s,\half(\pm r-2n-1)\mp\half(k+2)s,k)\!
    \Bigm|\!n\in\oZ,~r,s\in\oN,~k\in\cP\Bigr\}\,.
    \label{oX}
  \end{equation}

  {\rm II.}~In the $O_c$ case, the module $\mR_{j,\Lambdach(n,j), k}$
  contains precisely one submodule (which is isomorphic to a Verma
  module) generated from singular vector~\req{chargedsl2}.

  {\rm III.}~In the $O_r$ case, the module $\mR_{j, \Lambda(r, s, j,
    k), k}$ contains precisely one submodule (which is isomorphic to a
  relaxed Verma module).  Expressions~\req{sigmaminus}
  and~\req{sigmaplus} evaluate as elements of the relaxed Verma module
  $\mR_{j, \Lambda(r, s, j, k), k}$ and satisfy the relaxed-\hw{}
  conditions
  \begin{equation}\new
    \begin{array}{l}
      J^+_{\geq1}\,\ket{\Sigma^\pm(r,s,j,k)}=
      J^0_{\geq1}\,\ket{\Sigma^\pm(r,s,j,k)}=
      J^-_{\geq1}\,\ket{\Sigma^\pm(r,s,j,k)}=0\,,\\
      J^-_{0}J^+_{0}\,\ket{\Sigma^\pm(r,s,j,k)}=
      \Lambda^\pm(r,s,j,k)\,\ket{\Sigma^\pm(r,s,j,k)}\,,\\
      J^0_{0}\,\ket{\Sigma^\pm(r,s,j,k)}=
     (j \pm r s)\,\ket{\Sigma^\pm(r,s,j,k)}\,,
    \end{array}\label{hwsing}
  \end{equation}
  where
  \begin{equation}
    \Lambda^\pm(r,s,j,k)=
    \fourth(r + (k + 2) s - 2 r s \mp(1 + 2 j))
    (r + (k + 2) s + 2 r s \pm(1 + 2 j))\,.
    \label{Lambdaplusminus}
  \end{equation}
  Each of the singular vectors $\Sigma^\pm(r,s,j,k)$ generates the
  entire relaxed Verma submodule; in particular, $J^\pm_0$-descendants
  of~\req{sigmaminus} and~\req{sigmaplus} are on the same extremal
  subdiagram and coincide up to numerical factors whenever they are in
  the same grade:
  \begin{equation}\kern-10pt\new
    \begin{array}{rcll}
      c_+(i,j,k)\,(J^+_0)^i\,\ket{\Sigma^-(r,s,j,k)}\kern-5pt&=&\kern-5pt
      c_-(i,j,k)\,(J^-_0)^{2rs-i}\,\ket{\Sigma^+(r,s,j,k)}\,,
      &0\leq i\leq2rs\,,\kern-30pt\\
      c_+(i,j,k)\,(J^+_0)^i\,\ket{\Sigma^-(r,s,j,k)}\kern-5pt&=&\kern-5pt
      c_-(i,j,k)\,(J^+_0)^{i-2rs}\,\ket{\Sigma^+(r,s,j,k)}\,,
      &i>2rs\,,\\
      c_+(i,j,k)\,(J^-_0)^{-i}\,\ket{\Sigma^-(r,s,j,k)}\kern-5pt&=&\kern-5pt
      c_-(i,j,k)\,(J^-_0)^{-i+2rs}\,\ket{\Sigma^+(r,s,j,k)}\,,
      &i<0\,,
    \end{array}
    \label{compare}
  \end{equation}
  where the numerical coefficients $c_\pm(i,j,k)$ are ($r$- and
  $s$-dependent) polynomials in~$j$ and~$k$.
\end{thm}

Thus, \req{sigmaminus} and~\req{sigmaplus} are, in general, different
representatives of the same singular vector, since they generate the
same submodule. This is not necessarily so in higher codimensions. To
see what can happen there, consider again diagram~\req{twofloors}.
There, the Verma modules that contain singular vectors are shown as
disconnected pieces, since they are related to the Verma module by
mappings with complex powers of $J^\pm_0$.  However, when one or both
of the roots of \req{mucondition} becomes an integer, one or both of
these Verma modules are no longer disconnected from the extremal
diagram of the relaxed Verma module.  Then the Verma \hw{} conditions
are encountered in the extremal diagram and restrict the possibilities
to travel over the states in the lower floor in~\req{twofloors}, since
some of the states in~\req{compare} might vanish.  For example, it may
be impossible to obtain the $\Sigma^+$ representative of the singular
vector as a $J^+_0$-descendant of $\Sigma^-$, or/and to obtain
$\Sigma^-$ as a $J^-_0$-descendant of $\Sigma^+$.  These cases in
codimension $\geq2$ describe simultaneous occurrence of several
singular vectors, and will be considered in what follows.  We now
proceed to codimension~$2$.

\paragraph{Codimension-$2$ cases.} In codimension~$2$, we
have three cases of a simultaneous occurrence of two singular vectors,
which are described in the three following Theorems.  The following
Lemma governs the appearance of Verma submodules in a relaxed Verma
module.  If one puts $\Lambda^\pm(r,s,j,k)$ in~\req{Lambdaplusminus}
equal to $\Lambdach(N,j \pm r s)$ (see~Eq.~\req{Lambdach}), one
discovers that the condition $\Lambda^\pm(r,s,j,k)=\Lambdach(N,j \pm r
s)$ implies a similar relation for $\Lambda(r,s,j,k)$, and therefore
charged singular vectors in the lower floor occur synchronously with
charged singular vectors in the top floor of the extremal diagram. In
this way, we arrive at
\begin{lemma}\label{chargemass:lemma}
  Let $\mR$ be the relaxed Verma module $\mR\equiv\mR_{j,\Lambda,k}$.
  \begin{enumerate}\addtolength{\parskip}{-6pt}
  \item Let $\mR\supset\mR'$ and $\mR\supset\mC$, where $\mR'$ is a
    relaxed Verma submodule and $\mC$ is a (twisted) Verma module
    generated from a charged singular vector in $\mR$, and let $\mC$
    be maximal in the sense that $\mR\supset\mC''\supset\mC$, where
    $\mC''$ is any other (twisted) Verma submodule, implies
    $\mC''=\mC$, Then $\mR'\cap\mC=\mC'\neq\{0\}$, where $\mC'$ is a
    (twisted) Verma module generated from a charged singular vector in
    $\mR'$.
    
  \item Conversely, if $\mR\supset\mR'$, where $\mR'$ is a relaxed
    Verma module and $\mR'\supset\mC'$, where $\mC'$ is a submodule
    generated from a charged singular vector in $\mR'$, then
    $\mR\supset\mC\supset\mC'$, where $\mC$ is a submodule generated
    from a charged singular vector in $\mR$, and
    $\mR\supset\mC''\supset\mC\Longrightarrow\mC''=\mC$.

  \item If \,$\mR\supset\mC'$, where $\mC'$ is a (twisted) Verma
    module, there exists a (twisted) Verma submodule $\mC\subset\mR$
    such that the embedding is given by a charged singular vector
    and $\mR\supset\mC''\supset\mC\Longrightarrow\mC''=\mC$.
  \end{enumerate}
\end{lemma}
This Lemma is a necessary tool in the proof of the following three
Theorems. Consider first the occurrence of two different relaxed
singular vectors.

We call a submodule {\it primitive\/} if it is not a sum of two other
submodules. 
\begin{thm}\label{twomassive} The \hw{} of the module
  $\mR_{j,\Lambda,k}$ belongs to the set $O_{rr}$ if and only if
  $\Lambda=\Lambda(r,s,j,k)$ with
  $$\new (r,s,j,k)\in\oN\times\oN\times\oC\times\oQ\setminus{}
  \Bigl(\oX(\oQ)\bigcup\,\oY\Bigr)
  $$
  where $\oY$ and $\oX(\oQ)$ are as in~\req{oY} and~\req{oX}.
  Then,
  \begin{enumerate}\addtolength{\parskip}{-4pt}
    
  \item any primitive submodule in
    $\mR_{j,\Lambda(r,s,j,\frac{p}{q}),\frac{p}{q}}$ is a relaxed
    Verma module.  It is generated from
    $\ket{\Sigma^-(r,s,j,\frac{p}{q})}$ or, equivalently, from
    $\ket{\Sigma^+(r,s,j,\frac{p}{q})}$ with the same $r$ and~$s$;

  \item for any primitive submodule
    $\mR'\subset\mR_{j,\Lambda(r,s,j,\frac{p}{q}),\frac{p}{q}}$, there
    exists a singular vector $e\equiv\cE\cdot
    \ketSL{\jminus(r,s,\frac{p}{q}),\frac{p}{q}}$ in the Verma
    module~$\mM_{\jminus(r,s,\frac{p}{q}),\frac{p}{q}}$ such that
    $\mR'$ is generated from a relaxed singular vector
    \begin{equation}
      \label{outofe}
      (\Jminus_0)^{\jminus(r, s, k) - j + r + r s}\,\cE\,
      (\Jminus_0)^{-\jminus(r, s, k) + j}\, \ketSL{j, \Lambda(r, s, j, k), k}\,;
    \end{equation}

  \item conversely, for any singular vector $e$ in
    $\mM_{\jminus(r,s,\frac{p}{q}),\frac{p}{q}}$, of the form
    $e=\ket{{\rm MFF}(r,s,\frac{p}{q})}^+$, $r\geq1$, $s\geq2$, or
    $e=\ket{{\rm MFF}(r,s,\frac{p}{q})}^-$, $r,s\geq1$, there exists a
    relaxed Verma submodule in
    $\mR_{j,\Lambda(r,s,j,\frac{p}{q}),\frac{p}{q}}$ generated by the
    relaxed singular vector~\req{outofe};

  \item two different singular vectors $e_1$ and $e_2$ in
    $\mM_{\jminus(r,s,\frac{p}{q}),\frac{p}{q}}$ correspond in this
    way to the same relaxed Verma submodule in $\mR$ if and only if
    one of the $e_i$ is the ${\rm MFF}^+(a\geq1,1,\frac{p}{q})$
    singular vector in the module generated from the other.
  \end{enumerate}
\end{thm}
In this case, therefore, the structure of the relaxed Verma module
$\mR_{j,\Lambda(r,s,j,\frac{p}{q}),\frac{p}{q}}$ is determined by the
structure of the Verma
module~$\mM_{\jminus(r,s,\frac{p}{q}),\frac{p}{q}}$.

\medskip

Now, consider
the case where both Verma modules shown in~\req{twofloors} actually
belong to the relaxed extremal diagram. It is described in the
following Theorem.\nopagebreak

\begin{thm}\label{twocharge} The \hw{} of the
  relaxed Verma module~$\mR_{j, \Lambda, k}$ belongs to the set
  $O_{cc}$ of the list of degeneration patterns if and only if
  $j=\jcc(n,m)$, $\Lambda=\Lambdacc(n,m)$, where
  \begin{equation}
    \jcc(n,m)=-\half(1 + m + n)\,,\quad\Lambdacc(n,m)=-mn\,,
  \end{equation}
  and
  \begin{equation}
    (n,m,k)\in
    \Bigl((-\oN)\times\oN_0\times(\oC\setminus\oQ)\Bigr)\bigcup\,
    \Bigl\{(n,m,-\frac{p}{q}-2\!\Bigm|\!n\in-\oN,~m\in\oN_0,~
    p,q\in\oN,~1\leq m-n\leq q\Bigr\}\,.
  \end{equation}
  Then the relaxed Verma module~$\mR_{j, \Lambda, k}$ contains the
  Verma submodule $\mC_-\approx\mM_{\half(n - m -1),k}$ generated from
  the charged singular vector $\ket{C(n, j, k)}$ and the twisted Verma
  submodule $\mC_+\approx\smM_{\half(1 + m - n+k),k;1}$ generated from
  the charged singular vector $\ket{C(m,j,k)}$. The maximal submodule
  is $\mC_+\cup\mC_-$, and $\mC_+\cap\mC_-=\emptyset$.
\end{thm}

\smallskip

The third possibility in codimension~2, i.e.\ the case $O_{rc}$ from
the list, is when one of the states~\req{toverma} belongs to the
module $\mR_{j,\Lambda,k}$ (hence this module contains a charged
singular vector) and at the same time one of states~\req{toverma}
admits a singular vector. This case is more involved and is analyzed
in what follows.\pagebreak[3]

According to Lemma~\ref{chargemass:lemma}, the Verma submodule growing
out of the charged singular vector on the top floor of the extremal
diagram necessarily has a submodule generated from a singular vector
${\rm MFF}^\pm(r,s,k)$, and this submodule coincides with the
submodule built on the charged singular vector in the lower floor of
the extremal diagram,~e.\,g.
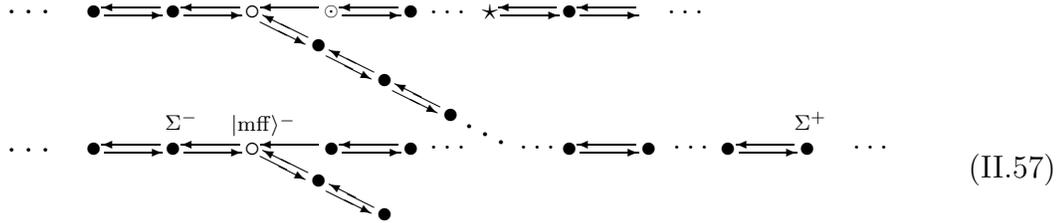
\begin{equation}
  \unitlength=1pt
  \begin{picture}(250,70)
    \put(-30,62){\large $\ldots$}
    \put(0,0){
      \put(0,60){$\bullet$}
      \put(28,64.5){\vector(-1,0){22}}
      \put(7,61.5){\vector(1,0){22}}
      }
    \put(30,0){
      \put(0,60){$\bullet$}
      \put(28,64.5){\vector(-1,0){22}}
      \put(7,61.5){\vector(1,0){22}}
      }
    \put(60,0){
      \put(0,60){$\circ$}
      \put(28,64.5){\vector(-1,0){22}}
      }
    \put(90,0){
      \put(0,60){\raisebox{1.5pt}{${\scriptscriptstyle\odot}$}}
      \put(28,64.5){\vector(-1,0){22}}
      \put(7,61.5){\vector(1,0){22}}
      \put(30,60){$\bullet$}
      }
    \put(130,62){\ldots}
    \put(150,0){
      \put(0,60){$\star$}
      \put(28,64.5){\vector(-1,0){22}}
      \put(7,61.5){\vector(1,0){22}}
      }
    \put(180,0){
      \put(0,60){$\bullet$}
      \put(28,64.5){\vector(-1,0){22}}
      \put(7,61.5){\vector(1,0){22}}
      }
    \put(220,62){\ldots}
%
%
    \put(66,59){\vector(2,-1){17}}
    \put(84,53){\vector(-2,1){17}}
    \put(85,47){$\bullet$}
    \put(25,-13){%
      \put(66,59){\vector(2,-1){17}}
      \put(84,53){\vector(-2,1){17}}
      \put(85,47){$\bullet$}
      }
    \put(50,-26){%
      \put(66,59){\vector(2,-1){17}}
      \put(84,53){\vector(-2,1){17}}
      \put(85,47){$\bullet$}
      }
    \put(43,-36){%
      \put(100,52){\large $\cdot$}
      \put(106,49){\large $\cdot$}
      \put(112,46){\large $\cdot$}
      }
%
    \put(0,-51){
      \put(66,59){\vector(2,-1){17}}
      \put(84,53){\vector(-2,1){17}}
      \put(85,47){$\bullet$}
      \put(25,-13){%
        \put(66,59){\vector(2,-1){17}}
        \put(84,53){\vector(-2,1){17}}
        \put(85,47){$\bullet$}
        }
      }
    \put(-30,10){\large $\ldots$}
    \put(0,-52){
      \put(0,60){$\bullet$}
      \put(28,64.5){\vector(-1,0){22}}
      \put(7,61.5){\vector(1,0){22}}
      \put(30,0){
        \put(0,60){$\bullet$}
        \put(28,64.5){\vector(-1,0){22}}
        \put(7,61.5){\vector(1,0){22}}
        \put(30,60){$\circ$} 
        \put(25,72){${}_{\ket{\rm mff}^-}$}
        }
      \put(60,0){
        \put(28,64.5){\vector(-1,0){22}}
        \put(30,60){$\bullet$}
        }
      \put(90,0){
        \put(28,64.5){\vector(-1,0){22}}
        \put(7,61.5){\vector(1,0){22}}
        \put(30,60){$\bullet$}
        }
      \put(130,63){\ldots}
      \put(164,63){\ldots}
      \put(180,0){
        \put(0,60){$\bullet$}
        \put(28,64.5){\vector(-1,0){22}}
        \put(7,61.5){\vector(1,0){22}}
        \put(30,60){$\bullet$}
        }
      \put(222,63){\ldots}
      \put(240,0){
        \put(0,60){$\bullet$}
        \put(28,64.5){\vector(-1,0){22}}
        \put(7,61.5){\vector(1,0){22}}
        \put(30,60){$\bullet$}
        }
      \put(290,63){\ldots}
      }
    \put(268,14){${}^{\Sigma^+}$}
    \put(30,14){${}^{\Sigma^-}$}
  \end{picture}
\end{equation}

\bigskip

\noindent
As we saw in~\req{hwsing}, the $\Sigma^-$ representative of the
relaxed singular vector is at the distance of \ $rs$ arrows on the
left from the `center' (the $\star$), while $\Sigma^+$ is at the same
distance on the right. It may happen that $\Sigma^-$ would be inside
the Verma submodule. In that case, the entire extremal diagram of the
submodule cannot be generated by $J^\pm_0$-descendants of~$\Sigma^-$.
Unless we are in codimension 3, however, the entire diagram of the
submodule would still be generated from $\Sigma^+$.  In particular,
following the arrows from $\Sigma^+$ to $\Sigma^-$ into the Verma
submodule makes the two representatives proportional. This is
described more precisely in the following Theorem.

\begin{thm}\label{thm:codim2}
  The \hw{} of the relaxed Verma module $\mR_{j,\Lambda,k}$ belongs to
  the set $O_{rc}$ of the list of degeneration patterns if and only if
  $j=\jrc^{\sigma}(r,s, n, k)$ and $\Lambda=\Lambdarc^{\sigma}(r,s, n,
  k)$, where $\sigma\in\{+,-\}$,
  \begin{equation}\new
    \begin{array}{l}
      \jrc^\pm(r, s, n, k)=-n-\half\pm\half(r-(k+2)s)\,,\\
      \Lambdarc^\pm(r, s, n, k)=-n^2 \pm n(r-(k+2)s)
    \end{array}
    \label{bars}
  \end{equation}
  and
  \begin{equation}
    (\sigma,r,s,n,k)\in
  \Bigl(\{\pm\}\times\oN\times\oN\times\oZ\times(\oC\setminus\oQ)\Bigr)
  \bigcup\,\oA(\oC\setminus\oQ)\bigcup\,\oB\,,
  \end{equation}
  where
  \begin{equation}
  \label{oAoB}\new\kern-20pt
  \begin{array}{rcl}
    \oA(\cP)\kern-6pt&=&\kern-6pt
    \Bigl\{(\{+\},r,0,n,k)\!\Bigm|\!r\in\oN,~n\in-\oN,~k\in\cP\Bigr\}
    \bigcup\,
    \Bigl\{(\{-\},r,0,n,k)\!\Bigm|\!r\in\oN,~n\in\oN_0,~k\in\cP\Bigr\},
    \kern-60pt\\
    \oB\kern-6pt&=&\kern-6pt
    \Bigl\{(\{+\},r,s,n,-\frac{p}{q}-2)\!\Bigm|\!n\in-\oN,~p,q\in\oN,~
    1\leq r\leq p,~0\leq s\leq q-1\Bigr\}
    \bigcup\,\\
    {}&{}&\Bigl\{(\{-\},r,s,n,-\frac{p}{q}-2)\!\Bigm|\!n\in\oN_0,~p,q\in\oN,~
    1\leq r\leq p,~0\leq s\leq q-1\Bigr\}\,.
  \end{array} 
\end{equation}
Then, the relaxed Verma module $\mR_{j,\Lambda,k}$ contains a relaxed
Verma submodule $\mR'$ generated from the singular vector
$\ket{\Sigma^+(r,s,\jrc^\pm(r, s, n, k),k)}$ (if $n<0$) or
$\ket{\Sigma^-(r,s,\jrc^\pm(r, s, n, k),k)}$ (if $n\geq0$); the module
$\mR_{j,\Lambda,k}$ contains also a Verma submodule
$\mC\approx\mM_{\jpm(r,s,k),k}$ (if $n<0$) or a twisted Verma
submodule $\mC\approx\smM_{\jpm(r,s,k),k;1}$ (if $n\geq0$) generated
from the charged singular vector $\ket{C(n, j, k)}$ in either case.
The maximal submodule in $\mR_{j,\Lambda,k}$ is $\mC\cup\mR'$.  The
intersection $\mC'\equiv\mC\cap\mR'$ is a Verma module (if $n<0$) or a
twisted Verma module (if $n\geq0$) generated from a charged singular
vector $\ket{c}\in\mR'$, which is also a singular vector in~$\mC$:
  \begin{equation}
    \ket{c}=\left\{\kern-5pt\new\begin{array}{rcll}
        \ket{\rm mff}^\pm\kern-5pt&=&\kern-5pt
        {\cal MFF}^\pm(r, s + \half \pm \half, k)\,
        \ket{C(n, \jrc^\pm(r, s, n, k), k)}\,,&n<0\,;\\
        \ket{\rm mff}^{\mp,1}\kern-5pt&=&\kern-5pt
        {\cal MFF}^{\mp,1}(r, s + \half \pm \half, k)\,
        \ket{C(n, \jrc^\pm(r, s, n, k), k)}\,,&n\geq0\,.
      \end{array}\right.
    \label{innerhw}
  \end{equation}
\end{thm}

\begin{rem} When the intersection $\mC'=\mC\cap\mR'$ is generated by
  the singular vector ${\cal MFF}^-(r, s, k)\ket{C(n, \jrc^-(r, s, n,
    k), k)}$ with $n<0$ and $|n|\leq r(s+1)$, the relaxed singular
  vector $\ket{\Sigma^-(r, s, \jrc^-(r, s, n, k),k)}$ belongs to the
  Verma submodule $\mC'$ and thus cannot generate the entire relaxed
  submodule $\mR$:
  \begin{equation}
    \ket{\Sigma^-(r,s,\jrc^-(r, s, n, k),k)}=
    (\Jminus_0)^{r + r s + n}\,\ket{c}\,.
    \label{sigmafromv}
  \end{equation}
  The relative positions of $\Sigma^-$ and $\Sigma^+$ in the extremal
  diagram are as follows:\footnote{At the point marked with a circle,
    two lines do not `intersect', i.e., they lead to different vectors
    in the same grade. The same applies to the diagrams below, which
    will not be stated explicitly however.}
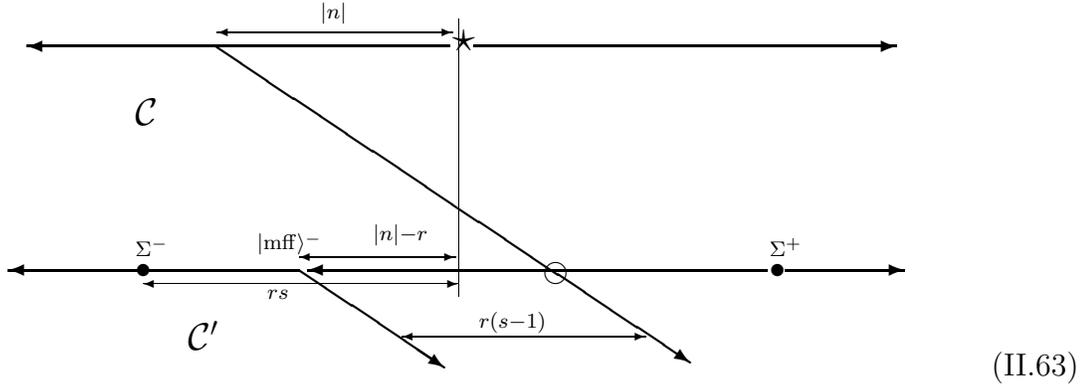
  \begin{equation}
    \unitlength=1pt
    \begin{picture}(250,140)
      \put(0,40){
        {\thicklines
          \put(138,85){\vector(1,0){160}}
          \put(129,85){\vector(-1,0){160}}
          \put(40,85){\vector(3,-2){180}}
          \put(72,0){\vector(-1,0){110}}
          \put(72,0){\vector(3,-2){55}} 
          \put(249,0){$\vector(-1,0){174}$} 
          \put(256,0){$\vector(1,0){45}$}
          }
        \put(10,55){\large$\mC$}
        \put(30,-30){\large$\mC'$}
        \put(130,83){\Large$\star$}
        {\linethickness{.1pt}
          \put(132.4,95){\line(0,-1){105}}
          }
        \put(10,1){${}^{\Sigma^-}$}
        \put(10,-3){$\bullet$}
        \put(56,10){${}_{\ket{\rm mff}^-}$}
        \put(250,1){${}^{\Sigma^+}$}
        \put(250,-3){$\bullet$}
        \put(169,-1){\circle{8}}
        \linethickness{.2pt}
        \put(13,-5){\vector(1,0){119}}
        \put(15,-5){\vector(-1,0){2}}
        \put(60,-9){${}_{rs}$}
        \put(42,90){\vector(1,0){88}}
        \put(43,90){\vector(-1,0){2}}
        \put(80,91){${}^{|n|}$}
        \put(113,-25){\vector(1,0){90}}
        \put(113,-25){\vector(-1,0){2}}
        \put(140,-20){${}_{r(s-1)}$}
        \put(100,7){${}^{|n|-r}$}
        \put(72,5){\vector(1,0){59}}
        \put(74,5){\vector(-1,0){2}}
        }
    \end{picture}
    \label{onesided}
  \end{equation}

  \bigskip

  \noindent
  where length $|n| - r$ shown as if positive.  Under the conditions
  of the Theorem, $\Sigma^+$ still has infinitely many descendants
  with respect to both $J^-_0$ and~$J^+_0$.
\end{rem}

\paragraph{Going `past the \hw'.} The above concludes the analysis of
codimension-2 cases. In the subsequent analysis of the codimension~3,
however, it will be very useful to know more subtle properties of the
chosen representatives of singular vectors.  As we saw
in~\req{onesided}, descendants of $\Sigma^-$ may not cover a part of
the extremal subdiagram because of the {\it Verma\/} \hw{} conditions
satisfied at $\ket{\rm mff}^-$. In applications, however, one needs
to know whether or not a given {\it Verma\/} submodule in a relaxed
Verma module is actually generated from a charged singular vector in
some relaxed Verma submodule.\footnote{Both cases can occur, and the
  distinction between the two possibilities is crucial for the correct
  classification of degenerate cases, and also, for the construction
  of embedding diagrams, etc.}

Thus, given a Verma \hw{} state $\ket{X}$, which satisfies
$J^+_0\ket{X}=0$, we are interested in whether there exists a state
$\ket{Y}$ {\it in the relaxed Verma module\/} such that
$\ket{X}=J^-_0\ket{Y}$, or, to put it differently, whether
$(J^-_0)^{-1}\ket{X}$ can be defined as an element of the relaxed
Verma module.

We now answer this question for vectors \req{sigmaminus} and
\req{sigmaplus}. First of all, whenever $\Sigma^-$ is in the Verma
submodule $\mC'$ generated by ${\rm MFF}^-(r,s,k)$, we consider
instead of $\Sigma^-$ the \hw{} vector of that submodule:
\begin{equation}
  \ket{{\widehat\Sigma}^-(r,s,\jrc^-(r, s, n, k),k)}\equiv
  \ket{\rm mff}^-=
  {\cal MFF}^-(r, s, k)\,
\ketSL{\jrc^-(r, s, n, k),\Lambdarc^-(r, s, n, k), k|{n}}
\end{equation}
of which $\Sigma^-$ is a $J^-_0$-descendant.

The vector $\ket{{\widehat\Sigma}^-(r,s,\jrc^-(r, s, n, k),k)}$ is the
farthest right one can reach by acting with $J^+_0$ on the original
singular vector $\ket{\Sigma^-(r,s,\jrc^-(r, s, n, k),k)}$. To map it
farther right, to the outside of the Verma submodule~$\mC'$, one
employs the operator $(J^-_0)^{-1}$, which is understood as one of the
`continued' operators from \req{sl2weylaction}--\req{properties}:
\begin{equation}
  \ket{{\widetilde\Sigma}^-(r,s,\jrc^-(r, s, n, k),k)} =
  (J^-_0)^{-1}\,\ket{{\widehat\Sigma}^-(r,s,\jrc^-(r, s, n, k),k)}\,.
  \label{tildeSigma}
\end{equation}
This can be pictured as
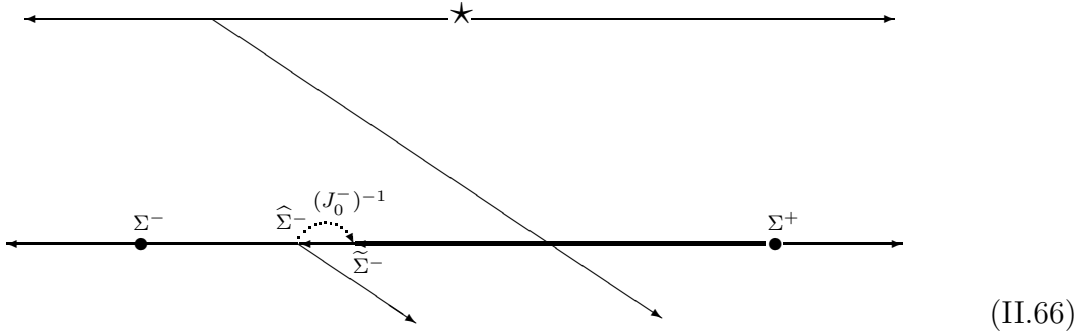
\begin{equation}
  \unitlength=1pt
  \begin{picture}(250,120)
    \put(0,30){
      {
        \put(138,85){\vector(1,0){160}}
        \put(129,85){\vector(-1,0){160}}
        \put(40,85){\vector(3,-2){170}}
        \put(72,0){\vector(-1,0){110}}
        \put(72,0){\vector(3,-2){45}} 
        \put(249,0){$\vector(-1,0){176}$} 
        \put(256,0){$\vector(1,0){45}$}
        }
      \put(130,83){\Large$\star$}
      \put(10,1){${}^{\Sigma^-}$}
      \put(10,-3){$\bullet$}
      \put(64,10){${}_{{\widehat\Sigma}^-}$}
      \put(93,-7){${}_{{\widetilde\Sigma}^-}$}
      \put(250,1){${}^{\Sigma^+}$}
      \put(250,-3){$\bullet$}
      \linethickness{.7pt}
      \bezier{12}(70,0)(83,16)(94,0)
      \put(91.5,4){\vector(1,-2){2}}
      \put(78,17){${}_{(J^-_0)^{-1}}$}
      \linethickness{1.2pt}
      \put(249,0){$\vector(-1,0){155}$} 
      }
  \end{picture}
\end{equation}

\bigskip

\noindent
The conditions for $\ket{{\widetilde\Sigma}^-(r,s,\jrc^-(r, s, n,
  k),k)}$ to exist are given by the following Lemma. Let
\begin{equation}
  f(r,s,n,k)=\left\{
    \begin{array}{ll}
      \prod\limits_{m=0}^{2r-n}(r-m-n-(k+2)s)\,,&1\leq n\leq2r\,,\\
      1\,,&n\geq2r+1\,.
    \end{array}
  \right.\label{f}
\end{equation}

\begin{lemma}\label{lemma:continue}In the relaxed Verma module
  $\mR_{\jrc^-(r, s, n, k), \Lambdarc^-(r, s, n, k),k}$ with $n<0$ and
  $|n|\leq r(s+1)$,\nopagebreak

  {\rm I.}~the vector
  \begin{equation}
    \ket{{\widetilde\Sigma}^-(r,s,\jrc^-(r, s, n, k),k)}=
    (J^-_0)^{-1}\,{\cal MFF}^-(r,s,k)
    \ketSL{\jrc^-(r, s, n, k), \Lambdarc^-(r, s, n, k), k|{n}}
    \label{continued}
  \end{equation}
  exists if and only if $f(r,s,|n|,k)\neq0$, in which case it is a
  relaxed singular vector;

  {\rm II.}~Under the conditions of Theorem~\ref{thm:codim2}, a
  $J^+_0$-descendant of $\ket{\Sigma^+(r,s,\jrc^-(r, s, n, k),k)}$ is
  proportional to vector~\req{continued}:
  \begin{equation}
    (J^-_0)^{r s - n - r - 1}
    \ket{\Sigma^+(r,s,\jrc^-(r, s, n, k),k)}=
    (-1)^r a(r,s,n,k)\,\ket{{\widetilde\Sigma}^-(r,s,\jrc^-(r, s, n, k),k)}\,,
  \end{equation}
  where
  \begin{eqnarray}
    a(r,s,n,k)\kern-4pt&=&\kern-4pt
    {\prod\limits_{i=0}^{r(s+1)-1}
      (\Lambdarc^-(r,s,n,k)-i(i+1)-2i\,\jrc^-(r,s,n,k))
      \over
      \prod\limits_{i=0}^{2r-1}(n - i - r(s-1))}\\
    {}\kern-4pt&=&\kern-4pt
    \prod_{i=0}^{r(s+1)-1}(i-r-n+(k+2)s) \prod_{i=0}^{r(s-1)-1}(n-i)\,.
    \nonumber
  \end{eqnarray}
\end{lemma}

To prove the Lemma, we observe that the action of $(J^-_0)^{-1}$ is
given by~\req{properties} and by Eqs.~\req{somevanish} rewritten~as
\begin{equation}\new
  \begin{array}{rcll}
    (J^-_0)^{-1}\ketSL{j,\Lambda,k|m}&=&
    \ketSL{j,\Lambda,k|m+1}\,,&m\leq-1\,,\\
    (J^-_0)^{-1}\ketSL{j,\Lambda,k|m}&=&
    {1\over \Lambda-m(m+1)-2mj}\ketSL{j,\Lambda,k|m+1}\,,&m\geq0\,.
  \end{array}\label{inverse}
\end{equation}
This applies unless the denominator in the last formula vanishes.  A
simple analysis of the relative $J^0_0$-charge of
$\ket{{\widehat\Sigma}^-(r,s,\jrc^-(r, s, n, k),k)}$ shows that
$f(r,s,-n,k)$ is precisely the function responsible for the existence
of $\ket{{\widetilde\Sigma}^-(r,s,\jrc^-(r, s, n, k),k)}$: to
evaluate~\req{tildeSigma}, one uses Eqs.~\req{properties} and then, as
negative powers of $J^-_0$ reach the \hw{} state, one
applies~\req{inverse} the appropriate number of times; the relevant
factors from the denominators are precisely the above~$f(r,s,-n,k)$,
whence the lemma follows.

Thus, whenever state~\req{tildeSigma} exists, it is proportional to a
$J^-_0$-descendant of $\Sigma^+$.

\paragraph{Codimension $3$.} The codimension-$3$ cases are more
involved than the previous ones. Let us begin with the case where a
further degeneration occurs in Theorem~\ref{twomassive}.
\begin{thm}\label{thm:3rat}
  The \hw{} of the relaxed Verma module $\mR_{j,\Lambda,k}$ belongs to
  the set $O_{rrc}$ of the list of degeneration patterns if and only
  if $j=\jrc^{\sigma}(r,s, n, k)$ and $\Lambdarc^+(r,s, n, k)$ as
  in~\req{bars}, where
  \begin{equation}
    (\sigma,r,s,n,k)\in
    \Bigl(
    \{\pm\}\times\oN\times\oN\times\oZ\times\oQ\bigcup\,\oA(\oQ)
    \Bigr)
    \setminus\,\oB \setminus\,\oD\,,
  \end{equation}
  with $\oA(\cP)$ and $\oB$ as in~\req{oAoB} and
  \begin{equation}
    \oD=\Bigl\{(\sigma,r,s,n,\frac{p}{q})\!\Bigm|
    \!r,s,q\in\oN,~p,n\in\oZ,~|r-s(\frac{p}{q}+2)|
    \in\{|n|,|n|+1,|n|+2,\ldots\}\Bigr\}\,.
  \end{equation}
  Then,
  \begin{enumerate}\addtolength{\parskip}{-4pt}
  \item $\mR_{j,\Lambda, k}$ contains a submodule $\mC$ generated from
    the charged singular vector $\ket{C(n, j, \frac{p}{q})}$; if
    $n<0$, then $\mC\approx\mM_{j+n,\frac{p}{q}}$, where
    $\mM_{j+n,\frac{p}{q}}$ is a Verma module with at least two
    singular vectors; if $n\geq0$,
    $\mC\approx\mM_{j+n+1+\frac{p}{2q},\frac{p}{q};1}$.
    
  \item Any other primitive submodule in $\mR_{j,\Lambda,k}$ is
    isomorphic either to a relaxed Verma module or, depending on
    whether $n<0$ or $n\geq0$, to a Verma module or a twisted Verma
    module with the twist parameter $\theta=1$, respectively.
    Further,
    \begin{enumerate}
    \item\label{nneg} if $n<0$, each Verma submodule
      $\mC'\subset\mR_{j,\Lambda,k}$ is a submodule in $\mC$ and,
      moreover, there exists a relaxed Verma submodule $\mR'$ such
      that $\mC'$ is generated from a charged singular vector
      in~$\mR'$.

    \item\label{npos} if $n\geq0$, each twisted Verma submodule
      $\mC'\subset\mR_{j,\Lambda,k}$ is a submodule in $\mC$ and,
      moreover, there exists a relaxed Verma submodule $\mR'$ such
      that $\mC'$ is generated from a charged singular vector
      in~$\mR'$.

    \item for any relaxed Verma submodule
      $\mR'\subset\mR_{j,\Lambda,k}$, the intersection
      $\mC'=\mR'\cap\mC$ is non-empty.  If, then, $\ket{v}$ is the
      \hw{} vector of $\mC'$, there exists a vector
      $\widetilde{\ket{v}}\in\mR_{j,\Lambda,k}$, defined as
      \begin{equation}
        \widetilde{\ket{v}}=\left\{\kern-4pt\new
          \begin{array}{ll}
            (J^+_0)^{-1}\,\ket{v}\,,&n\geq0\,,\\
            (J^-_0)^{-1}\,\ket{v}\,,&n<0\,,
          \end{array}\right.
        \label{tildev}
      \end{equation}
      that satisfies relaxed \hw{} conditions and generates the
      relaxed Verma submodule $\mR'$.
    \end{enumerate}
    
  \item For any primitive submodule $\mC'\subset\mC$, there exists a
    relaxed Verma submodule $\mR'$ such that
    \begin{enumerate}\addtolength{\parskip}{-4pt}

    \item $\mC'\subset\mR'\cap\mC$;

    \item $\mR'$ is generated from a vector $\widetilde{\ket{v}}$ such
      that the vector
      \begin{equation}
        \ket{v}=\left\{\kern-4pt\new\begin{array}{ll}
            J^+_0\,\widetilde{\ket{v}}\,,&n\geq0\,,\\
            J^-_0\,\widetilde{\ket{v}}\,,&n<0
          \end{array}\right.
      \end{equation}
      generates a Verma module (if $n<0$) or a twisted Verma module
      (if $n\geq0$) that either coincides with $\mC'$ or contains
      $\mC'$ as a submodule generated from the singular vector
      $\ket{{\rm MFF}^+(a,1,k)}$ (if $n<0$) or $\ket{{\rm
          MFF^{-,1}}(a,1,k)}$ (if $n\geq0$), where $a\in\oN$.
    \end{enumerate}
    Whenever $s=0$, we have $\mR'=\mR_{j,\Lambda,k}$.
  \end{enumerate}
\end{thm}
Thus, the structure of $\mR_{j,\Lambda,k}$ is completely determined by
the structure of its Verma
submodule~$\mC=\mM_{\jminus(r,s,\frac{p}{q}),\frac{p}{q}}\subset
\mR_{j,\Lambda,k}$ (if $n<0$) or
$\mC=\smM_{\jplus(r,s,\frac{p}{q}),\frac{p}{q};1}\subset
\mR_{j,\Lambda,k}$ (if $n\geq0$).

\bigskip

The case $O_{rcc}$ of the list of possible degenerations is the most
involved one, and we describe it in the following Theorem.
\begin{thm}\label{thm:overlap0} The \hw{} of the module
  $\mR_{j,\Lambda,k}$ belongs to the set $O_{rcc}$ of the list of
  degeneration patterns if and only if
  \begin{equation}
    j=\jcc(n,m)\,,\quad\Lambda=\Lambdacc(n,m)\,,\quad-n,m+1\in\oN\,,\quad
    k+2=\frac{r\pm(m-n)}{s}\,,\quad r,s\in\oN.
  \end{equation}
  In this case, the module $\mR_{j,\Lambda,k}$ contains a Verma
  submodule $\mC_-$ generated from the charged singular vector
  $\ket{C(n,\jcc(n,m), k)}$ and a Verma submodule $\mC_+$ (with the
  twist parameter $\theta=1$) generated from the charged singular
  vector $\ket{C(m,\jcc(n,m), k)}$.  Each of the modules $\mC_-$ and
  $\mC_+$ admits at least one singular vector. Moreover, a singular
  vector $\ket{{\rm MFF}^\pm(a, b, k)}$ exists in the module $\mC_-$
  if and only if the singular vector $\ket{{\rm MFF}^{\mp,1}(a, b,
    k)}$ with the same $a,b\in\oN$ exists in~$\mC_+$.
  
  Any other primitive submodule $\mN\subset\mR_{j,\Lambda,k}$ is one
  of the following:
  \begin{enumerate}\addtolength{\parskip}{-4pt}
  \item $\mN$ is a Verma submodule, such that $\mN\subset\mC_-$;

  \item $\mN$ is a twisted Verma submodule with the twist parameter~1,
    such that $\mN\subset\mC_+$;

  \item $\mN$ is a relaxed Verma module, in which case the
    intersections $\mN\cap\mC_-=\mC_-'$ and $\mN\cap\mC_+=\mC_+'$ are
    non-empty. Whenever $\mC'_+$ is generated from a singular vector
    $\ket{{\rm MFF}^\pm(a,b,k)}$, $\mC'_-$ is generated from singular
    vector $\ket{{\rm MFF}^\mp(a,b,k)}$ with the same $a,b\in\oN$ (and
    the above~$k$).
  \end{enumerate}
\end{thm}

Next, we give explicit constructions of singular vectors that generate
submodules described in Theorem~\ref{thm:overlap0}:
\begin{thm}\label{thm:overlap}
  Under the conditions of Theorem~\ref{thm:overlap0},
  \begin{enumerate}\addtolength{\parskip}{-4pt}

  \item\label{1} whenever $k+2=\frac{m - n + r}{s}$, the module
    $\mR_{j,\Lambda,k}$ contains a relaxed submodule $\mR'$ such that
    its intersections with $\mC_-$ and $\mC_+$ are generated from the
    following vectors:
    \begin{equation}\new
      \begin{array}{rcl}
        \ket{{\rm mff}}^+&=&{\cal MFF}^+(r, s + 1, k)\,(\Jminus_0)^{-n}\,
        \ketSL{\jcc(n,m), \Lambdacc(n,m), k}\,,\\
        \ket{{\rm mff}}^{-,1}&=&
        {\cal MFF}^{-,1}(r, s + 1, k)\,(\Jplus_0)^{m+1}\,
        \ketSL{\jcc(n,m), \Lambdacc(n,m), k}\,.
      \end{array}
    \end{equation}
    respectively, and $\mR'$ itself is generated from any of the states
    \begin{equation}
      (J^-_0)^{-1}\,\ket{{\rm mff}}^+\qquad\mbox{\sl or}\qquad
      (J^+_0)^{-1}\,\ket{{\rm mff}}^{-,1}\,.
    \end{equation}
    The states $\ket{{\rm mff}}^+$ and $\ket{{\rm mff}}^{-,1}$ are
    charged singular vectors in $\mR'$; after taking the quotient of
    $\mR'$ with respect to $\ket{{\rm mff}}^+$ and $\ket{{\rm
        mff}}^{-,1}$, the states remaining on the top level furnish an
    irreducible representation of the Lie algebra $\SL2$ of
    dimension~$m-n+2r$;

  \item\label{II-i} whenever $k+2=\frac{-m + n + r}{s}$ and
    $2r-m+n\leq-1$, the module $\mR_{j,\Lambda,k}$ contains a relaxed
    Verma submodule $\mR'$ such that its intersections with $\mC_-$
    and $\mC_+$ are generated from the vectors
    \begin{equation}\new\begin{array}{rcl}
        \ket{{\rm mff}}^-&=&{\cal MFF}^-(r, s, k)\,(\Jminus_0)^{-n}\,
        \ketSL{\jcc(n,m), \Lambdacc(n,m), k}\\
        \ket{{\rm mff}}^{+,1}&=&{\cal MFF}^{+,1}(r, s, k)\,(\Jplus_0)^{m+1}\,
        \ketSL{\jcc(n,m), \Lambdacc(n,m), k}
      \end{array}
      \label{mffpm}
    \end{equation}
    respectively, and $\mR'$ itself is generated from any of the
    states
    \begin{equation}
      (J^-_0)^{-1}\,\ket{{\rm mff}}^-\qquad\mbox{\sl or}\qquad
      (J^+_0)^{-1}\,\ket{{\rm mff}}^{+,1}\,.
    \end{equation}
    The vectors $\ket{{\rm mff}}^-$ and $\ket{{\rm mff}}^{+,1}$ are
    charged singular vectors in $\mR'$; after taking the quotient with
    respect to $\ket{{\rm mff}}^+$ and $\ket{{\rm mff}}^{-,1}$, the
    states remaining on the top level of $\mR'$ furnish an irreducible
    representation of the Lie algebra $\SL2$ of dimension~$m-n-2r$;

  \item\label{II-ii} whenever $k+2=\frac{-m + n + r}{s}$ and
    $2r-m+n\geq0$, the module $\mR_{j,\Lambda,k}$ contains submodules
    $\mC'_-\subset\mC_-$ and $\mC'_+\subset\mC_+$ that are generated
    from singular vectors~\req{mffpm} respectively.  There does not
    exist a relaxed Verma submodule $\mR'\subset\mR_{j,\Lambda,k}$ in
    which either $\ket{{\rm mff}}^-$ or $\ket{{\rm mff}}^{+,1}$ would
    be a charged singular vector.

    Moreover, whenever $2r-m+n\geq1$, the states
    $\;(J^+_0)^i\,\ket{{\rm mff}}^{+,1}\;$ and
    $\;(J^-_0)^{2r-m+n-1-i}\,\ket{{\rm mff}}^-$,
    $i=0,\ldots,2r-m+n-1$, satisfy the relaxed \hw{} conditions, are
    in the same grade for each $i$ and are linearly independent.  The
    module \,$\mC'_-$ contains the singular vector $\ket{{\rm
        MFF}^+(2r-m+n,1,\frac{-m + n + r}{s}-2)}$, \ while the module
    \,$\mC'_+$ contains the singular vector \ $\ket{{\rm
        MFF}^{-,1}(2r-m+n,1,\frac{-m + n + r}{s}-2)}$.  After taking
    the quotients with respect to the submodules generated from these
    two singular vectors, the states remaining in the top level of the
    submodule $\mC'_-\bigcup\mC'_+$ arrange into a direct sum of two
    irreducible representations of the $\SL2$ Lie algebra of the
    dimension~$2r-m+n$ each.

  \end{enumerate}
\end{thm}

Parts \ref{1} and~\ref{II-i} of Theorem~\ref{thm:overlap} follow from
Lemma~\ref{lemma:continue}. This can be illustrated as
\begin{equation}
  \unitlength=1pt
  \begin{picture}(250,130)
    \put(-20,30){
      {\thicklines
        \put(138,85){\vector(1,0){180}}
        \put(129,85){\vector(-1,0){170}}
        \put(20,85){\vector(3,-2){170}}  
        \put(290,85){\vector(-3,-2){170}}  
        \put(102,0){\vector(-1,0){140}}
        \put(102,0){\vector(3,-2){44}} 
        \put(208,0){\vector(-3,-2){44}} 
        \put(208,0){\vector(1,0){100}} 
        }
      \put(10,60){\Large$\mC_-$}
      \put(300,60){\Large$\mC_+$}
      \put(50,-20){\Large$\mC_-'$}
      \put(240,-20){\Large$\mC_+'$}
      \put(130,83){\Large$\star$}
      \put(10,1){${}^{\Sigma^-}$}
      \put(10,-3){$\bullet$}
      \put(83,-8){${}_{\ket{\rm mff}^\pm}$}
      \put(201,-8){${}_{\ket{\rm mff}^{\mp,1}}$}
      \put(250,1){${}^{\Sigma^+}$}
      \put(250,-3){$\bullet$}
      \bezier{60}(76,0)(140,0)(204,0)
      \linethickness{.2pt}
      \put(104,6.5){\vector(1,0){104}}
      \put(104,6.5){\vector(-1,0){2}}
      \put(133,7){${}^{m+1-n\pm2r}$}
      \put(24,90){\vector(1,0){108}}     
      \put(24,90){\vector(-1,0){2}}
      \put(70,91){${}^{-n}$}
      \put(137,90){\vector(1,0){153}}     
      \put(137,90){\vector(-1,0){2}}
      \put(200,91){${}^{m+1}$}
      }
  \end{picture}
  \label{gap}
\end{equation}

\smallskip

\noindent
where the upper signs are to be taken in case \ref{1} of the Theorem,
and the lower signs, in case~\ref{II-i}.

In case \ref{II-ii} of the Theorem, the appearance of singular vectors
$\ket{{\rm MFF}^+(2r\!-\!m\!+\!n,1,\frac{n + r - m}{s}\!-\!2)}$ and
$\ket{{\rm MFF}^{-,1}(2r-m+n,1,\frac{n + r - m}{s}-2)}$ in the modules
$\mC'_-$ and $\mC'_+$ respectively follows by a direct analysis of the
relevant \hw s.  This can be illustrated~by the extremal diagram
\begin{equation}
  \unitlength=1pt
  \begin{picture}(250,120)
    \put(-10,20){
      {\thicklines
        \put(138,85){\vector(1,0){180}}
        \put(129,85){\vector(-1,0){170}}
        \put(90,85){\vector(3,-2){163}}  
        \put(210,85){\vector(-3,-2){163}}  
        \put(175,-.6){\vector(3,-2){37}} 
        \put(130,.6){\vector(-3,-2){37}} 
        \put(130,.6){\vector(1,0){190}} 
        \put(175,-.6){\vector(-1,0){190}} 
        \put(193,.6){\vector(-3,-2){37}} 
        \put(110,-.6){\vector(3,-2){37}} 
        }
      \put(130,83){\Large$\star$}
      \put(65,1){${}^{\Sigma^-}$}
      \put(65,-3.6){$\bullet$}
      \put(225,1){${}^{\Sigma^+}$}
      \put(225,-2.4){$\bullet$}
      \put(163,8){${}_{\ket{{\rm mff}}^-}$}
      \put(122,8){${}_{\ket{{\rm mff}}^{+,1}}$}
      \linethickness{.2pt}
      \put(130,90){\vector(-1,0){40}}
      \put(130,90){\vector(1,0){2}}
      \put(110,92){${}^{-n}$}
      \put(131,-5){\vector(1,0){45}}
      \put(131,-5){\vector(-1,0){2}}
      \put(128,-16){${}^{2r-m-1+n}$}
      \put(137,90){\vector(1,0){72}}
      \put(137,90){\vector(-1,0){2}}
      \put(166,92){${}^{m+1}$}
      }
  \end{picture}
  \label{overlap2}
\end{equation}

\medskip

\noindent
where $\ket{{\rm mff}}^{+,1}$ lies on the upper of the two overlapping
lines, and $\ket{{\rm mff}}^-$, on the lower one.

\begin{rem}
  As we see from the Theorem, whenever two different singular vectors
  appear in the same grade, these are necessarily descendants of
  singular vectors in the corresponding (twisted) {\it Verma\/}
  submodules.  Thus, understanding by the relaxed singular vector
  proper the entire floor in the extremal diagram, we can say that two
  different relaxed singular vectors in the same grade never occur.
\end{rem}

\medskip

This concludes the list of our constructions on the $\tSL2$ side.  The
above formulation was given in such a way as to facilitate comparison
with the $\N2$ superconformal algebra; as we will see later, all of
the above objects and constructions have appropriate $\N2$ analogues.

\subsection{Auxiliaries}
We now introduce two free-field systems that will be needed in
Sec.~\ref{sec:3}.
\subsubsection{Fermions (ghosts)}\lvm
We will need a fermionic system (the so-called ghosts, or a $bc$
system), defined in terms of the operator products as
\begin{equation}
  B(z)\,C(w)={1\over z-w}
\end{equation}
with the \emt
\begin{equation}
  T^{\rm GH}=-B\,\d C\,.
\end{equation}
We use the mode expansions $B(z)=\sum_{n\in\oZ}B_nz^{-n-1}$ and
$C(z)=\sum_{n\in\oZ}C_nz^{-n}$.

Denote by $\Omega$ the module generated from the vacuum $\ketGH0$
defined by the conditions
\begin{equation}
  C_{\geq1}\ketGH0=B_{\geq0}\ketGH0=0\,.
\end{equation}
The thus defined vacuum is an $sl_2$-invariant state \cite{[FMS]};
along with this one, we can choose other highest-weight states,
$\ketGH\lambda$, which belong to the same module and are determined by
\begin{equation}
  C_{\geq1-\lambda}\ketGH\lambda=B_{\geq\lambda}\ketGH\lambda=0\,.
  \label{lambdavac}
\end{equation}
Then, as is easy to see,
\begin{equation}\new\begin{array}{rcl}
    (BC)_0\,\ketGH\lambda&=&-\lambda\,\ketGH\lambda\,,\\
    L^{\rm GH}_0\ketGH\lambda&=&\half\lambda(\lambda-1)\,\ketGH\lambda\,.
  \end{array}
\end{equation}

The character of $\Omega$ is given by
\begin{equation}
  \chi^{\rm GH}(z,q)=
  \Tr_\Omega^{\phantom{y}}(q^{L^{\rm GH}_0}\,z^{-(BC)_0})
  =\prod_{i=0}^\infty(1+zq^i)\,\prod_{i=1}^\infty(1+z^{-1}q^i)\,.
\end{equation}

We can connect the states $\ketGH\lambda$ with different $\lambda$ by
means of operators $c(\mu,\nu)$ and $b(\mu,\nu)$ which are the
products of fermionic modes
\begin{equation}
  c(\mu,\nu)=\prod_{n=1}^{\nu-\mu+1}C_{\mu+n}\,,\qquad
  b(\mu,\nu)=\prod_{n=1}^{\nu-\mu+1}B_{\nu+n}\,,\qquad
  \nu-\mu+1\in\oN\,.
\end{equation}
These $c(\mu,\nu)$ and $b(\mu,\nu)$ map a vector $\ketGH\lambda$ as
follows
\begin{equation}
  c(-\lambda-\ell+1,-\lambda)\,:\,\ketGH\lambda\mapsto
  \ketGH{\lambda+\ell}\,,\quad
  b(-\lambda-\ell,\lambda-1)\,:\,\ketGH\lambda\mapsto
  \ketGH{\lambda-\ell}\,,\quad\ell\in\oN\,.
\end{equation}
All the necessary properties of the objects $c(\mu,\nu)$ and
$b(\mu,\nu)$ can be derived when $\nu-\mu+1$ is a positive integer and
then generalized to the case of arbitrary $\mu$ and $\nu$. For
example,
\begin{equation}
  c(\mu,\nu-1)c(\nu,\nu_1)=c(\mu,\nu_1)\,,\qquad
  b(\mu,\nu-1)b(\nu,\nu_1)=b(\mu,\nu_1)\,.
\end{equation}

\subsubsection{The `Liouville' scalar\label{subsubsec:Liouville}}\lvm
In this subsection we introduce a `Liouville' scalar which will be
used to invert the KS mapping.  This is just a free scalar, called
`Liouville' for the signature of its operator product:
\begin{equation}
  \phi(z)\phi(w)=-\ln(z-w)\,.
\end{equation}
We define vertex operators $\psi=e^\phi$ and $\spsi=e^{-\phi}$ with
the mode expansions
\begin{equation}
  \psi(z)=\sum_{n=-\infty}^\infty{\psi_n\over z^{n}}\,,\qquad
  \spsi(z)=\sum_{n=-\infty}^\infty{\spsi_n\over z^{n-1}}\,,
\end{equation}
and
\begin{equation}
  \d\phi(z)=\sum_{n=-\infty}^\infty{\phi_n\over z^{n+1}}\,,\quad{\rm with}\quad
  \left[\phi_n,\phi_m\right]=-n\delta_{n+m,0}\,.
  \label{psipsi}
\end{equation}
We take the energy-momentum tensor to be
\begin{equation}
  T_\phi=-\half\d\phi\d\phi+\half\d^2\phi\,,
  \label{pseudoemt}
\end{equation}
then $\psi$ and $\spsi$ have the dimensions $0$ and $-1$ respectively.

The $\psi$ and $\spsi$ operators are naturally represented as
intertwining operators on a direct sum
\begin{equation}
  \Xi=\bigoplus_{n\in\oZ}\mH_n\,,
  \label{Hminus}
\end{equation}
where $\mH_n$ is a Verma module over the Heisenberg algebra
\req{psipsi} with the highest-weight vector $\ket{n}_\phi$
\begin{equation}\new
  \begin{array}{l}
    \phi_m\ket{n}_\phi=0\,,\quad m\geq1\,,\qquad
    \psi_m\ket{n}_\phi=0\,,\quad m\geq n+1\,,\qquad
    \spsi_m\ket{n}_\phi=0\,,\quad m\geq -n+2\,,\\
    \phi_0\ket{n}_\phi=-n\ket{n}_\phi\,.
  \end{array}
\end{equation}
Note also that the highest-weight vectors $\ket{n}_\phi$ can be
realized as the operators $\ket{n}_\phi\leadsto e^{n\phi}(0)$.

\subsection{$\N2$\label{subsec:N2}}\lvm
In the description of our second main ingredient, the $\N2$ algebra,
we follow refs.~\cite{[ST2],[ST3]}, particularly as regards the
singular vectors. Other (non-exhaustive) references on the $\N2$
superconformal algebra are~\cite{[Ade],[BFK],[SS],[LVW],[EHy]}.

\subsubsection{The $\N2$ superconformal algebra}\lvm
The $\N2$ superconformal algebra is spanned by Virasoro generators
$L_m$, two fermionic fields $ G^\pm_r$, and a $U(1)$ current $ H_n$
(and a central element $\Ctop$). The nonvanishing commutation
relations are:
\begin{equation}\new
  \begin{array}{ll}
    \commut{L_n}{L_m}=(n-m)L_{n+m}+\frac{\Ctop}{12}(n^3-n)\delta_{n+m}\,,&
    \commut{ H_n}{ H_m}=\frac{\Ctop}{3}n\delta_{n+m}\,,\\
    \commut{L_n}{ H_m}=-m H_{n+m}\,,\quad
    \commut{L_n}{ G^\pm_r}=(\half n-r) G^\pm_{n+r}\,,&
    \commut{ H_n}{ G^\pm_r}=\pm G^\pm_{n+r}\,,\\
    \{ G^+_r,\, G^-_s\}=2L_{r+s}+(r-s) H_{r+s}
    +\frac{\Ctop}{3}(r^2-\frac{1}{4})\delta_{r+s}\,,& n\,,\
    m\in\oZ\,,\quad r,s\in\oZ+\half\,.
  \end{array}
  \label{N2untw}
\end{equation}
We find it more convenient to work with the version of this algebra
obtained by redefining the generators as
\begin{equation}
  L_n=L_n|_{\req{N2untw}}+\half(n+1)H_n\,,\quad
  G_r=G^+_{r+{1\over2}}\,,\quad
  Q_r=G^-_{r-{1\over2}}\,.
\end{equation}
This redefinition affects two things: the modding of the fermions and
the choice of the Virasoro generators. The latter is due to the
freedom of adding a derivative of the $U(1)$ current to the
energy-momentum tensor, and is nothing but a change of basis in the
algebra.  As to the different moddings, they label different members
of a family of isomorphic algebras related by the spectral
flow~\cite{[SS],[LVW]} (see Eq.~\req{U} below). Thus any statement
found for one of these algebras does immediately translate to any
other one. In order to be specific in our constructions below, we need
to choose the concrete representative from that family as well as the
modding of the generators. These will be
\begin{equation}\new
  \begin{array}{lclclcl}
    \L[L_m,L_n\R]&=&(m-n)L_{m+n}\,,&\qquad&[ H_m, H_n]&=
    &\frac{\Ctop}{3}m\delta_{m+n,0}\,,\\
    \L[L_m, G_n\R]&=&(m-n) G_{m+n}\,,&\qquad&[ H_m, G_n]&=& G_{m+n}\,,
    \\
    \L[L_m, Q_n\R]&=&-n Q_{m+n}\,,&\qquad&[ H_m, Q_n]&=&- Q_{m+n}\,,\\
    \L[L_m, H_n\R]&=&\multicolumn{5}{l}{-n H_{m+n}+\frac{\Ctop}{6}(m^2+m)
      \delta_{m+n,0}\,,}\\
    \L\{ G_m, Q_n\R\}&=&\multicolumn{5}{l}{2L_{m+n}-2n H_{m+n}+
      \frac{\Ctop}{3}(m^2+m)\delta_{m+n,0}\,,}
  \end{array}\qquad m,~n\in\oZ\,.
  \label{topalgebra}
\end{equation}
Denote this algebra as $\cA$. Its generators $L_m$, $ Q_m$, $ H_m$,
and $ G_m$ will be called the Virasoro generators, the `BRST'
current,\footnote{$Q(z)$ {\it is\/} the BRST current when the algebra
  \req{topalgebra} is realized on the bosonic string
  world-sheet\cite{[GS2],[BLNW]}.} the $U(1)$ current, and the spin-2
fermionic current respectively.  As a Lie superalgebra, $\cA$ is
generated by the following set of elements:
\begin{equation}\new\begin{array}{lll}
    e_1=\half Q_0\,,\quad& e_2= G_1\,,& e_3= H_1\,,\\
    h_1=L_0\,,& h_2=L_0+ H_0+\frac{\Ctop}{3}\,,\quad&h_3=\frac{\Ctop}{3}\,,\\
    f_1= G_0\,,& f_2=\half Q_{-1}\,,& f_3= H_{-1}\,.
  \end{array}\label{generatedby}
\end{equation}
The elements $h_i$, $i=1,2,3$ generate the Cartan subalgebra $\frh$.

Now we define the {\it topological\,\footnote{The name is inherited
    from the non-critical bosonic string, where matter vertices can be
    dressed into $\N2$ primaries that satisfy the highest-weight
    conditions \req{tophw}; in that context, the algebra
    \req{topalgebra} is viewed as a topological algebra.} Verma
  modules\/} over the $\N2$ algebra; the highest-weight vector in any
such module satisfies the annihilation and eigenvalue conditions
\begin{eqnarray}
  Q_{\geq0}\,\ket{h, t}_{\rm top}&=& G_{\geq0}\,\ket{h, t}_{\rm top}~{}={}~
  L_{\geq1}\,\ket{h, t}_{\rm top}~{}={}~
  H_{\geq1}\,\ket{h, t}_{\rm top}~{}={}~0\,,\label{tophw}\\
  H_0\,\ket{h, t}_{\rm top}&=&h\,\ket{h, t}_{\rm top}\,,\qquad
  L_0\,\ket{h, t}_{\rm top}=0\,,\qquad
  \Ctop\,\ket{h,t}_{\rm top}=\frac{3(t-2)}{t}\,\ket{h,t}_{\rm top}
\end{eqnarray}
(the last equation being simply a parametrization of the central
charge in terms of a new parameter, $t$; later this parameter will be
identified with $k+2$, where $k$ is the $\tSL2$ level).  Accordingly,
the topological Verma module $\mV_{h, t}$ is freely generated by
\begin{equation}
  L_{-m}\,,~m\in\oN\,,\qquad
  H_{-m}\,,~m\in\oN\,,\qquad
  Q_{-m}\,,~m\in\oN\,,\qquad
  G_{-m}\,,~m\in\oN
  \label{topver}
\end{equation}
from the {\it topological \hw\ vector\/} $\ket{h, t}_{\rm top}$.

The extremal diagram of a topological Verma module is of the form
\begin{equation}
  \unitlength=1.00mm
  \begin{picture}(140,40)
    \put(50.00,5.00){
      \put(00.00,00.00){$\bullet$}
      \put(10.00,20.00){$\bullet$}
      \put(10.00,20.00){$\bullet$}
      \put(20.00,30.00){$\bullet$}
      \put(29.70,20.00){$\bullet$}
      \put(40.00,00.00){$\bullet$}
      \put(9.70,19.00){\vector(-1,-2){8}}
      \put(19.70,29.70){\vector(-1,-1){7}}
      \put(22.00,29.70){\vector(1,-1){7}}
      \put(32.00,19.00){\vector(1,-2){8}}
      \put(00.00,13.00){${}_{G_{-2}}$}
      \put(11.00,28.00){${}_{G_{-1}}$}
      \put(27.00,28.00){${}_{Q_{-1}}$}
      \put(37.00,13.00){${}_{Q_{-2}}$}
      \put(19.00,34.00){${}_{\ket{h,t}_{\rm top}}$}
      \put(00.50,-06.00){$\vdots$}
      \put(40.50,-06.00){$\vdots$}
      }
  \end{picture}
  \label{newdiagram2}
\end{equation}
where the bigrading in the plane of the diagram is by (charge, level).

\subsubsection{Automorphisms of the $\N2$ algebra}
\paragraph{Spectral flow transform.}
When applied to the generators of the algebra $\cA$,
Eqs.~\req{topalgebra}, the spectral flow transformation $\cU_\theta$
acts as
\begin{equation}\new
  \begin{array}{rclcrcl}
    L_n&\mapsto&L_n+\theta H_n+\frac{\Ctop}{6}(\theta^2+\theta)
    \delta_{n,0}\,,&{}&
    H_n&\mapsto& H_n+\frac{\Ctop}{3}\theta\delta_{n,0}\,,\\
    Q_n&\mapsto& Q_{n-\theta}\,,&{}& G_n&\mapsto& G_{n+\theta}\,.
  \end{array}\label{U}
\end{equation}
This gives an isomorphic algebra $\cA_\theta$, whose generators
$L^\theta_n$, $ Q^\theta_n$, $H^\theta_n$ and $ G^\theta_n$ can be
taken as the RHSs of \req{U}. For~$\theta\in\oZ$, the spectral flow is
an {\it auto\/}morphism of the algebra $\cA$. For arbitrary $\theta$,
we obtain a family of algebras such that any two of its members,
$\cA_{\theta_1}$ and $\cA_{\theta_2}$, are related by an isomorphism
of the type of \req{U} with $\theta=\theta_1-\theta_2$. This family
includes the Neveu--Schwarz and Ramond $\N2$ algebras, as well as the
algebras in which the fermion modes range over $\pm\theta+\oZ$,
$\theta\in\oC$.

\paragraph{The involutive automorphism.} The $\N2$ algebra has the
involutive automorphism
\begin{equation}\new
  \begin{array}{ll}
    G_n\rightarrow Q_{n+1}\,,& Q_n\rightarrow G_{n-1}\,,\\
    H_n\rightarrow- H_n\,,&L_n\rightarrow L_n-(n+1) H_n\,.
  \end{array}
  \label{I}
\end{equation}
Together with automorphisms~\req{U} for $\theta\in\oZ$,
transformations~\req{I} span the group~$\oZ_2\semi\oZ$ isomorphic to
the group generated by the canonical involution and the spectral flow
of the affine $\SL2$ algebra.

\subsubsection{Twisted topological Verma modules}\lvm
We now define {\it  the twisted topological Verma modules\/}
$\smV_{h,t;\theta}$.  This will be a module over the algebra
$\cA_\theta$ (see \req{U}), with the {\it twisted topological
highest-weight vector\/} $\ket{h,t;\theta}_{\rm top}$ defined by
\begin{equation}\new
  \begin{array}{rclcrcl}
    L_m\ket{h,t;\theta}_{\rm top}&=&0\,,\quad m\geq1\,,\quad&
    Q_\lambda\ket{h,t;\theta}_{\rm top}&=&0\,,
    &\lambda\in-\theta+\oN_0\\
    H_m\ket{h,t;\theta}_{\rm top}&=&0\,,\quad m\geq1\,,&
    G_\nu\ket{h,t;\theta}_{\rm top}&=&0\,,&\nu\in\theta+\oN_0
  \end{array}
  \quad\theta\in\oZ\,;
  \label{gentophwint}
\end{equation}
in addition, the `Cartan' generators act on the state
$\ket{h,t;\theta}_{\rm top}$ as
\begin{equation}\new
  \begin{array}{rcl}
    ( H_0+\frac{\Ctop}{3}\theta)\,\kettop{h,t;\theta}&=&
    h\,\kettop{h,t;\theta}\,,\\
    (L_0+\theta H_0+\frac{\Ctop}{6}(\theta^2+\theta))
    \,\kettop{h,t;\theta}&=&0\,,\\
    \Ctop\,\kettop{h,t;\theta}&=&\frac{3(t-2)}{t}\,\kettop{h,t;\theta}\,.
  \end{array}
  \label{genhw}
\end{equation}
The $\theta=0$ case of \req{gentophwint} describes the `ordinary'
topological Verma modules introduced above.

\medskip

Next, we need the concept of terminating fermionic chains.
\begin{dfn}
  Let $F$ denote either $Q$ or $G$ (the fermionic $\N2$ generators),
  and $\ket{X}$ be an element of a module over the $\N2$ algebra. Fix
  also an integer $n$. We say that the fermionic $F$-chain terminates
  on $\ket{X}$, and write
  $\ldots\,F_{n-3}\,F_{n-2}\,F_{n-1}\,F_{n}\,\ket{X}=0$ if
  $$
  \exists N\in\oZ,\quad N\leq n~:~
  F_{N}\,F_{N+1}\,\ldots\,F_{n}\,\ket{X}=0\,.
  $$
\end{dfn}
Now we are in a position to give the following
\begin{dfn}
  An $\N2$ module $\smU$ is an object of the topological $\N2$
  category $\TOP$ whenever the following conditions are satisfied:
  \begin{enumerate}
    \addtolength{\parskip}{-6pt}
  \item $H_0$ and $L_0$ are diagonalizable operators, i.e. $\smU$ can
    be decomposed into a direct sum of intersections of $H_0$- and
    $L_0$- eigenspaces;
  \item The action with all $H_p$, $p\in\oN$, and $L_p$, $p\in\oN$, on
    any element $\ket{X}\in\smU$ spans a finite-dimensional space;
  \item for any element $\ket{X}$ of $\smU$,
    \begin{equation}
      \forall n\in\oZ\qquad
      \new\begin{array}{ll}
        {\it either}&
        \ldots\,Q_{n-3}\,Q_{n-2}\,Q_{n-1}\,Q_{n}\,\ket X=0\\
        {\it or}    &
        \ldots\,G_{-n-4}\,G_{-n-3}\,G_{-n-2}\,G_{-n-1}\,\ket X=0
      \end{array}\label{terminate}
    \end{equation}
  \end{enumerate}
  Morphisms are standard homomorphisms between $\N2$-modules.
\end{dfn}
Condition \req{terminate} will be called the condition of terminating
fermionic chains.\pagebreak[3]

Note finally that the characters of twisted topological Verma modules
are given by
\begin{equation}\new
  \begin{array}{rcl}
    \chi_{h,\theta}^{\N2}(z,q)&=&
    \Tr_{\smV_{h,t;\theta}}^{\phantom{y}}
    (q^{L_0}\,z^{ H_0})\\
    {}&=&
    q^{-\theta h + \Ctop(\theta^2-\theta)/6}\,z^{h-\Ctop\theta/3}\,\,
    {
      \prod\limits_{i=-\theta+1}^\infty(1+z q^i)\,
      \prod\limits_{i=\theta+1}^\infty(1+z^{-1} q^i)
      \over
      \prod\limits_{i=1}^\infty(1-q^i)^2}
  \end{array}
\end{equation}

\subsubsection{Topological $\N2$ singular vectors}\lvm
Here, we describe singular vectors in the topological Verma modules
introduced above; later, we will see that these are isomorphic to the
$\tSL2$ singular vectors, as suggested in~\cite{[S-sing]}.  We
introduce two operators $g(\mu,\nu)$ and $q(\mu,\nu)$, with
$\mu,\nu\in\oC$, that represent the action of two `$\N2$ Weyl group'
generators $\ssr1$ and $\ssr2$ when $\mu$ and $\nu$ are special
(see~\cite{[ST2],[ST3]} for the details): the operators $g(\mu,\nu)$
and $q(\mu,\nu)$ act on the plane $t=\const$ as
\begin{equation}\new
  \begin{array}{l}
    g(ht+\theta-1,\theta-1)\,:\,\kettop{h,t;\theta}\mapsto
    \kettop{\frac{2}{t}-h,t;ht+\theta-1}\,,\\
    q(-(h+1)t-\theta+1,-\theta-1)\,:\,\kettop{h,t;\theta}\mapsto
    \kettop{\frac{2}{t}-2-h,t;(h+1)t+\theta-1}\,.
  \end{array}
  \label{n2weylaction}
\end{equation}

The action of operators $g(a,b)$ and $q(a,b)$ on the highest-weight
vectors can be extended to the topological Verma module generated from
these highest-weight vectors~\cite{[ST3]}.  This allows us to
construct singular vectors in topological Verma modules.
\begin{thm}\mbox{}\nopagebreak

  {\rm I.}~A topological singular vector exists in the topological
  Verma module $\mV_{h,t}$ if and only if either $h=\htop^+(r,s,t)$ or
  $h=\htop^-(r,s,t)$, where
  \begin{equation}\new
    \begin{array}{ll}
      \htop^+(r,s,t)=-\frac{r-1}{t}+s-1\,,\\
      \htop^-(r,s,t)=\frac{r+1}{t}-s\,,
    \end{array}\qquad r,s\in\oN\,.
    \label{n2singcon}
  \end{equation}

  {\rm II. (\cite{[ST2]})}~All singular vectors in the topological
  Verma module $\mV_{\htop^\pm(r,s,t),t}$ are given by the explicit
  construction:
  \begin{eqnarray}
    \ket{E(r,s,t)}^+\kern-7pt&=&\kern-5pt
    g(-r,(s-1)t-1)\,q((1-s)t,r-1-t)\ldots\cdot{}\nonumber\\
    {}&{}&\qquad{}\cdot
    g((s-2)t-r,t-1)\,q(-t,r-1-(s-1)t)\nonumber\\
    {}&{}&\qquad\qquad
    {}\cdot g((s-1)t-r,-1)\,\kettop{\hplus(r,s,t),t}\,,\label{Tplus}\\
    \ket{E(r,s,t)}^-\kern-7pt&=&\kern-5pt
    q(-r, (s-1) t - 1)\,g((1-s)t, r - t - 1)\ldots\cdot{}\nonumber\\
    {}&{}&\qquad{}\cdot
    q((s-2) t - r, t-1)\,g(-t, r - 1 - (s-1) t)\nonumber\\
    {}&{}&{}\qquad\qquad{}\cdot
    q((s-1) t - r, -1)\,\kettop{\hminus(r,s,t),t}\label{Tminus}
  \end{eqnarray}
  where $r,s\in\oN$ and the factors in the first two lines of each
  formula are
  \begin{eqnarray}
    g(-r - t - m t + s t, -1 + m t)\,q(-m t, r - 1 + m t - s t)\,,
    && s-1\geq m\geq1
    \label{plusfact}\\
  \noalign{\noindent\mbox{and}}
    q(-r - t - m t + s t, -1 + m t)\,g(-m t, r - 1 + m t - s t)\,,
    && s-1\geq m\geq1
    \label{minusfact}
  \end{eqnarray}
  respectively.
\end{thm}
Part~I can be deduced from~\cite{[BFK]}.  Singular vectors
\req{Tminus}, \req{Tplus} satisfy \hw{} conditions \req{gentophwint}
with $\theta=\mp r$ respectively.

\subsubsection{Massive $\N2$ Verma modules}\lvm
We now introduce $\N2$ Verma modules of a different type, in which the
\hw{} states satisfy the following annihilation and eigenvalue
conditions:
\begin{eqnarray}
  Q_{\geq1}\,\ket{h,\ell, t}&=& G_{\geq0}\,\ket{h,\ell, t}~{}={}~
  L_{\geq1}\,\ket{h,\ell, t}~{}={}~
  H_{\geq1}\,\ket{h,\ell, t}~{}={}~0\label{upper}\\
  H_0\,\ket{h,\ell, t}&=&h\,\ket{h,\ell, t}\,,\qquad
  L_0\,\ket{h,\ell, t}~{}={}~\ell\,\ket{h,\ell, t}\,.\label{Cartan}
\end{eqnarray}
The module freely generated from $\ket{h,\ell, t}$ by the remaining
$\N2$ generators will be denoted as~$U_{h,\ell,t}$.  Because of the
property to have a dimension $\ell$ generally different from zero,
$\ket{h,\ell, t}$ are called the massive \hw\ states and
$U_{h,\ell,t}$, accordingly, {\it the massive $\N2$ modules}.

The twists of massive Verma modules are introduced by imposing the
highest-weight conditions (as before, $\ctop=\frac{3(t-2)}{t}$)
\begin{equation}\new
  \begin{array}{rclrcl}
    L_m\ket{h,\ell,t;\theta}\kern-6pt&=&\kern-6pt0\,,\quad m\geq1\,,&
    Q_\lambda\ket{h,\ell,t;\theta}\kern-6pt&=&\kern-6pt
    0\,,\quad\lambda\in-\theta+\oN\\
    H_m\ket{h,\ell,t;\theta}\kern-6pt&=&\kern-6pt0\,,\quad m\geq1\,,&
    G_\nu\ket{h,\ell,t;\theta}\kern-6pt&=&\kern-6pt
    0\,,\quad\nu=\theta+\oN_0\\
    (H_0+\frac{\ctop}{3}\theta)\,\ket{h,\ell,t;\theta}
    \kern-6pt&=&\kern-6pt
    h\,\ket{h,\ell,t;\theta}\,,&
    (L_0+\theta H_0+\frac{\ctop}{6}(\theta^2+\theta))
    \,\ket{h,\ell,t;\theta}\kern-6pt&=&\kern-6pt
    \ell\ket{h,\ell,t;\theta}\,,
  \end{array}\label{twistedmassivehw}
\end{equation}

Extremal diagrams of massive Verma modules have the following `fat'
form for $\theta=0$ (see~\cite{[ST3]} for the details)
\begin{equation}
  \unitlength=1.00mm
  \begin{picture}(140,41)
    \put(50.00,5.00){
      \put(00.00,00.00){$\bullet$}
      \put(10.00,20.00){$\bullet$}
      \put(10.00,20.00){$\bullet$}
      \put(20.00,30.00){$\bullet$}
      \put(30.00,30.00){$\bullet$}
      \put(40.00,20.00){$\bullet$}
      \put(50.00,00.00){$\bullet$}
      \put(01.00,03.00){\vector(1,2){8}}
      \put(10.50,18.50){\vector(-1,-2){8}}
      \put(11.50,23.00){\vector(1,1){7}}
      \put(19.70,29.00){\vector(-1,-1){7}}
      \put(22.20,31.50){\vector(1,0){7}}
      \put(28.70,29.95){\vector(-1,0){7}}
      \put(33.00,30.00){\vector(1,-1){7}}
      \put(39.00,22.00){\vector(-1,1){7}}
      \put(43.00,19.00){\vector(1,-2){8}}
      \put(49.50,03.00){\vector(-1,2){8}}
      \put(00.00,13.00){${}_{Q_2}$}
      \put(07.00,07.00){${}^{G_{-2}}$}
      \put(09.00,26.50){${}_{Q_{1}}$}
      \put(16.00,21.00){${}^{G_{-1}}$}
      \put(23.90,33.50){${}_{Q_{0}}$}
      \put(24.50,25.50){${}^{G_{0}}$}
      \put(37.00,27.50){${}_{Q_{-1}}$}
      \put(32.50,22.00){${}^{G_{1}}$}
      \put(47.00,13.00){${}_{Q_{-2}}$}
      \put(41.00,07.00){${}^{G_{2}}$}
      \put(-20.00,00.00){${}_{\ket{h_{-2},\ell_{-2},t;-2}}$}
      \put(-10.00,21.00){${}_{\ket{h_{-1},\ell_{-1},t;-1}}$}
      \put(11.00,32.00){${}_{\ket{h,\ell,t}}$}
      \put(33.50,32.00){${}_{\ket{h_{1},\ell_{1},t;1}}$}
      \put(43.00,21.00){${}_{\ket{h_{2},\ell_{2},t;2}}$}
      \put(53.00,00.00){${}_{\ket{h_{3},\ell_{3},t;3}}$}
      \put(00.50,-06.00){$\vdots$}
      \put(50.50,-06.00){$\vdots$}
      }
  \end{picture}
  \label{massdiagramdouble}
\end{equation}
The states in the diagram are freely generated from $\ket{h,\ell,t}$
by the $G_{-1}\,G_{-2}\,\ldots$ and $Q_{0}\,Q_{-1}\,\ldots$ arrows in
accordance with the definition of massive Verma modules.  The arrows
mapping back towards $\ket{h,\ell,t}$ are described in the following
lemma:
\begin{lemma} The mappings
  \begin{equation}\new
    \begin{array}{rcll}
      \ket{h_\theta,\ell_\theta,t;\theta}&
      \mapsto&\ket{h_{\theta+1},\ell_{\theta+1},t;\theta+1}\,,&\theta<0\,,\\
      \ket{h_\theta,\ell_\theta,t;\theta}&
      \mapsto&\ket{h_{\theta-1},\ell_{\theta-1},t;\theta-1}\,,&\theta>0
    \end{array}
  \end{equation}
  are implemented by $Q_{-\theta}$ and $G_{\theta-1}$ respectively:
  \begin{equation}\new
    \begin{array}{rcll}
      Q_{-\theta}\,\ket{h,\ell,t;\theta}&=&
      2\ell\ket{h-\frac{2}{t},\ell+h-\frac{2}{t},t;\theta+1}\,,&
      \theta<0\,,\\
      G_{\theta-1}\,\ket{h,\ell,t;\theta}&=&
      2(\ell-h)\ket{h+\frac{2}{t},\ell-h,t;\theta-1}\,,&\theta>0
    \end{array}\label{mapback}
  \end{equation}
\end{lemma}

It is worth mentioning that, while $\ket{h,\ell,t}$
in~\req{massdiagramdouble} satisfies the untwisted \hw{} conditions
($\theta=0$ in~\req{twistedmassivehw}), the other states in the
diagram satisfy the twisted massive \hw{}
conditions~\req{twistedmassivehw} with $\theta$ running over all
integers (from $-\infty$ to $+\infty$ as one moves from left to
right).

It follows from \req{mapback} that, as soon as one of the factors on
the right-hand sides vanishes, the respective state would satisfy the
twisted {\it topological\/} \hw{} conditions~\req{tophw}.  At these
`topological points'$\!$, the extremal diagram branches, e.g.,
\begin{equation}
  \unitlength=1.00mm
  \begin{picture}(140,66)
    \put(30.00, 00.00){
      \put(00.00,00.00){$\bullet$}
      \put(10.00,30.00){$\bullet$}
      \put(20.00,50.00){$\bullet$}
      \put(20.00,50.00){$\bullet$}
      \put(30.00,60.00){$\bullet$}
      \put(40.00,60.00){$\bullet$}
      \put(50.00,50.00){$\bullet$}
      \put(60.00,30.00){$\bullet$}
      \put(70.00,00.00){$\bullet$}
      \put(50.00,40.00){$\bullet$}
      \put(40.00,40.00){$\bullet$}
      \put(30.00,30.00){$\bullet$}
      \put(20.00,10.00){$\bullet$}
      \put(11.00,33.00){\vector(1,2){8}}
      \put(20.50,48.50){\vector(-1,-2){8}}
      \put(00.50,03.50){\vector(1,3){8}}
      \put(09.90,27.50){\vector(-1,-3){8}}
      \put(21.50,53.00){\vector(1,1){7}}
      \put(29.70,59.00){\vector(-1,-1){7}}
      \put(32.20,61.50){\vector(1,0){7}}
      \put(38.70,59.95){\vector(-1,0){7}}
      \put(43.00,60.00){\vector(1,-1){7}}
      \put(49.00,52.00){\vector(-1,1){7}}
      \put(53.00,49.00){\vector(1,-2){8}}
      \put(63.00,28.00){\vector(1,-3){8}}
      \put(69.50,03.80){\vector(-1,3){8}}
      \put(59.30,31.00){\vector(-1,1){7.7}}
      \put(52.50,39.40){\vector(1,-1){7.7}}
      \put(42.10,41.60){\vector(1,0){8}}
      \put(49.55,39.98){\vector(-1,0){8}}
      \put(31.80,32.50){\vector(1,1){8}}
      \put(40.20,39.00){\vector(-1,-1){8}}
      \put(30.20,28.50){\vector(-1,-2){8}}
      \put(20.70,12.30){\vector(1,2){8}}
      \put(10.00,43.00){${}_{Q_2}$}
      \put(17.00,37.00){${}^{G_{-2}}$}
      \put(19.00,56.50){${}_{Q_{1}}$}
      \put(26.00,51.50){${}^{G_{-1}}$}
      \put(33.90,63.50){${}_{Q_{0}}$}
      \put(34.50,55.50){${}^{G_{0}}$}
      \put(47.00,58.00){${}_{Q_{-1}}$}
      \put(42.00,53.00){${}^{G_{1}}$}
      \put(57.00,43.00){${}_{Q_{-2}}$}
      \put(69.00,15.00){${}_{Q_{-3}}$}
      \put(61.00,12.00){${}^{G_{3}}$}
      \put(53.00,31.00){${}^{G_{1}}$}
      \put(44.00,35.50){${}^{G_{0}}$}
      \put(44.00,43.50){${}_{Q_{0}}$}
      \put(31.00,38.00){${}_{Q_{1}}$}
      \put(36.00,31.00){${}^{G_{-1}}$}
      \put(24.00,62.00){${}^{\ket{h,\ell,t}}$}
      \put(00.50,-06.00){$\vdots$}
      \put(70.50,-06.00){$\vdots$}
      \put(20.50,04.00){$\vdots$}
      }
  \end{picture}
  \label{branchdiag}
\end{equation}

\bigskip

\noindent
The inner diagram in \req{branchdiag} corresponds to an $\N2$ {\it
  subrepresentation\/}, which is identified with a singular vector.
By iteratively applying formulas~\req{mapback}, we arrive at
\begin{thm}\label{thm:charged}
  A massive Verma module $U_{h,\ell,t}$ contains a twisted topological
  Verma submodule if and only if $\ell=\theel_{\rm ch}(r,h,t)$, where
  \begin{equation}
    \theel_{\rm ch}(r,h,t)=r(h+\frac{r-1}{t})\,,\quad r\in\oZ;
    \label{3rdconditions}
  \end{equation}
  The corresponding singular vector reads
  \begin{equation}
    \ket{E(r,h,t)}_{\rm ch}=\left\{\kern-4pt\new\begin{array}{ll}
        Q_{r}\,\ldots\,Q_0\,\ket{h,\theel_{\rm ch}(r,h,t),t}&r\leq0\,,\\
        G_{-r}\,\ldots\,G_{-1}\,\ket{h,\theel_{\rm ch}(r,h,t),t}\,,&r\geq1.
      \end{array}\right.
    \label{thirdE}
  \end{equation}
  It satisfies twisted topological \hw{} conditions \req{tophw} with
  $\theta=-r$.
\end{thm}
(these are singular vectors of the `charged' series from
\cite{[BFK]}).

\medskip

The remaining singular vectors in massive Verma modules can be
constructed as follows (we only give the `untwisted' case, that of
twisted massive modules follows by applying the spectral flow
transform).  As follows from the analysis of the Ka\v c
determinant~\cite{[BFK]}, the necessary condition for a massive
singular vector to exist in the module $\mU_{h,\ell,t}$ is
$\ell=\theel(r,s,h,t)$, where
  \begin{equation}
    \theel(r,s,h,t) = \half h -
    \half rs - \frac{1}{4t} + \frac{r^2}{4t} - \fourth h^2 t +
    \fourth s^2
    t\,,\qquad r,s\in\oN\,.
    \label{ellmassiveN2}
  \end{equation}
The necessary and sufficient condition as well as
explicit formulae for the singular vectors are given in the following
\begin{thm}\mbox{}\nopagebreak

  {\rm I. (\cite{[ST4]})} For $\ctop\neq3$, a massive singular vector
  in $\mU_{h,\ell,t}$ can be constructed if and only if
  $\ell=\theel(r,s,h,t)$, where
  $(r,s,h,t)\in(\oN\times\oN\times\oC\times\oC)\setminus\oX$ with
  \begin{equation}
    \oX=\Bigl\{(r,s,h,t)\Bigm|st-r=m-n,\;ht-1=-m-n,\;r+st=0,\quad
    n\in\oN,\;m\in-\oN_0\Bigr\}\,.
  \end{equation}

  {\rm II. (\cite{[ST3]})} The massive singular vectors in
  $\mU_{h,\theel(r,s,h,t),t}$ are of the form
  \begin{equation}\new
    \begin{array}{rcl}
      \ket{S(r,s,h,t)}^-\kern-6pt&=&\kern-6pt
      g(-rs,r+\theta^-(r,s,h,t)-1)\,
      \cE^{-,\theta^-(r,s,h,t)}(r,s,t)\cdot{}
      \\{}&{}&\qquad\qquad\qquad\qquad{}\cdot
      g(\theta^-(r,s,h,t),-1)\,\ket{h,\ell(r,s,h,t),t}\,,
      \\
      \ket{S(r,s,h,t)}^+\kern-6pt&=&\kern-6pt
      q(1-rs,r-\theta^+(r,s,h,t)-1)\,
      \cE^{+,\theta^+(r,s,h,t)}(r,s,t)\cdot{}\\
      {}&{}&
      \qquad\qquad\qquad\qquad
      {}\cdot q(-\theta^+(r,s,h,t),0)\,
      \ket{h,\ell(r,s,h,t),t}\,,
    \end{array}\label{Sgen}
  \end{equation}
  where
  \begin{equation}\new
    \begin{array}{rclcl}
      \theta^-(r, s, h, t)
      &=&\frac{t}{2}(h-\hminus(r,s, t))\,,\\
      \theta^+(r, s, h, t)
      &=&\frac{t}{2}(h-1-\hplus(r,s, t))
    \end{array}\label{theta1theta2}
  \end{equation}
  and $\cE^{\pm,\theta}(r,s,t)$ denotes the spectral flow transform
  \req{U} of the topological singular vector operator
  $\cE^{\pm}(r,s,t)$.  The two expressions \req{Sgen} in the general
  position belong to the same extremal diagram and generate the same
  massive submodule.
\end{thm}
This construction translates in the obvious way to the twisted case.

\medskip

Degenerate cases corresponding to simultaneous occurrence of several
$\N2$ singular vectors can be described similarly to the $\tSL2$ case.
An essential point is to consider on equal footing all states from
each extremal diagram.  Restricting oneself to only top-level
representatives of extremal diagrams leads to additional complications
in describing the structure of $\N2$ modules. In a number of
degenerate cases, the entire $\N2$ submodule is {\it not\/} generated
by the top-level representative of the extremal diagram of the
submodule. This happens already in the case where a charged singular
vector $\ket{E(n,h,t)}_{\rm ch}$ (see Theorem~\ref{thm:charged}) and a
massive singular vector labelled by two integers~$r$ and~$s$ coexist
in a massive $\N2$ module and $r>n$.  Describing this case only in
terms of top-level representatives of each of the singular vectors
would lead one to assume that {\it sub\/}singular vectors exist in the
module, since the chosen representative of the singular vector does
not generate the entire submodule.

Let us describe in more detail the case where two charged singular
vectors coexist with a massive singular vector. This is directly
analogous to the case considered in Part~\ref{II-ii} of
Theorem~\ref{thm:overlap}:
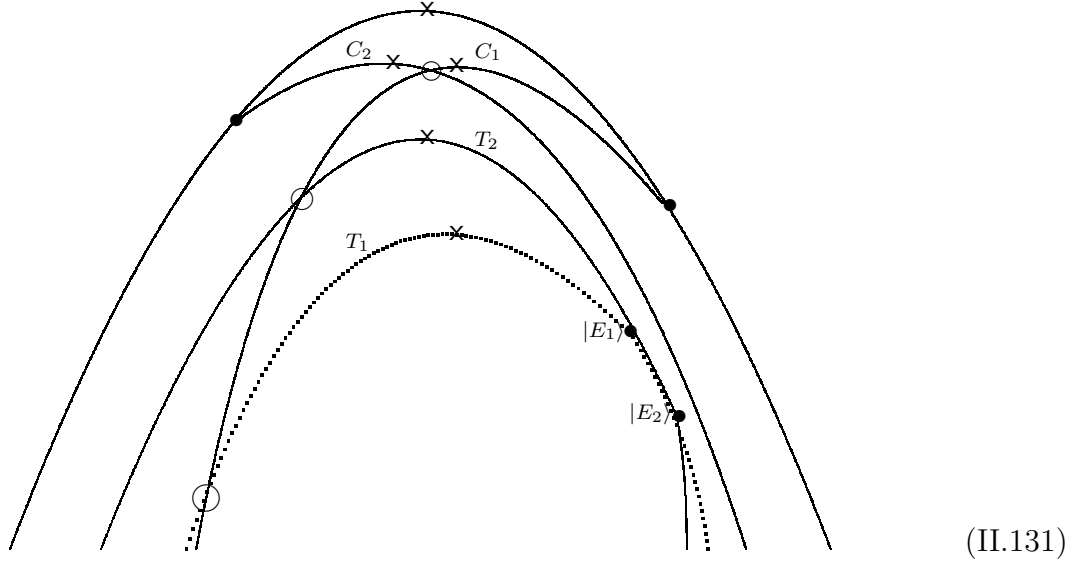
\begin{equation}
 \message{please wait...}
  \unitlength=.8pt
  \begin{picture}(400,280)
    \bezier{2000}(2,1)(196,510)(390,1)
    \put(196.2,253){${\ssf x}$}
    \put(161,229){$^{C_2}$}
    \put(201.5,227){\circle{9}}
    \put(180,228){${\ssf x}$}
    \put(222,228){$^{C_1}$}
    \put(222,186){$^{T_2}$}
    \put(161,138){$^{T_1}$}
    \put(210,226.5){${\ssf x}$}
    \put(196.2,192.5){${\ssf x}$}
    \put(210,147){${\ssf x}$}
    \put(105.5,200){$\bullet$}
    \bezier{1500}(108.5,203)(250,310)(350,1)
    \put(310.5,160){$\bullet$}
    \bezier{1500}(90,1)(155,350)(310.5,164.5)
    \put(140.5,166.5){\circle{10}}
    \put(95,25){\circle{12}}
    \put(25,0){
      \put(267,100){$\bullet$}
      \put(248,96){${}^{\ket{E_1}}$}
      \put(290,60){$\bullet$}
      \put(270,57){${}^{\ket{E_2}}$}
      \bezier{200}(292.5,62)(297,35)(297,1)
      \bezier{1000}(20,1)(160,355)(293,62)
      {\linethickness{1pt}
        \bezier{100}(60,1)(140,235)(271,101)
        \bezier{30}(271,101)(300,55)(306.5,1)
        }
      }
  \end{picture}
\message{...done}
\end{equation}
In this extremal diagram, the crosses denote the top-level
representatives, the filled dots are twisted topological \hw{} states,
and circles indicate that different curves do not intersect, i.e.
correspond to linearly independent states at the overlapping points.
$C_1$ and $C_2$ are extremal diagrams of submodules built on the
charged singular vectors, while $T_1$ and $T_2$ are topological
submodules in the modules corresponding to $C_1$ and $C_2$
respectively.  Submodules $T_1$ and $T_2$ are generated by topological
singular vectors $\ket{E_1}$ and $\ket{E_2}$ respectively.  The pairs
of states lying on the section of the $T_1$ and $T_2$ curves between
$\ket{E_1}$ and $\ket{E_2}$ give linearly independent singular vectors
in the same grade.  For the values of the parameters assumed in the
diagram, choosing the top-level representative of $T_1$ and $T_2$
would conceal the fact that there exist linearly independent singular
vectors in the same grade. By an analysis parallel to the one of
Sect.~\ref{subsub:relaxed}, two linearly independent singular vectors
in the same grade exist whenever $m-n\leq2r$, where $m\geq1$ and
$n\leq-1$ label the two charged singular vectors and $r$ is the first
from the pair of integers $(r,s)$ that label the massive singular
vector.  The top-level representatives would be inside the overlapping
section of the two curves only when both $m$ and $|n|$ are $\leq r$;
however these restrictions have nothing to do with the structure of
submodules that applies whenever $m+|n|\leq 2r$, and is an artifact of
the choice of top-level representatives. Let us note also that in the
case where two linearly independent singular vectors exist in the same
grade, each of these singular vectors belongs to a topological
submodule; this is completely similar to the $\tSL2$ case discussed
above (see Part~\ref{II-ii} of Theorem~\ref{thm:overlap}). For more
details and a systematic account of degenerations of $\N2$ Verma
modules, we refer the reader to~\cite{[ST4]}.  

\bigskip

The above condition for the fermionic chains to terminate is {\it
  not\/} satisfied for the (twisted) massive Verma modules.  Instead,
a fermionic chain terminates whenever there are two or more arrows
along the same line, e.g.
\begin{equation}
  \unitlength=1pt
  \thicklines
  \begin{picture}(160,60)
    \put(40,39.5){\Large$\cdot$}
    \put(46,41){\vector(3,-1){22}}
    \put(38,44){\vector(-3,1){22}}
    \put(69,29){\Large$\cdot$}
    \put(11,49){\Large$\cdot$}
    \put(-6,52){${}^{G_{0}}$}
    \put(25,46){${}^{G_{1}}$}
    \put(50,38){${}^{Q_{-1}}$}
    \put(10,52.5){\vector(-1,0){25}}
    \put(73,30.5){\vector(3,-2){28}}
    \put(85,21){${}^{Q_{-2}}$}
    \put(102,-2){$\vdots$}
    \put(-22,47){$\cdot$}
    \put(-25,46){$\cdot$}
    \put(-28,45){$\cdot$}
  \end{picture}
\end{equation}
which is formalized in the definition of the category of (twisted)
massive highest-weight type modules:

\begin{dfn}
  An $\N2$ module $\smW$ is an object of the category $\MHW$ whenever
  \begin{enumerate}
    \addtolength{\parskip}{-6pt}
  \item $H_0$ and $L_0$ are diagonalizable operators, i.e. $\smU$ can
    be decomposed into a direct sum of intersections of $H_0$-
    $L_0$-eigenspaces;
  \item the action with all $H_p$, $p\in\oN$, and $L_p$, $p\in\oN$, on
    any element $\ket{X}\in\smU$ spans a finite-dimensional space;
  \item for any element $\ket{X}$ of $\smU$,
    \begin{equation}
      \forall n\in\oZ\qquad
      \new\begin{array}{ll}
        {\rm either}&\ldots\,Q_{n-3}\,Q_{n-2}\,Q_{n-1}\,Q_{n}\,\ket X=0\\
        {\rm or}    &\ldots\,G_{-n-3}\,G_{-n-2}\,G_{-n-1}\,G_{-n}\,\ket X=0
      \end{array}
    \end{equation}
  \end{enumerate}
  Morphisms are standard homomorphisms between $\N2$-modules.
\end{dfn}

\section{Kazama--Suzuki and the related mappings\label{sec:3}}
\subsection{The Kazama--Suzuki mapping}\lvm
We use the simplest KS construction associated with the coset
$\tSL2_k\oplus u(1)/u(1)$.  The KS `numerator' $u(1)$ algebra
fermionizes into a couple of spin-1 ghosts, denoted in the following
as $BC$, which allows us to build up the $\N2$ algebra generators in
the standard way \cite{[DvPYZ],[KS]}: The odd generators $Q$ and $G$
are given by
\begin{equation}
  Q=CJ^+\,,\qquad
  G=\frac{2}{k+2}BJ^-\,,
  \label{QGsl}
\end{equation}
while the $U(1)$ current and the \emt\ of the $\N2$ algebra take the
form
\begin{equation}
  H=\frac{k}{k+2}BC-\frac{2}{k+2}J^0\,,\label{Hsl}
\end{equation}
\begin{equation}
  T=\frac{1}{k+2}(J^+J^-)-\frac{k}{k+2}B\d C-\frac{2}{k+2}BCJ^0\,.
  \label{Tsl}
\end{equation}

\begin{lemma}
  The modes of currents \req{QGsl}, \req{Hsl}, and \req{Tsl} satisfy
  the algebra \req{topalgebra}. The eigenvalue of the central element
  in each representation is expressed through the eigenvalue of the
  $\tSL2$ central element as
  \begin{equation}\ctop={3k\over k+2}\,.\label{ctopk}
  \end{equation}
  Thus, Eqs.~\req{QGsl}, \req{Hsl}, and \req{Tsl} determine a mapping
  \begin{equation}
    \FKS:\cA\to\cU\tSL2_k\tensor\,[BC]\,,\label{frttosl(2)}
  \end{equation}
  where $\cA$ is the $\N2$ algebra \req{topalgebra} and $\cU$ denotes
  the universal enveloping (and $[BC]$ is the free fermion theory).
\end{lemma}

The KS mapping does also produce a bosonic current
\begin{equation}
  I^+=\sqrt{\frac{2}{k+2}}(BC+J^0)\label{Heisenberg}
\end{equation}
whose modes commute with the $\N2$ generators \req{QGsl}--\req{Tsl}.
Expanding as $I^+(z)=\sum_{n\in\oZ}I^+_nz^{-n-1}$, we get a Heisenberg
algebra
\begin{equation}
  [I^+_m,\,I^+_n]=m\delta_{m+n,0}\,.
\end{equation}
The \emt\ of this $U(1)$ current is introduced as
\begin{equation}
  T^+=\half\,(I^+)^2-\frac{1}{\sqrt{2(k+2)}}\d\,I^+.
\end{equation}

We define a module $\mH_p^+$ over the Heisenberg algebra by the
highest-weight conditions
\begin{equation}
  I_{\geq1}\,\ket{p}^+=0\,,\qquad I_0\,\ket{p}^+=p\,\ket{p}^+.
  \label{Heishw}
\end{equation}
Its character is
\begin{equation}
  \chi_{p,k}^+(z,q)=
  \Tr_{\mH_p^+}^{\phantom{y}}(q^{L^+_0}\,z^{I^+_0})=
  z^{p}\,
  {q^{{1\over2} p(p+\sqrt{{2\over k+2}})}
    \over\prod\limits_{i=1}^\infty(1-q^i)}\,.
\end{equation}
\begin{lemma}\label{KSidentities}
  Under the KS mapping, we have the identities
  \begin{eqnarray}
    T_{\rm Sug}+T_{\rm GH}&=&T + T^+\\
    \noalign{\noindent
      \mbox{(where $T$ is the \emt~\req{Tsl}), and}}\nonumber
    J^0-BC&=&-2 H+\frac{k-2}{\sqrt{2(k+2)}}\,I^+\,.
  \end{eqnarray}
\end{lemma}

\subsection{The `anti'-Kazama--Suzuki mapping}\lvm
A mapping in the inverse direction to the KS mapping can be defined as
follows. It is associated with the `coset' $\cA\tensor u(1)/u(1)$
where the numerator $u(1)$ is represented in terms of the
antifermionic system~$\psi=e^\phi$, $\spsi=e^{-\phi}$ from
section~\ref{subsubsec:Liouville}.  Then we can identify the affine
$\SL2$ generators as
\begin{equation}\new
  \begin{array}{l}
    J^+= Q\psi\,,\qquad J^-=\frac{3}{3-\ctop}\, G\spsi\,,\\
    J^0=-\frac{3}{3-\ctop}\,H+\frac{\ctop}{3-\ctop}\,\d\phi\,.
  \end{array}
  \label{invKS}
\end{equation}
\begin{lemma}
  For $\ctop\neq3$, generators~\req{invKS} close to the
  algebra~\req{sl2algebra} of the level $k={2\ctop\over3-\ctop}$,
  where $\ctop$ is the $\N2$ central charge (the eigenvalue of the
  central element). Thus, Eqs.~\req{invKS} determine a mapping
  \begin{equation}
    \FKS^{-1}:\tSL2\to\cU\,\cA\tensor[\psi\,\psi^*]\,.
  \end{equation}
\end{lemma}

The Sugawara \emt\ \req{twistedS} takes for the
construction~\req{invKS} the form
\begin{equation}
  T^{\rm Sug}=T+\half\d H+\frac{k+2}{4}H H
  \frac{k+2}{2}H\d\phi+\frac{k}{4}\d\phi\d\phi
\end{equation}
We also have a free scalar, with signature $-1$, whose modes commute
with the $\tSL2$ generators~\req{invKS}:
\begin{equation}
  I^-=\sqrt{\frac{k+2}{2}}(H-\d\phi)\,.
  \label{dF}
\end{equation}
The modes $I^-_n$ introduced as $I^-(z)=\sum^{\infty}_{n=-\infty}I^-_n
z^{-n-1}$ generate a Heisenberg algebra
\begin{equation}
  \L[I^-_n,\,I^-_m\R]=-n\delta_{m+n,0}
  \label{LHeisen}\,.
\end{equation}
This current has the \emt{}
\begin{equation}
  T^-=-\half \left(I^-\right)^2-\frac{1}{\sqrt{2(k+2)}}\d I^-.
\end{equation}
The module $\mH^-_q$ is defined as a Verma module over the Heisenberg
algebra~\req{LHeisen} with the highest-weight vector defined by
\begin{equation}
  I^-_n\ket{q}^-=0\,,\quad n\geq1\,,\qquad I^-_0\ket{q}^-=q\ket{q}^-.
\end{equation}

\begin{lemma}
  Under the anti-KS mapping we have the identities
  \begin{equation}
    T+T_\phi=T^{\rm Sug}+T^-
  \end{equation}
  (where $T$ is the \emt\ of the $\N2$ algebra and $T_\phi$ is the
  \emt\ of Liouville system~\req{pseudoemt}), and
  \begin{equation}
    -2H+\d\phi=J^0+\L(\frac{k}{2}-1\R)\sqrt{\frac{k+2}{2}}I^-.
  \end{equation}
\end{lemma}

\subsection{Composing the KS and anti-KS mappings}\lvm
Let us consider compositions of the KS and anti-KS mappings.
\begin{lemma}\label{lemma:sl2back}
  The composition $F_{\rm KS}^{-1}\circ F_{\rm KS}$ maps the $\tSL2$
  algebra into the $\tSL2$ algebra in the tensor product
  $\cU\,\tSL2\tensor[BC]\tensor[\psi\,\psi^*]$ represented by the
  currents
  \begin{equation}\new
    \begin{array}{l}
      \bar J^+=J^+e^{\phi}\,C\,,\qquad\bar J^-=J^-e^{-\phi}B\,,\\
      \bar J^0=J^0+\frac{k}{2}(\d\phi-BC)\,.
    \end{array}
    \label{sl2back}
  \end{equation}
\end{lemma}
For the \emt s, we then have
\begin{equation}
  \bar T^{\rm Sug}=T^{\rm Sug}+\frac{k}{4}((\d\phi)^2
  -2\d\phi\,BC + \d B\,C - B\,\d C) +
  J^0(\d\phi-BC)
\end{equation}
A remarkable feature of these formulas is the appearance of a {\it
  null current\/} $BC-\d\phi$; the fact that it is null is crucial for
establishing that the generators on the RHSs close to the $\tSL2$
algebra.

\begin{rem}
  The formulas of the lemma can be given a coset-like interpretation
  by noticing that the image of $\tSL2$ under the above embedding is
  characterized by the fact that it commutes with (the modes of) the
  currents
  \begin{equation}\new\begin{array}{rcl}
      I^+&=&\sqrt{\frac{2}{k+2}}(J^0+BC)\,,\\
      F^-&=&\sqrt{\frac{2}{k+2}}(J^0-\frac{k}{2}BC+\frac{k+2}{2}\d\phi)
    \end{array}
  \end{equation}
  (these currents do also commute with each other and are matter-like
  and Liouville-like, respectively).
\end{rem}

The same happens with the $\N2$ algebra under the action of~$F_{\rm
  KS}\circ F_{\rm KS}^{-1}$:
\begin{lemma}\label{lemma:N2back}
  The composition $F_{\rm KS}\circ F_{\rm KS}^{-1}$ maps the $\N2$
  algebra into the $\N2$ algebra in the tensor product
  $\cU\,\cA\tensor[BC]\tensor[\psi\,\psi^*]$ represented by the
  currents
  \begin{equation}\new
    \begin{array}{rcl}
      \bar Q&=&Q e^{\phi}C\,,\qquad\bar G= G e^{-\phi}B\,,\\
      \bar H&=&H+\frac{k}{k+2}(BC-\d\phi)\,,\\
      \bar T&=&T+H(BC-\d\phi)+\frac{k}{2(k+2)}
      \L((\d\phi)^2-2\d\phi\,BC+\d^2\phi - 2 B\,\d C\R)\,.
    \end{array}\label{N2back}
  \end{equation}
\end{lemma}

\begin{rem}
  We also find two currents that commute with the `target' $\N2$
  algebra~\req{N2back}:
  \begin{equation}\new
    \begin{array}{rcl}
      I^-&=&\sqrt{\frac{t}{2}}(H-\d\phi)\,,\\
      F^+&=&\sqrt{\frac{t}{2}}(H-\frac{2}{t}BC-\frac{t-2}{t}\d\phi)
    \end{array}
  \end{equation}
  (these currents are Liouville-like and matter-like respectively, and
  commute with each other).
\end{rem}

\subsection{KS and anti-KS mappings at the level of
  representations\label{subsec:isomorphism}}
\subsubsection{The Verma module case}\lvm
The following theorem is essentially a reformulation of
Lemmas~\ref{lemma:sl2back} and~\ref{lemma:N2back}:
\begin{thm}\label{weaker}\mbox{}\nopagebreak

  {\rm I.}~The composition of the KS and anti-KS mappings induces an
  isomorphism of $\tSL2$ representations
  \begin{equation}
    \smM_{j,k;\theta}\tensor\Omega\tensor\Xi\,{}\approx\bigoplus_{n,m\in\oZ}\,
    \smM_{j,k;\theta+n-m}\tensor
    \mH^+_{\sqrt{\frac{2}{k+2}}(j-\frac{k}{2}\theta-m)}\tensor
    \mH^-_{-\sqrt{\frac{2}{k+2}}(j-\frac{k}{2}(\theta-m)-\frac{k+2}{2}n)}
    \label{slton2}
  \end{equation}
  where on the LHS the $\tSL2$ action is given by the RHSs
  of~\req{sl2back}, while on the RHS the $\tSL2$ algebra acts naturally
  on the modules $\smM_{j,k;\theta}$ and trivially, on the Heisenberg
  modules.

  {\rm II.}~The composition of the anti-KS mapping and the direct KS
  mapping induces an isomorphism of $\N2$ representations
  \begin{equation}
    \smV_{h,t;\theta}\tensor\Xi\tensor\Omega\approx\bigoplus_{n,m\in\oZ}
    \smV_{h,t;\theta+m-n}\tensor
    \mH^-_{\sqrt{\frac{t}{2}}(h-\frac{t-2}{t}\theta+n)}\tensor
    \mH^+_{-\sqrt{{t\over2}}(h-\frac{t-2}{t}(\theta-n)+\frac{2}{t}m)}
  \end{equation}
  where on the LHS the $\N2$ algebra acts by generators~\req{N2back},
  while on the RHS we have the natural action of $\N2$ on the modules
  $\smV_{h,t;\theta}$ and the trivial action on the Heisenberg
  modules.
\end{thm}

\begin{rem}
  The theorem remains valid when the $\tSL2$ or $\N2$ modules are no
  longer Verma modules but rather arbitrary modules from the category
  of highest-weight modules.
\end{rem}

In fact, one has a stronger result, which follows from the analysis of
each of the mappings $\FKS$ and $\FKS^{-1}$ separately:
\begin{thm}\label{mainthm}\mbox{}\nopagebreak

  {\rm I.}~The KS mapping induces an isomorphism of $\N2$
  representations
  \begin{equation}
    \smM_{j,k;\theta}\tensor\Omega\,{}\approx \bigoplus_{m\in\oZ}\,
    \smV_{{-2j\over k+2},k+2;m-\theta}\tensor
    \mH^+_{\sqrt{{2\over k+2}}(j-\frac{k}{2}\theta-m)}
    \label{idspaces}
  \end{equation}
  where on the LHS the $\N2$ algebra acts by
  generators~\req{QGsl}--\req{Tsl}, while on the RHS it acts naturally
  on the twisted topological Verma modules $\smV_{{-2j\over
      k+2},k+2;m-\theta}$ .

  {\rm II.}~The anti-KS mapping induces an isomorphism of $\tSL2$
  representations
  \begin{equation}
    \smV_{h,t;\theta}\tensor\Xi\approx\bigoplus_{n\in\oZ}\,
    \smM_{-\frac{t}{2}h,t-2;n-\theta}\tensor
    \mH^-_{\sqrt{\frac{t}{2}}(h-\frac{t-2}{t}\theta+n)}
    \label{idspaces2}
  \end{equation}
  where on the LHS the $\tSL2$ algebra acts by generators~\req{invKS},
  while on the RHS it acts naturally on the twisted Verma modules
  $\smM_{-\frac{t}{2}h,t-2;n-\theta}$.  
\end{thm}

\begin{rem}
  Observe that for $k\in\oZ$, each of formulas \req{idspaces} and
  \req{idspaces2} translates the periodicity under
  $\theta\mapsto\theta+2$ on the $\tSL2$ side (which is known to hold
  for integrable representations) into the periodicity under
  $\theta\mapsto\theta+k+2$ on the $\N2$ side (which, too, is a
  well-known property of the corresponding $\N2$ representations and
  conformal models, where it is observed, for example, in the
  Landau--Ginzburg formulation).
\end{rem}

\begin{prf} of the Theorem.
  It is easy to see that the LHS of~\req{slton2} contains each state
  $\ketSL{j,k}\tensor\ketGH{\theta}$ with all $\theta\in\oZ$.
  Further, we see that each state $\ketSL{j,k}\tensor\ketGH{\theta}$
  satisfies the $\N2$- and Heisenberg annihilation conditions,
  Eqs.~\req{tophw} and \req{Heishw}, respectively. Therefore, each
  state $\ketSL{j,k}\tensor\ketGH{\theta}$ can be rewritten in the
  form
  \begin{equation}
    \ketSL{j,k}\tensor\ketGH{\theta}=
    \kettop{\frac{-2j}{k+2},k+2;\theta}
    \tensor\ket{\sqrt{\frac{2}{k+2}}(j-\theta)}^+,\quad\theta\in\oZ\,.
    \label{idhw}
  \end{equation}
  
  Thus, as $\theta$ runs over the integers, every $\tSL2$ \hw{} state
  gives rise to infinitely many states satisfying twisted topological
  \hw\ conditions \req{gentophwint}--\req{genhw}. From any of these
  states we can generate the twisted topological Verma module
  $\smV_{\frac{-2j}{k+2},k+2;\theta}$.  All the states in every such
  module have a fixed value of the $I^+_0$ charge
  (see~\req{Heisenberg}), and this charge is different for modules
  with different $\theta$. Thus the modules
  $\smV_{\frac{-2j}{k+2},k+2;\theta}$ do not overlap for different
  $\theta\in\oZ$. Therefore, we have the mapping
  \begin{equation}
    \smM_{j,k;\theta}\tensor\Omega\longleftarrow\bigoplus_{m\in\oZ}\,
    \smV_{{-2j\over k+2},k+2;m-\theta}\tensor
    \mH^+_{\sqrt{{2\over k+2}}(j-\frac{k}{2}\theta-m)}
    \label{work1}
  \end{equation}
  whose kernel may be different from zero only as a result of the
  vanishing of some singular vectors in the modules~$\smV_{{-2j\over
      k+2},k+2;m-\theta}$.  In a similar way, we arrive at the mapping
  \begin{equation}
    \smV_{h,t;\theta}\tensor\Xi\longleftarrow\bigoplus_{n\in\oZ}\,
    \smM_{-\frac{t}{2}h,t-2;n-\theta}\tensor
    \mH^-_{\sqrt{\frac{t}{2}}(h-\frac{t-2}{t}\theta+n)}\,.
    \label{similar}
  \end{equation}

  Now, we tensor~\req{work1} with $\Xi$ and make use of~\req{similar}.
  Note that for the modules under consideration, tensor products
  preserve exactness.  Then the mapping \req{work1} gives rise to
  \begin{equation}
    \smM_{j,k;\theta}\tensor\Omega\tensor\Xi
    \longleftarrow
    \!\!\bigoplus_{m,n\in\oZ}\!\!
    \smM_{j,k;\theta+n-m}\tensor
    \mH^+_{\sqrt{\frac{2}{k+2}}(j-\frac{k}{2}\theta-m)}\tensor
    \mH^-_{-\sqrt{\frac{2}{k+2}}(j-\frac{k}{2}(\theta-m)-\frac{k+2}{2}n)}\,,
    \label{work3}
  \end{equation}
  which is an isomorphism by Theorem~\ref{weaker}, hence
  both~\req{work1} and~\req{similar} are isomorphisms.
\end{prf}\pagebreak[3]

\noindent{\bf Corollary~1.}~Singular vectors in the $\tSL2$ Verma
modules and in the topological $\N2$ Verma modules occur (or do not
occur) simultaneously.  Thus formulas~\req{n2singcon} for positions of
the topological singular vectors are re-derived by translating the
positions of the $\tSL2$ singular vectors under the KS mapping.

\medskip

\noindent{\bf Corollary~2.}~Let there exist a singular vector with the
$J^0_0$-charge equal to $r$ in the $\tSL2$ Verma module on the LHS
of~\req{work1}; then each topological $\N2$ Verma module
$\smV_{{-2j\over k+2},k+2;m-\theta}$ from the RHS of~\req{work1}
contains a topological singular vector, which satisfies \hw\
conditions~\req{gentophwint} with $\theta\rightarrow m-\theta+r$.

\begin{prf}
  Let us consider the tensor product~$\ket{S(r)}\tensor\ketGH\lambda$,
  where~$\ket{S(r)}$ denotes a singular vector with the charge $r$ in
  the $\tSL2$ Verma module from the LHS of~\req{work1} and
  $\ketGH\lambda$ is the ghost $\lambda$-vacuum~\req{lambdavac}.  Let
  us determine which module $\smV_{{-2j\over k+2},k+2;m-\theta}$ from
  the RHS of~\req{work1} the state~$\ket{S(r)}\tensor\ketGH\lambda$
  belongs to.  To this end, we calculate the eigenvalue of~$I^+_0$
  (see~\req{Heisenberg}) on the
  state~$\ket{S(r)}\tensor\ketGH\lambda$:
  $$
  I^+_0\,\ket{S(r)}\tensor\ketGH\lambda=
  \sqrt{\frac{2}{k+2}}(j+r-\frac{k}{2}-\lambda)\,
  \ket{S(r)}\tensor\ketGH\lambda\,.
  $$
  Therefore, the state~$\ket{S(r)}\tensor\ketGH{m+r}$ belongs to
  $\smV_{{-2j\over k+2},k+2;m-\theta}$ and, as follows easily from
  formulas~\req{lambdavac} and~\req{QGsl}, satisfies \hw\
  conditions~\req{gentophwint} with $\theta\rightarrow m-\theta+r$.
\end{prf}

\noindent
A constructive description of the correspondence between singular
vectors is considered in subsection~\ref{subsec:chains}.

\medskip

\noindent{\bf Corollary~3.}~The isomorphism established in the
Theorem implies an identity for the characters:
\begin{equation}
  \Tr_{\mM_{j,k}\tensor\Omega}^{\phantom{y}}
  \Bigl(q^{L^{\rm Sug}_0+L^{\rm GH}_0}\,z^{J^0_0-(BC)_0}\Bigr)=
  \Tr_{\bigoplus\limits_{\theta=-\infty}^\infty
    \smV_{{-2j\over k+2},k+2;\theta}
    \tensor\mH^+_{\sqrt{{2\over k+2}}(j-\theta)}}^{\phantom{y}}
  \Bigl(q^{L_0+L^+_0}\,z^{-2 H_0+{k-2\over\sqrt{2(k+2)}}I^+_0}\Bigr)
\end{equation}
Inserting the characters, given above, of each of the modules
involved, this can be rewritten as the following formal identity:
\begin{equation}\new
  \begin{array}{l}
    {\prod\limits_{i=0}^{\infty}(1 + z q^i)\,
      \prod\limits_{i=1}^{\infty}(1 + z^{-1} q^i) \over
      \prod\limits_{i=0}^{\infty}(1 - z^{-1} q^i)\,
      \prod\limits_{i=1}^{\infty}(1 - z q^i)}\\
    \qquad\qquad{} =
    {\prod\limits_{i=1}^{\infty}(1 + z^2 q^i)
      \prod\limits_{i=1}^{\infty}(1 + z^{-2} q^i)\over
      \prod\limits_{i=1}^{\infty}(1 - q^i)^2}\,
    \biggl(1 + (1+z^2)
    \sum_{r=1}^\infty\Bigl[{z^{-r}\over 1+z^2q^r}+{q^rz^{r-2}\over1+z^{-2}q^r}
    \Bigr]
    \biggr).
  \end{array}
\end{equation}

\subsubsection{The relaxed/massive Verma module case}\lvm
Now, we give an extension of theorems~\ref{weaker}--\ref{mainthm} to
the relaxed and massive Verma modules.
\begin{thm}\label{relaxedweaker}\mbox{}\nopagebreak

  {\rm I.}~The composition of the KS and anti-KS mappings induces an
  isomorphism of $\tSL2$ representations
  \begin{equation}
    \smR_{j,\Lambda,k;\theta}\tensor\Omega\tensor\Xi\,{}\approx
    \bigoplus_{n,m\in\oZ}\,
    \smR_{j,\Lambda,k;\theta+n-m}\tensor
    \mH^+_{\sqrt{\frac{2}{k+2}}(j-\frac{k}{2}\theta-m)}\tensor
    \mH^-_{-\sqrt{\frac{2}{k+2}}(j-\frac{k}{2}(\theta-m)-\frac{k+2}{2}n)}
    \label{relaxedslton2}
  \end{equation}
  where on the LHS the $\tSL2$ action is given by the RHSs
  of~\req{sl2back}, while on the RHS the $\tSL2$ algebra acts naturally
  on the modules $\smR_{j,\Lambda,k;\theta}$ and trivially, on the
  Heisenberg modules.

  {\rm II.}~The composition of the anti-KS mapping and the direct KS
  mapping induces an isomorphism of $\N2$ representations
  \begin{equation}
    \smW_{h,\ell,t;\theta}\tensor\Xi\tensor\Omega\approx\bigoplus_{n,m\in\oZ}
    \smW_{h,t,\ell;\theta+m-n}\tensor
    \mH^-_{\sqrt{\frac{t}{2}}(h-\frac{t-2}{t}\theta+n)}\tensor
    \mH^+_{-\sqrt{{t\over2}}(h-\frac{t-2}{t}(\theta-n)+\frac{2}{t}m)}
  \end{equation}
  where on the LHS the $\N2$ algebra acts by generators~\req{N2back},
  while on the RHS we have the natural action of $\N2$ on the modules
  $\smW_{h,\ell,t;\theta}$ and the trivial action on the Heisenberg
  modules.
\end{thm}

This allows us to obtain, similarly to the above,
\begin{thm}\label{relaxedmainthm}\mbox{}\nopagebreak

  {\rm I.}~The KS mapping induces an isomorphism of $\N2$
  representations
  \begin{equation}
    \smR_{j,\Lambda,k;\theta}\tensor\Omega\,{}\approx\bigoplus_{\lambda\in\oZ}\,
    \smU_{{-2j\over k+2},\frac{\Lambda}{k+2},k+2;\lambda-\theta}\tensor
    \mH^+_{\sqrt{{2\over k+2}}(j-\frac{k}{2}\theta-\lambda)}\,,
    \label{relaxedidspaces}
  \end{equation}
  where on the LHS the $\N2$ algebra acts by
  generators~\req{QGsl}--\req{Tsl}, while on the RHS it acts naturally
  on the twisted massive Verma modules $\smU_{{-2j\over
      k+2},\ell,k+2;\lambda-\theta}$.

  {\rm II.}~The anti-KS mapping induces an isomorphism of $\tSL2$
  representations
  \begin{equation}
    \smU_{h,\ell,t;\theta}\tensor\Xi\approx\bigoplus_{n\in\oZ}\,
    \smR_{-\frac{t}{2}h,t\ell,t-2;n-\theta}\tensor
    \mH^-_{\sqrt{\frac{t}{2}}(h-\frac{t-2}{t}\theta+n)}\,,
    \label{relaxedidspaces2}
  \end{equation}
  where on the LHS the $\tSL2$ algebra acts by generators~\req{invKS},
  while on the RHS it acts naturally on the twisted relaxed Verma
  modules $\smR_{-\frac{t}{2}h,t\ell,t-2;n-\theta}$.
\end{thm}
We could word by word repeat the proof of
theorems~\ref{weaker}--\ref{mainthm} in the case of relaxed Verma
modules.  Instead, we illustrate the main points by the following
diagram:
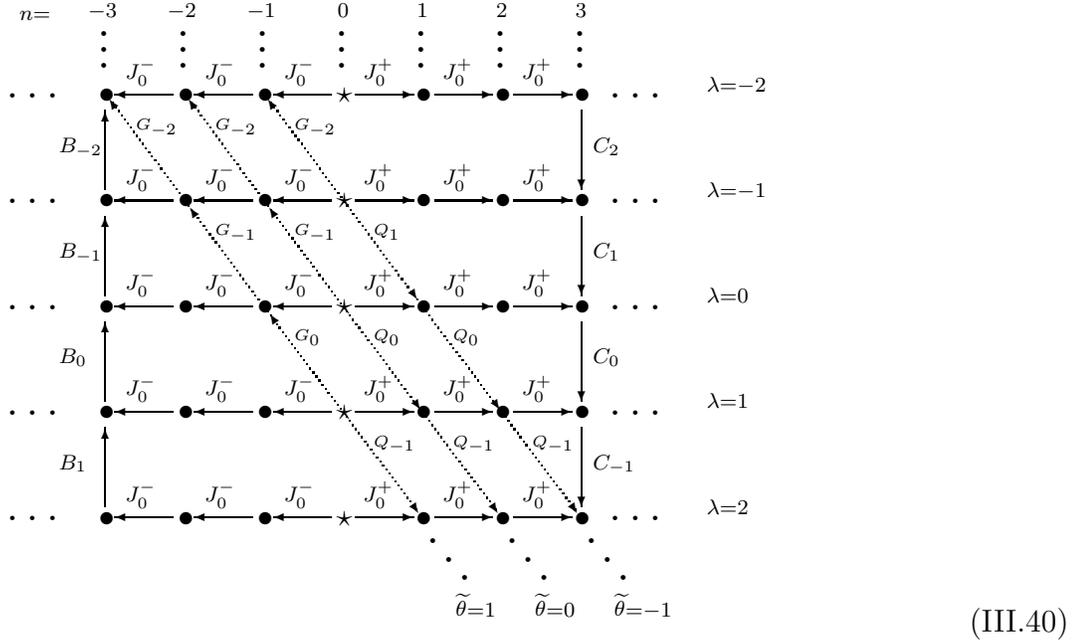
\begin{equation}
  \unitlength=1pt
  \begin{picture}(250,240)
    \put(0,120){
      \put(-30,107){${}^{n={}}$}
      \put(-10,95){
        \put(6,13){${}^{-3}$}
        \put(0,0){
          \put(9.5,-5){\Large $\cdot$}
          \put(9.5,1){\Large $\cdot$}
          \put(9.5,7){\Large $\cdot$}
          }
        }
      \put(20,95){
        \put(6,13){${}^{-2}$}
        \put(0,0){
          \put(9.5,-5){\Large $\cdot$}
          \put(9.5,1){\Large $\cdot$}
          \put(9.5,7){\Large $\cdot$}
          }
        }
      \put(50,95){
        \put(6,13){${}^{-1}$}
        \put(0,0){
          \put(9.5,-5){\Large $\cdot$}
          \put(9.5,1){\Large $\cdot$}
          \put(9.5,7){\Large $\cdot$}
          }
        }
      \put(80,95){
        \put(10.3,13){${}^{0}$}
        \put(0,0){
          \put(9.5,-5){\Large $\cdot$}
          \put(9.5,1){\Large $\cdot$}
          \put(9.5,7){\Large $\cdot$}
          }
        }
      \put(110,95){
        \put(10.3,13){${}^{1}$}
        \put(0,0){
          \put(9.5,-5){\Large $\cdot$}
          \put(9.5,1){\Large $\cdot$}
          \put(9.5,7){\Large $\cdot$}
          }
        }
      \put(140,95){
        \put(10.3,13){${}^{2}$}
        \put(0,0){
          \put(9.5,-5){\Large $\cdot$}
          \put(9.5,1){\Large $\cdot$}
          \put(9.5,7){\Large $\cdot$}
          }
        }
      \put(170,95){
        \put(10.3,13){${}^{3}$}
        \put(0,0){
          \put(9.5,-5){\Large $\cdot$}
          \put(9.5,1){\Large $\cdot$}
          \put(9.5,7){\Large $\cdot$}
          }
        }
%
      \put(2.4,7){\vector(0,1){30}}
      \put(187,17){${}^{C_{1}}$}
      \put(2.4,47){\vector(0,1){30}}
      \put(187,57){${}^{C_{2}}$}
      \put(2.4,-33){\vector(0,1){30}}
      \put(187,-23){${}^{C_{0}}$}
      \put(2.4,-73){\vector(0,1){30}}
      \put(187,-63){${}^{C_{-1}}$}
      \put(182.5,37){\vector(0,-1){30}}
      \put(-15,17){${}^{B_{-1}}$}
      \put(182.5,77){\vector(0,-1){30}}
      \put(-15,57){${}^{B_{-2}}$}
      \put(182.5,-3){\vector(0,-1){30}}
      \put(-15,-23){${}^{B_{0}}$}
      \put(182.5,-43){\vector(0,-1){30}}
      \put(-15,-63){${}^{B_{1}}$}
%
      \put(60,0){
        \put(-30,0){
          \put(154,-90){\Large $\cdot$}
          \put(160,-97){\Large $\cdot$}
          \put(166,-104){\Large $\cdot$}
          \put(165,-110){${}_{\widetilde\theta=-1}$}
          }
        \bezier{100}(5,80)(62.5,2)(120,-74)
        \put(5,80){\vector(-2,3){1}}
        \put(59.5,7){\vector(2,-3){1}}
        \put(90.7,-34.7){\vector(2,-3){1}}
        \put(120,-74){\vector(2,-3){1}}
        \put(44,30){${}_{{}^{Q_{1}}}$}
        \put(14,70){${}_{{}^{G_{-2}}}$}
        \put(74,-10){${}_{{}^{Q_{0}}}$}
        \put(104,-50){${}_{{}^{Q_{-1}}}$}
        }
      \put(30,0){
        \put(-30,0){
          \put(154,-90){\Large $\cdot$}
          \put(160,-97){\Large $\cdot$}
          \put(166,-104){\Large $\cdot$}
          \put(165,-110){${}_{\widetilde\theta=0}$}
          }
        \bezier{100}(5,80)(62.5,2)(120,-74)
        \put(5,80){\vector(-2,3){1}}
        \put(35.5,39){\vector(-2,3){1}}
        \put(90.7,-34.7){\vector(2,-3){1}}
        \put(120,-74){\vector(2,-3){1}}
        \put(44,30){${}_{{}^{G_{-1}}}$}
        \put(14,70){${}_{{}^{G_{-2}}}$}
        \put(74,-10){${}_{{}^{Q_{0}}}$}
        \put(104,-50){${}_{{}^{Q_{-1}}}$}
        }
      \put(0,0){
        \put(-30,0){
          \put(154,-90){\Large $\cdot$}
          \put(160,-97){\Large $\cdot$}
          \put(166,-104){\Large $\cdot$}
          \put(165,-110){${}_{\widetilde\theta=1}$}
          }
        \bezier{100}(5,80)(62.5,2)(120,-74)
        \put(5,80){\vector(-2,3){1}}
        \put(35.5,39){\vector(-2,3){1}}
        \put(65.7,-1.3){\vector(-2,3){1}}
        \put(120,-74){\vector(2,-3){1}}
        \put(44,30){${}_{{}^{G_{-1}}}$}
        \put(14,70){${}_{{}^{G_{-2}}}$}
        \put(74,-10){${}_{{}^{G_{0}}}$}
        \put(104,-50){${}_{{}^{Q_{-1}}}$}
        }
%
%
      \put(230,0){${}^{\lambda=0}$}
      \put(0,0){
        \put(-35,2){\Large $\ldots$}
        \put(0,0){$\bullet$}
        \put(10,5){${}^{J^-_0}$}
        \put(28,3){\vector(-1,0){22}}
        \put(30,0){$\bullet$}
        \put(40,5){${}^{J^-_0}$}
        \put(58,3){\vector(-1,0){22}}
        \put(60,0){$\bullet$}
        \put(70,5){${}^{J^-_0}$}
        \put(88,3){\vector(-1,0){22}}
%
        \put(90,0){$\star$}
        \put(100,5){${}^{J^+_0}$}
        \put(97,3){\vector(1,0){22}}
        \put(120,0){$\bullet$}
        \put(130,5){${}^{J^+_0}$}
        \put(127,3){\vector(1,0){22}}
        \put(150,0){$\bullet$}
        \put(160,5){${}^{J^+_0}$}
        \put(157,3){\vector(1,0){22}}
        \put(180,0){$\bullet$}
        \put(193,2){\Large $\ldots$}
        }
%
%
      \put(230,40){${}^{\lambda=-1}$}
      \put(0,40){
        \put(-35,2){\Large $\ldots$}
        \put(0,0){$\bullet$}
        \put(10,5){${}^{J^-_0}$}
        \put(28,3){\vector(-1,0){22}}
        \put(30,0){$\bullet$}
        \put(40,5){${}^{J^-_0}$}
        \put(58,3){\vector(-1,0){22}}
        \put(60,0){$\bullet$}
        \put(70,5){${}^{J^-_0}$}
        \put(88,3){\vector(-1,0){22}}
%
        \put(90,0){$\star$}
        \put(100,5){${}^{J^+_0}$}
        \put(97,3){\vector(1,0){22}}
        \put(120,0){$\bullet$}
        \put(130,5){${}^{J^+_0}$}
        \put(127,3){\vector(1,0){22}}
        \put(150,0){$\bullet$}
        \put(160,5){${}^{J^+_0}$}
        \put(157,3){\vector(1,0){22}}
        \put(180,0){$\bullet$}
        \put(193,2){\Large $\ldots$}
        }
%
      \put(230,-40){${}^{\lambda=1}$}
      \put(0,-40){
        \put(-35,2){\Large $\ldots$}
        \put(0,0){$\bullet$}
        \put(10,5){${}^{J^-_0}$}
        \put(28,3){\vector(-1,0){22}}
        \put(30,0){$\bullet$}
        \put(40,5){${}^{J^-_0}$}
        \put(58,3){\vector(-1,0){22}}
        \put(60,0){$\bullet$}
        \put(70,5){${}^{J^-_0}$}
        \put(88,3){\vector(-1,0){22}}
%
        \put(90,0){$\star$}
        \put(100,5){${}^{J^+_0}$}
        \put(97,3){\vector(1,0){22}}
        \put(120,0){$\bullet$}
        \put(130,5){${}^{J^+_0}$}
        \put(127,3){\vector(1,0){22}}
        \put(150,0){$\bullet$}
        \put(160,5){${}^{J^+_0}$}
        \put(157,3){\vector(1,0){22}}
        \put(180,0){$\bullet$}
        \put(193,2){\Large $\ldots$}
        }
%
%
      \put(230,80){${}^{\lambda=-2}$}
      \put(0,80){
        \put(-35,2){\Large $\ldots$}
        \put(0,0){$\bullet$}
        \put(10,5){${}^{J^-_0}$}
        \put(28,3){\vector(-1,0){22}}
        \put(30,0){$\bullet$}
        \put(40,5){${}^{J^-_0}$}
        \put(58,3){\vector(-1,0){22}}
        \put(60,0){$\bullet$}
        \put(70,5){${}^{J^-_0}$}
        \put(88,3){\vector(-1,0){22}}
%
        \put(90,0){$\star$}
        \put(100,5){${}^{J^+_0}$}
        \put(97,3){\vector(1,0){22}}
        \put(120,0){$\bullet$}
        \put(130,5){${}^{J^+_0}$}
        \put(127,3){\vector(1,0){22}}
        \put(150,0){$\bullet$}
        \put(160,5){${}^{J^+_0}$}
        \put(157,3){\vector(1,0){22}}
        \put(180,0){$\bullet$}
        \put(193,2){\Large $\ldots$}
        }
%
%
      \put(230,-80){${}^{\lambda=2}$}
      \put(0,-80){
        \put(-35,2){\Large $\ldots$}
        \put(0,0){$\bullet$}
        \put(10,5){${}^{J^-_0}$}
        \put(28,3){\vector(-1,0){22}}
        \put(30,0){$\bullet$}
        \put(40,5){${}^{J^-_0}$}
        \put(58,3){\vector(-1,0){22}}
        \put(60,0){$\bullet$}
        \put(70,5){${}^{J^-_0}$}
        \put(88,3){\vector(-1,0){22}}
%
        \put(90,0){$\star$}
        \put(100,5){${}^{J^+_0}$}
        \put(97,3){\vector(1,0){22}}
        \put(120,0){$\bullet$}
        \put(130,5){${}^{J^+_0}$}
        \put(127,3){\vector(1,0){22}}
        \put(150,0){$\bullet$}
        \put(160,5){${}^{J^+_0}$}
        \put(157,3){\vector(1,0){22}}
        \put(180,0){$\bullet$}
        \put(193,2){\Large $\ldots$}
        }
      }
  \end{picture}
  \label{lattice}
\end{equation}
This describes the case of $\theta=0$ in \req{relaxedidspaces} (which
is the $\tSl2$ `spectral' parameter, not to be confused with the one
labelling twisted $\N2$ modules that are also present in the diagram;
the $\N2$ spectral parameter is denoted by $\widetilde\theta$ here).
The diagram represents the tensor product of the relaxed Verma module
extremal diagram \req{floor} with a ghost extremal diagram. The latter
consists of ghost vacua~\req{lambdavac} taken in different pictures.
As we have chosen $\theta=0$, the $\tSL2$ arrows are horizontal, as
in~\req{floor}, while the ghost ones are shown vertical.  For
simplicity, the ghost ($B$ and $C$) arrows are shown explicitly only
in two columns.  Since the different pictures \req{lambdavac} in the
free-fermion system are all equivalent, all the vertical arrows are
invertible, however we have separated the $B$ and $C$ arrows in the
diagram, trying to keep it readable. Thus the~$\bullet$ dots
denote~$\ketSL{j,\Lambda,k|n}\tensor\ketGH{\lambda}$, the values of
$n$ being written in the upper row and those of $\lambda$, in the
right column.  The~$\star$s
denote~$\ketSL{j,\Lambda,k}\tensor\ketGH{\lambda}$.  Now, the dotted
lines are precisely the extremal diagrams~\req{massdiagramdouble} of
massive Verma modules~$\smU_{h,\ell,t;\widetilde\theta}$ from the RHS
of~\req{relaxedidspaces}, viewed from above (the grading with respect
to the level is in the direction orthogonal to the page).


Note also that the analogues of corollaries~1 and~2 from the previous
subsection are straightforward to formulate in the relaxed/massive
case.

\section{Equivalence between categories\label{sec:4}}
\subsection{Equivalence of Verma chain
  categories\label{subsec:chains}}\lvm It follows from the above
analysis that an equivalence between some categories of $\tSL2$ and
$\N2$ modules can only be established for those categories which
effectively allow for a factorization with respect to the spectral
flow.  These can be defined as follows.

  Let us consider the objects that are infinite chains
  $\left(\smM_{j,k;\theta}\right)_{\theta\in\oZ}$, where
  $\smM_{j,k;\theta}$ are twisted Verma modules.  As a morphisms
  between $\left(\smM_{j,k;\theta}\right)_{\theta\in\oZ}$ and
  $\left({\smM}_{j',k';\theta}\right)_{\theta\in\oZ}$, we take any
  Verma module morphisms $\smM_{j,k;\theta_1}\to
  {\smM}_{j',k';\theta_2}$.  Let us call this category the $\tSL2$
  Verma chain category~$\CVER$.

  On the $N=2$ side, the topological Verma chain category $\CTVER$ is
  defined similarly. Namely, one takes the chains of twisted
  topological Verma modules, every such chain comprising the modules
  that are the spectral flow transforms of a topological Verma module
  with all $\theta\in\oZ$.  Morphisms of the chains are defined
  similarly to the $\tSL2$ case.\footnote{It is important to note here
    that any submodule of a (twisted) topological Verma module is a
    twisted topological Verma module.}

  The meaning of the definition of morphisms of chains is that,
  obviously, given a morphism between any two modules, one spreads it
  over the entire chains by spectral flow transforms.

To define a functor relating such chains, we first construct
correspondences between individual modules in the chains:

  Given a topological Verma module $\smV_{h,t;\theta}$, and an
  arbitrary $\theta'\in\oZ$, take the Heisenberg modules
  $\mH^+_{-\sqrt{{2\over t}}(\frac{t}{2}j+\frac{t}{2}\theta'-\theta)}$
  and consider
  \begin{equation}
    \smV_{h,t;\theta}\tensor
    \mH^+_{-\sqrt{{2\over t}}(\frac{t}{2}j+\frac{t}{2}\theta'-\theta)}
    \oplus\bigoplus_{m\in\oZ,m\neq 0}\,
    \smV_{h,t;\theta+m}\tensor
    \mH^+_{-\sqrt{{2\over t}}(\frac{t}{2}j+\frac{t}{2}\theta'-\theta+m)}
    \label{FKS}
  \end{equation}
  where $\smV_{h,t;\theta+m}$ are the images of $\smV_{h,t;\theta}$
  under the spectral flow.  By theorem~\ref{weaker}, \req{FKS} is
  isomorphic to the tensor product of an $\tSL2$ Verma
  module~$\smM_{-\frac{t}{2}h,t-2;\theta'}$ with a ghost module.  We
  define the result of applying $F_{\rm KS}(\theta,\theta')$
  to~$\smV_{h,t;\theta}$ to be the
  module~$\smM_{-\frac{t}{2}h,t-2;\theta'}$:
  \begin{equation}
    F_{\rm KS}(\theta,\theta')\,:\,\smV_{h,t;\theta}
    \leadsto
    \smM_{-\frac{t}{2}h,t-2;\theta'}\,,\qquad\theta,\theta'\in\oZ\,.
    \label{FKSfunctor}
  \end{equation}

A similar definition can be given for
the correspondence
  \begin{equation}
    F^{-1}_{\rm KS}(\theta,\theta')\,:\,\smM_{j,k;\theta}
    \leadsto
    \smV_{-\frac{2}{k+2}j,k+2;\theta'}\,,\qquad\theta,\theta'\in\oZ\,.
    \label{IFKSM}
  \end{equation}
  Given a Verma module $\smM_{j,k;\theta}$ and $\theta'\in\oZ$, we
  construct the sum
  \begin{equation}
    \smM_{j,k;\theta}\tensor
    \mH^-_{-\sqrt{\frac{2}{k+2}}(j+\theta'-\frac{k+2}{2}\theta)}
    \oplus\bigoplus_{n\in\oZ,n\neq0}\,
    \smM_{j+\frac{k}{2}n,k;\theta+n}\tensor
    \mH^-_{-\sqrt{\frac{2}{k+2}}(j+\theta'-\frac{k+2}{2}\theta+n)}
    \label{IFKS}
  \end{equation}
  which, by theorem~\ref{weaker}, is isomorphic to the module
  $\smV_{-\frac{2}{k+2}j,k+2;\theta'}$ tensored with a module of
  antifermions.  This twisted topological $\N2$ module is by definition
  the result of applying $\FKS^{-1}(\theta,\theta')$
  to~$\smM_{j,k;\theta}$.

\medskip

While the correspondences from \req{FKSfunctor} and \req{IFKSM} depend
on the chosen $\theta$ and $\theta'$, the $\theta$-dependence
disappears when applied to the elements of $\CVER$ and $\CTVER$;
therefore $F_{\rm KS}(\,\cdot\,,\,\cdot\,)$ and $F^{-1}_{\rm
  KS}(\,\cdot\,,\,\cdot\,)$, which we denote again by $\FKS$ and
$\FKS^{-1}$, are candidates for the functors
$$\new\begin{array}{rclcl}
  \FKS&:&\CTVER&\leadsto&\CVER\\
  \FKS^{-1}&:&\CVER&\leadsto&\CTVER
\end{array}
$$
Evidently, the composition of $F_{\rm KS}$ and $F^{-1}_{\rm KS}$
maps each chain of twisted Verma modules into an isomorphic chain.
Therefore, $\FKS$ and $\FKS^{-1}$ would be the direct and inverse
functors, thus establishing the isomorphism of categories, once we
define how $F_{\rm KS}$ and $F^{-1}_{\rm KS}$ act on morphisms.

Recall that in Verma-module categories, morphisms are naturally
identified with singular vectors.  As we see from the isomorphisms of
section~\ref{subsec:isomorphism}, an $\tSL2$ singular vector exists in
a twisted Verma module $\smM_{j,k;\theta}$ (hence in all those with
$\theta\mapsto\theta+n$, $n\in\oZ$) if and only if a topological
singular vector exists in at least one (hence in each) twisted
topological Verma module~$\smV_{{-2j\over k+2},k+2;m}$, $m\in\oZ$,
from the RHS of~\req{idspaces}.  The following lemma gives an explicit
mapping between the `building blocks' of the corresponding singular
vectors:
\begin{lemma}\label{mapweyl}
  The KS mapping induces a correspondence between the `continued' \
  objects, $(J^-_{-\theta})^{\nu-\mu+1}$ and
  $(J^+_{\theta-1})^{\nu-\mu+1}$ on the one hand, and $g(\mu,\nu)$,
  $q(\mu,\nu)$ on the other hand, which act on the respective highest
  weights as shown in~\req{sl2weylaction} and~\req{n2weylaction}.  The
  correspondence reads
  $$\new
    \begin{array}{ll}
      F_{\rm KS}(\theta',\theta)\,:\,g(\mu,\nu)\mapsto
      (J^-_{-\theta})^{\nu-\mu+1}b(\mu,\nu)\,,&
      F_{\rm KS}(\theta',\theta)\,:\,q(\mu,\nu)\mapsto
      (J^+_{\theta-1})^{\nu-\mu+1}c(\mu,\nu)\,,\\
      F^{-1}_{\rm KS}(\theta,\theta')\,:\,(J^-_{-\theta})^{\nu-\mu+1}\mapsto
      g(\mu,\nu)e^{-(\nu-\mu+1)\phi}\,,&
      F^{-1}_{\rm KS}(\theta,\theta')\,:\,(J^+_{\theta-1})^{\nu-\mu+1}\mapsto
      q(\mu,\nu)e^{(\nu-\mu+1)\phi}\,.
    \end{array}
  $$
\end{lemma}
This leads us to
\begin{thm}
  The KS and anti-KS mappings give rise to identifications, which we
  denote again by $F_{\rm KS}(\theta,\theta')$ and $F^{-1}_{\rm
    KS}(\theta,\theta')$ respectively, between the twisted $\tSL2$
  singular vectors~\req{mffplus} in the Verma module
  $\smM_{j,k;\theta}$ and the $\N2$ singular vectors~\req{Tplus}
  and~\req{Tminus} $\ket{E(r,s,t)}^{\pm,\theta}$ in the twisted
  topological Verma modules $\smV_{-\frac{2}{k+2}j,k+2;\theta}$:
  \begin{equation}\new\begin{array}{l}
      F_{\rm KS}(\theta,\theta')\,:\,\ket{E(r,s,k+2)}^{\pm,\theta}
      \mapsto
      \ket{S_\pm^{\rm MFF}(r,s,k)}^{\theta'},\\
      F^{-1}_{\rm KS}(\theta,\theta')\,:\,
      \ket{S_\pm^{\rm MFF}(r,s,k)}^{\theta}
      \mapsto
      \ket{E(r,s,k+2)}^{\pm,\theta'},
    \end{array}\label{KSonvectors}
  \end{equation}
  where $\ket{E(r,s,t)}^{\pm,\theta}$ and $\ket{S_\pm^{\rm
      MFF}(r,s,k)}^{\theta'}$ denote the singular vectors transformed
  by the corresponding spectral flows.
\end{thm}

Evidently, $\FKS(\,\cdot\,,\,\cdot\,)$ and
$\FKS^{-1}(\,\cdot\,,\,\cdot\,)$ applied to the chains of the
respective Verma modules take morphisms (between chains) into
morphisms. Thus we have made $\FKS$ and $\FKS^{-1}$ into functors.
The results of this subsection can be summarized in the following
theorem:
\begin{thm}
  The functors $F_{\rm KS}$ and $F^{-1}_{\rm KS}$ are covariant
  functors which are inverse to each other and which therefore
  establish the equivalence between the Verma chain category~$\CVER$
  on the $\tSL2$ side and the topological Verma chain
  category~$\CTVER$ on the $\N2$ side.
\end{thm}

\subsection{Extending the equivalence to \hw-type categories}\lvm
We have seen in the previous subsection that the $\tSL2$ Verma, and
$\N2$ topological Verma chain categories are equivalent.  Now we would
like to extend the equivalence to larger categories.  Let us define
the \hw{} type chain category:
\begin{dfn}\label{CHW:def}
  The \hw{} type chain category $\CHW$ is the category whose objects
  are chains of modules from the category $\HW$ of $\tSL2$
  highest-weight type representations twisted by the spectral flow
  with integer parameters. A morphism between two such chains
  $(\cN_\theta)_{\theta\in\oZ}$ and $(\cN'_\theta)_{\theta\in\oZ}$ is
  any morphism $\cN_{\theta_1}\to\cN'_{\theta_2}$ (which is understood
  to be spread over the entire chains by the spectral flow).
\end{dfn}
\begin{rem}
  As an aside, note that the condition for any such chain to admit
  non-trivial automorphisms singles out the {\it minimal\/} $\SL2$
  models, and the same is true for the chains of $\N2$ modules.  Then,
  taking the factors with respect to the respective spectral flows, we
  would have the same number of fields in the $\SL2$ and $\N2$ minimal
  models. Thus, the minimal models of these algebras can be thought of
  as {\it periodical chains\/} of representations; we hope to
  demonstrate elsewhere the usefulness of such a viewpoint.
\end{rem}

On the $\N2$ side, these twisted categories mix into one category
because submodules of a (twisted) topological Verma module are the
twisted topological Verma modules with different~$\theta$.
\begin{dfn}\label{CTOP:dfn}
  The topological chain category $\CTOP$ is the category whose objects
  are chains of modules from the topological category $\TOP$ twisted
  by the spectral flow with integer parameters.  A morphism between
  two such chains $(\cU_\theta)_{\theta\in\oZ}$ and
  $(\cU'_\theta)_{\theta\in\oZ}$ is any morphism
  $\cU_{\theta_1}\to\cU'_{\theta_2}$ of $\N2$ modules.
\end{dfn}
\begin{thm}\label{CHWCTOP:thm}
  The functors $\FKS$ and $\FKS^{-1}$ establish the equivalence
  between categories $\CHW$ and $\CTOP$.
\end{thm}

\subsection{Equivalence of the relaxed and massive Verma chain
  categories}\lvm We define chains of relaxed Verma modules on the
$\tSL2$ side and chains of massive Verma modules on the $\N2$ side as
this was done in Sec.~\ref{subsec:chains} Then, we formulate the
theorem, whose proof is a consequence of Lemma~\ref{mapweyl}:
\begin{thm}
  The KS and anti-KS mappings gives rise to identifications, which we
  denote again by $F_{\rm KS}(\theta,\theta')$ and $F^{-1}_{\rm
    KS}(\theta,\theta')$ respectively, between
  \begin{enumerate}
  \item[\rm 1)] singular vectors~\req{sigmaminus} and~\req{sigmaplus}
    in the relaxed Verma module $\smR_{j,\Lambda,k;\theta}$ and $\N2$
    singular vectors~\req{Sgen}, $\ket{S(r,s,h,t)}^{\mp,\theta}$, in
    the twisted massive Verma modules
    $\smU_{-\frac{2}{k+2}j,\frac{\Lambda}{k+2},k+2;\theta}$:
    \begin{equation}\new
      \begin{array}{l}
        F_{\rm KS}(\theta,\theta')\,:\,\ket{\Sigma^\pm(r,s,j,k)}^\theta
        \mapsto
        \ket{S(r,s,\frac{-2j}{k+2},k+2)}^{\mp,\theta'},\\
        F^{-1}_{\rm KS}(\theta,\theta')\,:\,
        \ket{S(r,s,h,t)}^{\mp,\theta}
        \mapsto
        \ket{\Sigma^\pm(r,s,-\frac{t}{2}h,t-2)}^{\theta'},
      \end{array}
    \end{equation}
  \end{enumerate}
  and also between
  \begin{enumerate}
  \item[\rm 2)]~charged $\tSL2$ singular vectors~\req{chargedsl2} and
    charged $\N2$ singular vectors~\req{thirdE}:
    \begin{equation}\new
      \begin{array}{l}
        F_{\rm KS}(\theta,\theta')\,:\,\ket{C(r,j,k)}^\theta
        \mapsto
        \ket{E(r,h,t)}_{\rm ch}^{\theta'},\\
        F^{-1}_{\rm KS}(\theta,\theta')\,:\,
        \ket{E(r,h,t)}_{\rm ch}^{\theta}
        \mapsto
        \ket{C(r,j,k)}^{\theta'},
      \end{array}
    \end{equation}
    where $\ket{\Sigma^\pm(r,s,j,k)}^\theta$ and
    $\ket{S(r,s,h,t)}^{\mp,\theta'}$ denote the respective singular
    vectors transformed by the corresponding spectral flows.
  \end{enumerate}
\end{thm}
Again, as in Sec.~\ref{subsec:chains}, we can formulate
\begin{thm}
  The functors $F_{\rm KS}$ and $F^{-1}_{\rm KS}$ are covariant
  functors which are inverse to each other and which therefore
  establish the equivalence of the relaxed Verma chain
  category~$\CRVER$ on the $\tSL2$ side and the massive Verma chain
  category~$\CMVER$ on the $\N2$ side.
\end{thm}

\subsection{Extending the equivalence to the relaxed \hw-type
  categories}\lvm We now formulate an extension of the categorial
equivalence $F_{\rm KS}$ and $F^{-1}_{\rm KS}$ to the case of the
largest categories $\CRHW$ and $\CMHW$ in diagram~\req{square}.  The
definitions of the chain categories $\CRHW$ and $\CMHW$ can be
obtained by means of a step-by-step repetition of
Definitions~\ref{CHW:def} and~\ref{CTOP:dfn} and are omitted.  Then
the analog of Theorem~\ref{CHWCTOP:thm} reads
\begin{thm}
  The functors \ $\FKS$ \ and \ $\FKS^{-1}$ \ establish the
  equivalence between categories $\CRHW$ and $\CMHW$.
\end{thm}

\section{Concluding remarks\label{sec:5}}\lvm
We believe that the issues addressed in this paper will be essential
in a number of problems in the $\tSL2$ and $\N2$ representation
theories as well as in conformal field theory.

Comparing the $\tSL2$ and $\N2$ \hw{} representations, it may be
observed that the occurrence of infinitely many \hw{} states in
$\tSL2$ relaxed Verma modules is intuitively more `obvious' than in
its $\N2$ counterpart; all of the $\tSL2$ relaxed-\hw{} states are on
the same floor in the extremal diagram, whereas in the $\N2$ case this
is not so: extremal diagrams of untwisted massive $\N2$ modules
contain only two vectors at the top level, and it may be tempting to
assign to these a more fundamental status than to the other vectors
from the extremal diagram.  Working with the top-level representative
of $\N2$ extremal diagrams conceals the `topological' nature of some
of submodules in massive $\N2$ Verma modules (see~\cite{[ST4]} for a
detailed description of the structure of $\N2$ Verma modules).

As to the structure of relaxed-$\tSL2$ or massive $\N2$ Verma modules,
we have seen that a significant difference from the familiar Virasoro
or the usual-$\tSL2$ case is that (to use the $\N2$ language) a given
massive $\N2$ Verma module may contain submodules of two types, the
twisted topological and the massive ones. An important point is that
every submodule can be {\it freely\/} generated from a twisted
(topological) \hw{} vector (singular vectors~\req{Tplus} and
\req{Tminus} in the topological Verma modules and \req{thirdE} and
\req{Sgen} in massive Verma modules). From these results, it is not
too difficult to derive the complete classification of embedding
diagrams of the relaxed-$\tSL2$ and massive-$\N2$ Verma
modules~\cite{[SSi]}.  These follow (an extension of) the familiar
I-II-III pattern; the embedding structure can be described without
invoking subsingular vectors~\cite{[ST4]}, since the above singular
vectors generate maximal submodules (that subsingular vectors in
massive-$\N2$/relaxed-$\tSL2$ modules are superfluous can most clearly
be seen in the $\tSL2$ language).

Another problem is the derivation of fusion rules. In order to
completely define the fusion rules one has to fix, among other things,
the category of representations.  On the $\tSL2$ side, the fusion
rules have been obtained in \cite{[AY],[FM-fusion],[Andreev]}.
However, these do not include the twisted (spectral-flow-transformed)
$\tSL2$ representations, whose contribution is to be expected on
general grounds~\cite{[FM-fusion]}.  This is related to the fact that
only the quantum group $osp(1|2)_q$ has so far been observed as a
symmetry of $\tSL2$ fusion rules, while a larger quantum group
$\SSL21_q$ is to be expected. This is an indication that some modules
are still missing from the known fusion; we hope that we have
described the representation-theoretic part of the desired more
general construction. Then the complete $\tSL2$ fusion rules factored
with respect to the spectral flow would be isomorphic to complete
$\N2$ fusion rules factored over the $\N2$ spectral flow.

\paragraph{Acknowledgements.}  We wish to thank F.~Malikov and
V.~Sirota for useful remarks. AMS is also grateful to C.~Preitschopf
for discussions. The work of AMS and IYT was supported in part by RFFI
Grant 96-01-00725, the work of IYT was supported in part by a Landau
Foundation grant, and the work of AMS, by grant \#93-0633-ext from the
European Community.

\small 


\begin{thebibliography}{99}
  \parindent=0pt \parskip=-2pt

\bibitem{[AY]} H.~Awata and Y.~Yamada, {\it Fusion Rules for the
    Fractional Level sl(2) Algebra}, \MPLA7, 1185 (1992).

\bibitem{[FM-fusion]}B.~Feigin and F.~Malikov, {\it Modular
    Functor and Representation Theory of $\widehat{sl}_2$}, Cont.\
  Math.\ 202 ``{\sl Operads: Proceedings of Reneissance
    Conferences\/}".

\bibitem{[Ade]} M.~Ademollo, L.~Brink, A.~D'Adda, R.~D'Auria,
  E.~Napolitano, S.~Sciuto, E.~Del~Guidice, P.~Di~Vecchia, S.~Ferrara,
  F.~Gliozzi, R.~Musto, and R.~Pettorino, {\it Dual String With U(1)
    Color Symmetry,\/} \PLB62, 105 (1976);\\ M.~Ademollo, L.~Brink,
  A.~D'Adda, R.~D'Auria, E.~Napolitano, S.~Sciuto, E.~Del~Guidice,
  P.~Di~Vecchia, S.~Ferrara, F.~Gliozzi, R.~Musto, R.~Pettorino, and
  J.H.~Schwarz, {\it Dual String Models With Nonabelian Color and
    Flavor Symmetries\/}, \NPB111, 77 (1976).

\bibitem{[Andreev]} O.~Andreev, {\it Operator Algebra of the
    $SL(2)$ Conformal Field Theories\/},
  \PLB363, 166 (1995).

\bibitem{[PRY]} J.L.~Petersen, J.~Rasmussen, and M.~Yu, {\it Fusion,
    Crossing and Monodromy in Conformal Field Theory Based on $SL(2)$
    Current Algebra with Fractional Level}, \NPB481, 577-624 (1996).

\bibitem{[FGP]} P.~Furlan, A.Ch.~Ganchev, and V.B.~Petkova, {\it
    $A^{(1)}_1$ admissible representations--fusion transformations and
    local correlators}, \NPB491, 635 (1997).

\bibitem{[S-sing]} A.M.~Semikhatov, {\it The MFF singular vectors
    in topological conformal theories\/}, \MPLA9, 1867 (1994).

\bibitem{[TheBook]} V.G.~Ka\v c {\sl Infinite Dimensional Lie
    Algebras\/}, Cambridge University Press 1990.

\bibitem{[ST2]}A.M.~Semikhatov and I.Yu.~Tipunin, {\it Singular
    Vectors of the Topological Conformal Algebra\/}, \IJMPA11, 4597
  (1996).

\bibitem{[ST3]}A.M.~Semikhatov and I.Yu.~Tipunin, {\it All Singular
    Vectors of the $N\!=\!2$ Superconformal Algebra via the Algebraic
    Continuation Approach\/}, hep-th/9604176.

\bibitem{[BH]} K.~Bardak\c ci and M.B.~Halpern, Phys.\ Rev.\ D3, 2493
  (1971).

\bibitem{[BFK]} W.~Boucher, D.~Friedan, and A.~Kent, {\it
    Determinant formulae and Unitarity for the N=2 Superconformal
    Algebras in Two-Dimensions Or Exact Results on String
    Compactification}, \PLB172,  316 (1986).

\bibitem{[MFF]} F.G.~Malikov, B.L.~Feigin, and D.B.~Fuchs, {\it
    Singular Vectors in Verma Modules over Ka\v c--Moody Algebras},
  Funk.\ An.\ Prilozh.\ 20 N2, 25 (1986).

\bibitem{[ST4]}A.M.~Semikhatov and I.Yu.~Tipunin, {\it The
    Structure of Verma Modules over the $N=2$ Superconformal
    Algebra\/}, hep-th/9704111. Commun.\ Math.\ Phys., to appear.

\bibitem{[SS]}A.~Schwimmer and N.~Seiberg, {\it Comments on the
    $N=2$, $N=3$, $N=4$ Superconformal Algebras in Two-Dimensions},
  \PLB184 (1987) 191.

\bibitem{[LVW]}W.~Lerche, C.~Vafa, and N.P.~Warner, {\it Chiral Rings
    in $N=2$ Superconformal Theories}, \NPB324, 427 (1989).

\bibitem{[FS]}
  B.L.~Feigin and A.V.~Stoianovsky {\it Functional Models of
    Representations of Current Algebras and Semi-infinite Schubert
    Cells},
  Funk. An. i ego prilozh., 28(1), 68 (1994).

\bibitem{[S-sl21sing]} A.M.~Semikhatov, {\it Verma Modules,
    Extremal Vectors, and Singular Vectors on the Non-Critical $\N2$
    String Worldsheet}, hep-th/9610084.

\bibitem{[DvPYZ]} P.~Di~Vecchia, J.L.~Petersen, M.~Yu, and
  H.B.~Zheng, \PLB174, 280 (1986).

\bibitem{[KS]}Y.~Kazama and H.~Suzuki, \NPB321, 232 (1989).

\bibitem{[KK]}V.G.~Ka\v{c} and D.A.~Kazhdan, {\it Structure of
    Representations with Highest Weight of Infinite-Dimensional Lie
    Algebras}, Adv. Math.~34, 97 (1979).

\bibitem{[Mal]}F.~Malikov, Algebra i Analiz, 2 No.~2, 65 (1990).

\bibitem{[FMS]}D.H.~Friedan, E.J.~Martinec, and S.H.~Shenker,
  {\it Conformal Invariance, Supersymmetry and String Theory},
  \NPB271, 93 (1986).

\bibitem[EHY]{[EHy]}T.~Eguchi, S.~Hosono, and S.-K.~Yang, {\it Hidden
    Fermionic Symmetry in Conformal Topological Field Theories},
  Commun.  Math. Phys.  140, 159 (1991).

\bibitem{[GS2]} B.~Gato-Rivera and A.M.~Semikhatov, {\it
    $d\leq1\cup d \geq 25$ and W Constraints From BRST-Invariance in
    the $c\neq3$ Topological Algebra}, \PLB293, 72 (1992).

\bibitem{[BLNW]}M.~Bershadsky, W.~Lerche, D.~Nemeschansky, and
  N.P.~Warner, {\it Extended $N=2$ Superconformal Structure of Gravity
    and W Gravity Coupled to Matter}, Nucl. Phys. B401, 304--347
  (1993).

\bibitem{[SSi]} A.M.~Semikhatov and V.A.~Sirota, {\it Embedding
    Diagrams of $\N2$ and Relaxed-$\tSL2$ Verma Modules},
  hep-th/9712102.

\end{thebibliography}
\end{document}